\documentclass[a4paper,11pt]{article}
\pdfoutput=1 

\usepackage{jheppub} 
\usepackage{natbib}
\usepackage{times}
\usepackage{float}
\usepackage{amsmath}
\usepackage{graphicx}
\usepackage{amssymb}
\usepackage{slashed}
\usepackage{xcolor}
\usepackage{soul} 
\usepackage{multirow}
\usepackage{bbold}
\usepackage{booktabs}
\usepackage[T1]{fontenc} 
\usepackage[utf8]{inputenc}

\allowdisplaybreaks

\makeatletter
\g@addto@macro\bfseries{\boldmath}
\makeatother

\title{Charm Physics Confronts High-$p_T$ Lepton Tails}

\author[a]{Javier Fuentes-Mart\'in,}
\author[b]{Admir Greljo,}
\author[c, d]{Jorge Martin Camalich,}
\author[e]{Jos\'e David Ruiz-Alvarez}

\affiliation[a]{Physik-Institut, Universit\"{a}t Z\"{u}rich, CH-8057 Z\"{u}rich, Switzerland}
\affiliation[b]{CERN, Theoretical Physics Department, CH-1211 Geneva 23, Switzerland}
\affiliation[c]{Instituto de Astrof\'isica de Canarias, C/ V\'ia L\'actea, s/n
E38205 - La Laguna, Tenerife, Espa\~na}
\affiliation[d]{Universidad de La Laguna, Departamento de Astrof\'isica, La Laguna, Tenerife, Spain}
\affiliation[e]{Instituto de F\'{i}sica, Universidad de Antioquia, A.A. 1226 Medell\'{i}n, Colombia.}

\emailAdd{fuentes@physik.uzh.ch} 
\emailAdd{admir.greljo@cern.ch}
\emailAdd{jorge.martin.camalich@cern.ch} 
\emailAdd{jose.ruiz@cern.ch}

\preprint{CERN-TH-2020-047, ZU-TH 07/20}

\abstract{
We present a systematic survey of possible short-distance new-physics effects in (semi)leptonic charged- and neutral-current charmed meson decays. Using the Standard Model Effective Field Theory (SMEFT) to analyze the most relevant experimental data at low and high energies, we demonstrate a striking complementarity between charm decays and high invariant mass lepton tails at the LHC. Interestingly enough, high-$p_T$ Drell-Yan data offer competitive constraints on most new physics scenarios. Furthermore, the full set of correlated constraints from $K$, $\pi$ and $\tau$ decays imposed by $SU(2)_L$ gauge invariance is considered. 
The bounds from $D_{(s)}$ decays, high-$p_T$ lepton tails and $SU(2)_L$ relations chart the space of the SMEFT affecting semi(leptonic) charm flavor transitions.
}

\arxivnumber{2003.12421}

\begin{document}

\maketitle

\newpage

\section{Introduction}

Understanding flavor transitions in the up-quark sector may prove crucial for unraveling the flavor puzzle and unveiling physics beyond the Standard Model (SM). 
A promising line in this direction is the investigation of transitions involving charmed hadrons.  
The recent discovery of direct CP violation in $D$ mesons decays~\cite{Aaij:2019kcg} illustrates the maturity of this field and its potential to lead to new discoveries in the near future.
In fact, an unprecedented amount of data on charm decays is expected to be collected at BES~III~\cite{Ablikim:2019hff}, LHCb~\cite{Cerri:2018ypt} and Belle~II~\cite{Kou:2018nap} experiments. Could this experimental program provide a charming gateway to new physics?

Leptonic and semileptonic charmed meson decays are an important benchmark in this program. These are exploited to determine the CKM matrix elements~\cite{Amhis:2019ckw,Aoki:2019cca} and {have been shown to be} sensitive probes of New Physics (NP)~\cite{Barranco:2013tba,Fajfer:2015ixa,Fleischer:2019wlx}. On the other hand, the interpretation of hadron weak decays requires calculations of hadronic matrix elements in lattice QCD which in the charm sector are becoming available with increasing precision~\cite{Carrasco:2014poa,Bazavov:2017lyh,Aubin:2004ej,Na:2010uf,Na:2011mc,Lubicz:2017syv,Lubicz:2018rfs,Aoki:2019cca}. Neutral-current decays are a priori more sensitive to NP because of the strong GIM suppression of the short-distance contributions in the up-quark sector~\cite{Burdman:2001tf,Paul:2011ar,Cappiello:2012vg,deBoer:2015boa,Fajfer:2015mia,Bause:2019vpr,Fajfer:2015zea}. Nonetheless, these are typically dominated by long-distance hadronic effects, which are difficult to treat from first principles~\cite{Fajfer:2015zea,Silvestrini:2015kqa,Fajfer:1997bh,Fajfer:1998dv,deBoer:2017que}, hampering a direct theoretical interpretation of the data in terms of short-distance physics. In principle, both charged- and neutral-current decays could be affected by NP and the recent example of the $B$-meson  anomalies~\cite{Aaij:2013qta,Aaij:2014ora,Aaij:2015oid,Aaij:2017vbb,Aaij:2019wad,Lees:2012xj,Huschle:2015rga,Aaij:2015yra,Aaij:2017tyk,Aaij:2017uff,Abdesselam:2019dgh} prompts us to be open about the possible forms in which they could appear. 

These anomalies have also fostered a more direct interplay between the traditional program of flavor physics at low energies and searches of NP in high-$p_T$ tails at the LHC. Crossing symmetry allows one to connect univocally the decay and scattering amplitudes. If the NP scale is quite higher than the energies reached in the respective physical processes, this connection can be established model-independently using effective operators for the NP interactions. In high-energy proton-proton collisions, heavy flavors are virtually present in the initial states and can be produced in the final states. As required by unitarity arguments, above the electroweak (EW) scale the SM scattering  amplitudes drop with energy while effective NP contributions keep growing. This energy-growing effect can compensate for the lower statistics in the high-$p_T$ tails, and provide competitive probes to the traditional low-energy high-intensity program. This will become especially relevant with the upcoming high-luminosity phase at the LHC (HL-LHC)~\cite{Cerri:2018ypt}. 

The importance of combining low-energy data and high-$p_T$ LHC data to constrain flavor-changing interactions has been already pointed out for the three light quarks~\cite{Cirigliano:2012ab,Gonzalez-Alonso:2016etj,Cirigliano:2018dyk}, the bottom quark~\cite{Faroughy:2016osc,Greljo:2017vvb,Altmannshofer:2017poe,Greljo:2018tzh,Baker:2019sli} and lepton-flavor violating interactions~\cite{Bhattacharya:2018ryy,Angelescu:2020uug}, while there has not been a study devoted to the reach of this program in the charm sector. 

In this work, we fill this gap by providing a comprehensive study of the interplay between the analyses of charmed-meson (semi)leptonic decays and high-$p_T$ lepton tails at the LHC. In particular, we systematically explore the sensitivity of these experiments to possible short-distance NP in the charm sector using the Standard Model effective field theory (SMEFT)~\cite{Buchmuller:1985jz,Grzadkowski:2010es}. The SMEFT provides a theoretical framework to describe NP effects originating above the EW scale, which is well-motivated given the lack of direct observation of new resonances at the LHC, and the consistency of the observed properties of the Higgs boson with the SM. Using the SMEFT, we can establish a link between charm decays and the production of high-$p_T$ leptons at the LHC. Moreover, due to its manifest $SU(2)_L$ gauge invariance, this framework allows to establish correlations with kaon and tau physics.
 
The next four sections investigate, in steps, charged-current transitions. Namely, starting from the effective field theory setup in Section~\ref{sec:eft}, we study the set of constraints from charmed meson decays in Section~\ref{sec:D}, the production of monoleptons at high-$p_T$ LHC in Section~\ref{sec:LHC} and, finally, compare the two in Section~\ref{sec:interplay}. The analysis is then repeated for neutral-current transitions in Section~\ref{sec:NC}. Complementary constraints implied by $SU(2)_L$ gauge symmetry are derived in Section~\ref{sec:SU2}. We conclude in Section~\ref{sec:Conc}.

\section{Theoretical framework: \texorpdfstring{$c \to d^i\bar e^\alpha \nu^\beta$}{c->d(s)lnu}}
\label{sec:eft}

\subsection{The high-energy effective theory}

We focus on short-distance NP that can affect semileptonic charged-current charm transitions, particularly when charm number changes by one unit, $\Delta C =1$. Under the assumption of no new degrees of freedom below (or at) the electroweak scale, NP effects can be fully described employing the SMEFT. The relevant Lagrangian is
\begin{align} \label{eq:SMEFTlag}
\mathcal{L}_{\rm SMEFT} \supset \frac{1}{v^2}\sum_k \mathcal{C}_k\,\mathcal{O}_k \,,
\end{align}
where $v\approx246$~GeV is the SM Higgs vacuum expectation value and the Wilson coefficients (WCs) scale as $\mathcal C_k\propto v^2/\Lambda^2$, with $\Lambda$ being the scale of NP. We employ the \textit{Warsaw basis}~\cite{Grzadkowski:2010es} for operators of canonical dimension six, which is particularly suited for flavor physics as covariant derivatives and field strengths are reduced in favor of fermionic currents using the equations of motion. The most general set of semileptonic four-fermion SMEFT operators contributing to $c\to d^i\bar e^\alpha\nu^\beta$ transitions are
\begin{align}\label{eq:SMEFTop4F}
\begin{aligned}
\mathcal{O}_{lq}^{(3)}&=(\bar l_L\gamma_\mu\tau^I l_L)(\bar q_L\gamma^\mu\tau^I q_L)\,,\qquad\qquad&
\mathcal{O}_{ledq}&=(\bar l_L e_R)(\bar d_R q_L)\,,\\
\mathcal{O}_{lequ}^{(1)}&=(\bar l_L^p e_R)\epsilon_{pr}(\bar q_L^r u_R)\,,\qquad\qquad&
\mathcal{O}_{lequ}^{(3)}&=(\bar l_L^p\sigma_{\mu\nu} e_R)\epsilon_{pr}(\bar q_L^r\sigma^{\mu\nu} u_R)\,,
\end{aligned}
\end{align}
with $\sigma^{\mu\nu}=\frac{i}{2}[\gamma^\mu,\gamma^\nu]$, $\tau^I$ the Pauli matrices, $\epsilon_{pr}$ the Levi-Civita symbol and $\{p,r\}$ being $SU(2)_L$ indices.\footnote{The SM extended by a light right-handed neutrino ($\nu_R$) potentially accessible in charm decays would require supplementing the SMEFT with a new set of operators such as $(\bar l_L \nu_R)(\bar u_R q_L)$. For the full list see Eq.~(2.1) in Ref.~\cite{Greljo:2018ogz}.} The left-handed quark and lepton doublets are denoted by $q_L$ and $l_L$, respectively, while the right-handed singlets are $u_R$, $d_R$ and $e_R$. On the other hand, the SMEFT operators that modify the $W$ {couplings to quarks} read
\begin{align}\label{eq:SMEFTop2F}
\mathcal{O}_{\phi q}^{(3)}&=( \phi^\dagger\,i\!\stackrel{\leftrightarrow}{D^I}_{\!\!\!\mu}\phi)(\bar q_L \gamma^\mu \tau^I q_L)\,,\qquad\qquad&
\mathcal{O}_{\phi ud}&=(\tilde\phi^\dagger\,iD_\mu\phi)(\bar u_R\gamma^\mu d_R)\,,
\end{align}
where $\phi$ is the Higgs field and $D_\mu$ its covariant derivative. We neglect the chirality-flipping $W$ vertices of the type $\bar \psi \sigma^{\mu \nu} \psi \, \phi  F_{\mu \nu}$. Their effects are subleading relative to the operators in Eq.~\eqref{eq:SMEFTop2F} at low-energies, since they are charm mass suppressed, and to the operators in Eq.~\eqref{eq:SMEFTop4F} at high-$p_T$, due to their different high-energy behavior discussed in Section~\ref{sec:tails}. We also neglect all modifications to the leptonic $W$ vertices, since they are better probed in purely leptonic transitions.

Thus, the operators in Eqs.~\eqref{eq:SMEFTop4F} and \eqref{eq:SMEFTop2F} capture all leading effects in the SMEFT in semileptonic charm transitions. Unless stated otherwise, throughout this paper we work in the up-basis for the $SU(2)_L$ multiplets, where
\begin{align}
\label{eq:UpBasis}
q_L^i&=\begin{pmatrix} u_L^i \\ V_{ij}\,d_L^j\end{pmatrix}\,,&
l_L^\alpha&=\begin{pmatrix} \nu_L^\alpha \\ e_L^\alpha\end{pmatrix}\,,
\end{align}
with $V$ the CKM matrix, 
{and} use $i,j=1,2,3$ and $\alpha,\beta = 1,2,3$ to label quark and lepton flavor indices, respectively. We also use $\ell$ to denote the light leptons $e$ and $\mu$, but not $\tau$. The matching of the SMEFT to the low-energy effective theory is reported next, while we postpone the discussion of $SU(2)_L$ relations to Section~\ref{sec:SU2}.

\subsection{The low-energy effective theory}
\label{sec:loweft}

The low-energy effective Lagrangian involving $c\to d^i\bar e^\alpha\nu^\beta$ transitions can be written as
\begin{align}\label{eq:LagCC}
\mathcal{L}_{\rm CC}=-\frac{4G_F}{\sqrt{2}}V_{ci}\left[\big(1+\epsilon_{V_L}^{\alpha\beta i}\big)\,\mathcal{O}_{V_L}^{\alpha\beta i}+\epsilon_{V_R}^{\alpha\beta i}\,\mathcal{O}_{V_R}^{\alpha\beta i}+\epsilon_{S_L}^{\alpha\beta i}\,\mathcal{O}_{S_L}^{\alpha\beta i}+\epsilon_{S_R}^{\alpha\beta i}\,\mathcal{O}_{S_R}^{\alpha\beta i}+\epsilon_T^{\alpha\beta i}\,\mathcal{O}_T^{\alpha\beta i}\right] + \rm{h.c.},
\end{align}
where the effective operators read
\begin{align}\label{eq:OpsCC}
\begin{aligned}
\mathcal{O}_{V_L}^{\alpha\beta i}&=(\bar e_L^\alpha\gamma_\mu \nu_L^\beta)(\bar c_L\gamma^\mu d_L^i)\,, \qquad\qquad&
\mathcal{O}_{V_R}^{\alpha\beta i}&=(\bar e_L^\alpha\gamma_\mu \nu_L^\beta)(\bar c_R\gamma^\mu d_R^i)\,, \\
\mathcal{O}_{S_L}^{\alpha\beta i}&=(\bar e_R^\alpha \,\nu_L^\beta)(\bar c_R\, d_L^i)\,, & 
\mathcal{O}_{S_R}^{\alpha\beta i}&=(\bar e_R^\alpha \,\nu_L^\beta)(\bar c_L\, d_R^i)\,, \\
\mathcal{O}_T^{\alpha\beta i}&=(\bar e_R^\alpha\sigma_{\mu\nu}\nu_L^\beta)(\bar c_R\sigma^{\mu\nu}d_L^i)\,.
\end{aligned}
\end{align}
Note that mixed chirality tensor operators vanish by Lorentz invariance. The extraction of the CKM matrix in the SMEFT is a delicate exercise~\cite{Descotes-Genon:2018foz}. For our purposes here, $V_{cd}$ and $V_{cs}$ can be safely obtained by exploiting unitarity in the Wolfenstein parametrization,
\begin{align}
\begin{aligned}
V_{cd}&=-\lambda_c+\mathcal O (\lambda_c^5),\\
V_{cs}&=1-\lambda_c^2/2+\mathcal O(\lambda_c^4),
\end{aligned}
\end{align}
where $\lambda_c$ is the sine of the Cabibbo angle. We assume that any contribution of NP to the inputs of these unitarity relations is small compared to the precision achieved with charm weak transitions. For instance, $\lambda_c$ obtained from kaon decays receives strong constraints from the unitarity of the first row of the CKM matrix (see e.g. Ref.~\cite{Gonzalez-Alonso:2016etj}). Similarly, we neglect the effects of NP modifications to $G_F$ {as determined from muon decays}.

The tree-level matching conditions between the SMEFT in Eq.~\eqref{eq:SMEFTlag} and the low-energy Lagrangian in Eq.~\eqref{eq:LagCC} are
\begin{align}\label{eq:WCmatching}
\begin{aligned}
\epsilon_{V_L}^{\alpha\beta i}&=-\frac{V_{ji}}{V_{ci}}\,[\mathcal{C}_{lq}^{(3)}]_{\alpha\beta2j}+\delta_{\alpha\beta}\,\frac{V_{ji}}{V_{ci}}[\mathcal{C}_{\phi q}^{(3)}]_{2j}\,,  \qquad& 
\epsilon_{V_R}^{\alpha\beta i}&=\frac{1}{2V_{ci}}\,\delta_{\alpha\beta}\,[\mathcal{C}_{\phi ud}]_{2i} \,, \\
\epsilon_{S_L}^{\alpha\beta i}&=-\frac{V_{ji}}{2V_{ci}}\,[\mathcal{C}_{lequ}^{(1)}]^*_{\beta\alpha j2}\,, &
\epsilon_{S_R}^{\alpha\beta i}&=-\frac{1}{2V_{ci}}\,[\mathcal{C}_{ledq}]^*_{\beta\alpha i2}\,, \\
\epsilon_{T}^{\alpha\beta i}&=-\frac{V_{ji}}{2V_{ci}}\,[\mathcal{C}_{lequ}^{(3)}]^*_{\beta\alpha j2}\,,
\end{aligned}
\end{align}
where a sum over $j$ is implicitly assumed. Interestingly, the low-energy operator $\mathcal{O}_{V_R}^{\alpha\beta i}$ is generated in the SMEFT from an operator that modifies a chirality preserving $W$ vertex but not from a new four-fermion interaction, unlike other operators in Eq.~\eqref{eq:OpsCC}. 
On the contrary, $\mathcal{O}_{V_L}^{\alpha\beta i}$ receives contributions from both a modified $W$ vertex and a new four-fermion interaction, which cannot be disentangled at low energies. 

The relations in Eq.~\eqref{eq:WCmatching} hold at the matching scale $\mu = m_W$. The renormalization group equations (RGE) {induced by QCD and EW (QED) radiative effects} allow one to robustly correlate low- and high-$p_T$ data~\cite{Jenkins:2017dyc,Alonso:2013hga}. In particular, the RGE running from $\mu=1$~TeV down to $\mu=2$~GeV yields sizable effects in scalar and tensor operators~\cite{Gonzalez-Alonso:2017iyc},
\begin{align}\label{eq:RGEops}
\begin{aligned}
\epsilon_{S_L}(2\,\mathrm{GeV})&\approx 2.1\,\epsilon_{S_L}(\mathrm{TeV}) - 0.3\,\epsilon_{T}(\mathrm{TeV})\,, &&&
\epsilon_{S_R}(2\,\mathrm{GeV})&\approx 2.0\,\epsilon_{S_R}(\mathrm{TeV})\,,\\
\epsilon_{T}(2\,\mathrm{GeV})&\approx 0.8\,\epsilon_{T}(\mathrm{TeV})\,.
\end{aligned}
\end{align}
{Here, $\epsilon_{X}(\mathrm{TeV})$ refers to the corresponding combination of SMEFT WCs in Eq.~\eqref{eq:WCmatching}. Vector operators do not run under QCD, and the electromagnetic and EW running remains at the percent level. Similarly, other RGE-induced contributions, including the mixing with other SMEFT operators, do not receive large QCD enhancements and remain at the percent level. All these effects are below the level of precision of our studies, so we neglect them in the following.}

\section{Decays of charmed mesons}
\label{sec:D}

Leptonic and semileptonic decays $D_{(s)}\to \bar e^\alpha\nu$ and $D\to \pi(K) \bar \ell\nu$ follow from the Lagrangian in Eq.~(\ref{eq:LagCC}). This  captures the leading effects of any possible short-distance contribution to $c\to d^i \bar e^\alpha\nu^\beta$ flavor transitions, with the SM being a particular limit, $\epsilon^{\alpha\beta i}_{X,SM} = 0$ for all $X$. Hadronic matrix elements of the corresponding operators are constrained by Lorentz symmetry and invariance of QCD under parity. As a result, pure leptonic decays are sensitive only to axial ($\epsilon^{\alpha\beta i}_{A}=\epsilon^{\alpha\beta i}_{V_R}-   \epsilon^{\alpha\beta i}_{V_L}$) and pseudoscalar ($\epsilon^{\alpha\beta i}_{P}=\epsilon^{\alpha\beta i}_{S_R}-\epsilon^{\alpha\beta i}_{S_L}$) combinations of WCs.  On the other hand, the semileptonic decays are sensitive to vectorial ($\epsilon^{\alpha\beta i}_{V}=\epsilon^{\alpha\beta i}_{V_R}+\epsilon^{\alpha\beta i}_{V_L}$) and scalar ($\epsilon^{\alpha\beta i}_{S}=\epsilon^{\alpha\beta i}_{S_R}+\epsilon^{\alpha\beta i}_{S_L}$) combinations of WCs, and to the tensor WC ($\epsilon^{\alpha\beta i}_{T}$). 

The largest available phase space that can be achieved for the semileptonic decays is given by $m_{D^+}-m_{\pi^0}\simeq 1.735$ GeV. Note that this is smaller than the $\tau$ lepton mass, which makes the semitauonic $D$-meson decays kinematically forbidden. A similar conclusion follows for the decays of charmed baryons. In other words, the tauonic vector, scalar and tensor operators ($\mathcal O_{V,S,T}^{\tau\beta}$) are not directly accessible and, as we will see below, high-$p_T$ tails provide a unique probe of these operators. On the other hand, pure tauonic decays of $D_{(s)}$ are allowed.\footnote{The phase-space restriction is lifted for semitauonic decays of excited $D^*$ mesons. However, these predominantly decay electromagnetically or strongly and the branching fractions of weak decays are suppressed~\cite{Grinstein:2015aua,Khodjamirian:2015dda}. Furthermore, one could in principle access the tauonic tensor operator by measuring $D_{(s)} \to \tau \nu \gamma$ (see e.g. Ref.~\cite{Gonzalez-Alonso:2016etj} for the equivalent pion and kaon decays).}

In the following, we derive bounds on the WCs of the operators in Eq.~(\ref{eq:OpsCC}) from $D_{(s)}$-meson decays. First, we restrict ourselves to the lepton-flavor diagonal case ($\epsilon_X^{\alpha i}\equiv \epsilon_X^{\alpha\alpha i}$), which interferes with the SM and leads to the strongest bounds. The rate of the leptonic $D$ decays is
\begin{align}
{\rm BR}(D^+\to\bar e^\alpha\nu^\alpha)=\tau_{D^+}\frac{m_{D^+}m_\alpha^2 f_{D}^2G_F^2|V_{cd}|^2\beta_\alpha^4}{8\pi}\left|1-\epsilon_A^{\alpha d}+\frac{m_{D}^2}{m_\alpha (m_c+m_u)}\epsilon_P^{\alpha d}\right|^2, 
\end{align}
where $\beta_\alpha^2=1-m_\alpha^2/m_D^2$ and $\tau_{D^+}$ ($f_{D^+}$) is the $D^+$ lifetime (decay constant). This formula with obvious replacements also describes the leptonic $D_s$ decays. We use $f_{D}=212.0(7)$ MeV and $f_{D_s}=249.9(5)$ MeV, obtained from an average of lattice QCD simulations with two degenerate light quarks and dynamical strange and charm quarks in pure QCD~\cite{Aoki:2019cca,Carrasco:2014poa,Bazavov:2017lyh}. An important feature of the leptonic decays is that the axial contribution, such as the one predicted in the SM, is suppressed by $m_\alpha^2$ due to the conservation of angular momentum. On the contrary, pseudoscalar NP contributions are unsuppressed, and they receive strong constraints from searches and measurements of these decays.       

\begin{table}[t]
\centering
\begin{tabular}{c|c|ccccc}
\toprule
$P$&$\alpha$ & $\boldsymbol{{\rm BR}_{\rm SM}}$& $\boldsymbol{x_S}$ & $\boldsymbol{x_T}$ & $\boldsymbol{y_S}$ & $\boldsymbol{y_T}$\\
\midrule
\multirow{2}{*}{$\pi^-$ }& $e$ &$2.65(18)\cdot10^{-3}$&$1.12(10)\cdot10^{-3}$&$1.21(15)\cdot10^{-3}$&$2.74(22)$&$1.14(21)$\\
& $\mu$ &$2.61(17)\cdot10^{-3}$&$0.228(19)$&$0.23(3)$&$2.73(18)$&$1.15(22)$\\
\midrule
\multirow{2}{*}{$K^-$ }& $e$ &$3.48(26)\cdot10^{-2}$&$1.29(8)\cdot10^{-3}$&$1.18(11)\cdot10^{-3}$&$2.00(11)$&$0.69(8)$\\
& $\mu$ &$3.39(25)\cdot10^{-2}$&$0.251(16)$&$0.224(20)$&$2.00(11)$&$0.71(8)$\\
\bottomrule
\end{tabular}
  \caption{Coefficients of the parametrization in Eq.~(\ref{eq:3bodyparam}) obtained using lattice QCD results~\cite{Lubicz:2017syv,Lubicz:2018rfs} for the form factors.\label{tab:3bodycoeffs}}
\end{table}

In the case of semileptonic $D$ decays, the expressions for total rates are more involved as they contain kinematic integrals with form factors, which are functions of the invariant mass of the dilepton pair. The decay rate of the neutral $D$ meson can be parametrized as a function of the WCs,
\begin{equation}\label{eq:3bodyparam}
\frac{{\rm BR}(D\to P_i\,\bar \ell^{\alpha}\nu^\alpha)}{{\rm BR}_{\rm SM}}=\left|1+\epsilon_V^{\alpha i}\right|^2+2\,{\rm Re}\left[(1+\epsilon_V^{\alpha i})(x_S\,\epsilon_S^{\alpha i*}+x_T\,\epsilon_T^{\alpha i*})\right]+y_S\,|\epsilon_S^{\alpha i}|^2+y_T\,|\epsilon_T^{\alpha i}|^2,
\end{equation}
where $x_{S,T}$ and $y_{S,T}$ describe the interference between NP and SM and the quadratic NP effects, respectively, and $P_i=\pi,~K$ for $i=d,~s$. The numerical values of these parameters can be obtained using lattice QCD calculations of the form factors and performing the kinematic integrals. In Table~\ref{tab:3bodycoeffs} we show the values of these parameters for the $D^0\to \pi^-(K^-)\ell^+\nu$ decays using the lattice results from~\cite{Lubicz:2017syv,Lubicz:2018rfs}. The errors in the parametrization employed in these references have been propagated consistently. 

The limits on the WCs are determined by comparing these predictions to the PDG averages~\cite{Tanabashi:2018oca} of the experimental data on the branching fractions~\cite{Eisenstein:2008aa,Ablikim:2013uvu,Zupanc:2013byn,Ablikim:2016duz,delAmoSanchez:2010jg,Alexander:2009ux,Naik:2009tk,Onyisi:2009th,Abe:2005nq,Ablikim:2018evp,Widhalm:2006wz,Ablikim:2015ixa,Besson:2009uv,Ablikim:2018frk}. The results are shown in Table~\ref{tab:WCsCClowE} where one WC is fitted at a time setting the rest to zero.
The sensitivity to vectorial currents is at the few percent level, reflecting the precision achieved in the experimental measurements and in the calculation of the respective semileptonic form factors. Bounds on axial currents depend strongly on the lepton flavor due to the chiral suppression of their contributions to the leptonic-decay rates.
Thus, the electronic axial operators are poorly constrained while muonic ones are constrained down to a few percent. The difference between $cs$ and $cd$ transitions in the bounds on the tauonic axial contributions is a result of the different experimental precision achieved in the measurement of the corresponding decays.

Direct bounds on scalar and tensor operators stemming from semileptonic decays are rather weak, with almost $\mathcal O(1)$ contributions still allowed by the data. As shown in Table~\ref{tab:3bodycoeffs}, this is due to the fact that the interference of these operators with the SM is chirally suppressed (see e.g. Ref.~\cite{Gonzalez-Alonso:2016etj}) and the bound is on their quadratic contribution to the rates. Pseudoscalar contributions to the leptonic-decay rates are, on the other hand, chirally enhanced with respect to the SM contribution and, as a result, constrained down to the per-mille level for electronic and muonic channels. For the tauonic ones, the lepton-mass enhancement is absent, and the bounds are $\sim1\%$ ($cs$) or $\sim10\%$ ($cd$), depending again on the experimental uncertainties. 

\begin{table}[t]
\centering
\begin{tabular}{c|c|ccccc}
\toprule
$i$&$\alpha$ & $\boldsymbol{\epsilon_V^{\alpha i}}$ &$ \boldsymbol{\epsilon_A^{\alpha i}}$& $ \boldsymbol{\epsilon_S^{\alpha i}}$&$ \boldsymbol{\epsilon_P^{\alpha i}}$&$ \boldsymbol{\epsilon_T^{\alpha i}}$\\
\midrule
\multirow{3}{*}{$d$}& $e$& $[-0.02,~0.11]$ & $[-32,~34]$&$[-0.29,~0.29] $ & $[-0.005,~0.005] $&$[-0.5,~0.5] $ \\
& $\mu$ & $[-0.06,~0.07]$&  $[-0.013,~0.07] $& $[-0.33,~0.17]$&$[-0.0024,~0.0004]$& $[-0.6,~0.22]$\\
& $\tau$ & $-$ & $[-0.27,~0.21]$& $-$ & $[-0.11,~0.15]$ & $-$ \\
\midrule
\multirow{3}{*}{$s$ }& $e$& $[-0.07,~0.08] $ & $[-27,~29]$& $[-0.29,~0.29] $ & $[-0.005,~0.004] $&$[-0.5,~0.5] $ \\
& $\mu$ & $[-0.09,~0.06]$&$[-0.07,~0.02] $ &$[-0.4,~0.16] $ &$[-0.0007,~0.0022] $ & $[-0.9,~0.22] $\\
& $\tau$ &$-$ & $[-0.07,~0.014]$& $-$ & $[-0.008,~0.04]$ & $-$ \\
\bottomrule
\end{tabular}
  \caption{95\% CL ranges of the WCs of the charged-current operators obtained at the scale $\mu=2$ GeV from current experimental data on (semi)leptonic $D_{(s)}$-meson decays, assuming them to be real.\label{tab:WCsCClowE}}
\end{table}

From the model building perspective, at a scale $\Lambda > v$, the NP effects are naturally realized in terms of operators in the chiral basis. Models for which the dominant contribution is through scalar operators receive the strongest constraint from leptonic decays, unless some tuning between $\mathcal O_{S_L}$ and $\mathcal O_{S_R}$ is enforced.
In addition, scalar and tensor operators receive radiative contributions that rescale and mix them significantly when connecting the direct bounds in Table~\ref{tab:WCsCClowE} to the matching scale, cf. Eq.~(\ref{eq:RGEops}). Or, inversely, a model producing a tensor contribution at the matching scale will produce a scalar contribution at low energies that is then constrained by leptonic decays. This is illustrated in Table~\ref{tab:WCsCC1TeV} where we have expressed the low-energy bounds in terms of the WCs in the chiral basis at $\mu=1$ TeV. As expected, bounds on single scalar and tensor operators are dominated by the measurements of pure leptonic decays.    

Except for operators whose dominant contribution to the observables is already quadratic ($\mathcal O_{A,P}^{ei}$ and $\mathcal O_{S,T}^{\ell i}$), the limits in Table~\ref{tab:WCsCClowE} are weakened if NP does not interfere with the SM. This is the case when the neutrino flavor is $\beta\neq\alpha$, or when the WCs are imaginary. The bounds are relaxed typically by a factor $\sim 3-6$ over the symmetrized ranges shown in that table. However, for a few operators, namely $\mathcal O_{A,P}^{\mu s}$, $\mathcal O_{A,P}^{\tau s}$ and $\mathcal O_{V}^{e d}$, the worsening is by an order of magnitude. Therefore, in the absence of SM interference, the bounds from $D_{(s)}$ meson decays are weak except for the pseudoscalar operators, which can still be competitive with other constraints.

Improvements on purely muonic and tauonic branching fractions by a factor $\sim 2-3$ are expected from future measurements at BES III~\cite{Ablikim:2019hff} and Belle II~\cite{Kou:2018nap} (see detailed projections in  Ref.~\cite{Ablikim:2019hff}), while no projections for electronic decays have been provided. For semileptonic decays, the data samples are expected to increase by two orders of magnitude after the full 50 ab$^{-1}$ of integrated luminosity planned at Belle II~\cite{Kou:2018nap}, thus the precision will most likely be limited by systematic uncertainties. Moreover, going beyond $\sim1\%$ accuracy in the SM prediction of these decay modes will be challenging because of the precision required in the computation of the corresponding hadronic matrix elements, including radiative (QED) effects (see LQCD projections in Ref.~\cite{Cerri:2018ypt}). In summary, improvements of the bounds reported in Tables~\ref{tab:WCsCClowE} and \ref{tab:WCsCC1TeV} from the modes analyzed in this work will remain modest in the near future.

\begin{table}[t]
\centering
\begin{tabular}{c|c|cc}
\toprule
$i$&$\alpha$ & $ \boldsymbol{\epsilon_{S_L}^{\alpha i}~(-\epsilon_{S_R}^{\alpha i})\times10^3}$&$\boldsymbol{\epsilon_T^{\alpha i}\times10^2}$\\[4pt]
\midrule
\multirow{3}{*}{$d$}& $e$&$[-2.5,~2.7]$ & $[-1.6,~1.5]$\\
& $\mu$ &  $[-0.2,~1.2]$&$[-0.7,~0.13]$\\
& $\tau$ &  $[-70,~60]$ &$[-33,~44]$\\
\midrule
\multirow{3}{*}{$s$ }& $e$&  $[-2.0,~2.2]$& $[-1.3,~1.2]$\\
& $\mu$ & $[-1.1,~0.3]$ &$[-0.2,~0.6]$\\
& $\tau$ & $[-19,~4.0]$&$[-2.0,~12]$\\
\bottomrule
\end{tabular}
  \caption{95\% CL ranges of the WCs, assumed to be real, obtained from $D_{(s)}$-meson decays for scalar and tensor operators in the chiral basis at $\mu=1$ TeV. The ranges of $\epsilon_{S_R}^\alpha$ are those of $-\epsilon_{S_L}^\alpha$.  \label{tab:WCsCC1TeV}}
\end{table}

Finally, it is important to stress that we have restricted our analysis to decay channels for which precise measurements and accurate LQCD predictions of the form factors currently exist. Additional modes that can be considered are $D\to V\ell\nu$ decays ($V=\rho,~K^*$), for which modern lattice results do not exist~\cite{Fleischer:2019wlx}, or baryonic $\Lambda_c$ decays for which data is not very precise yet. In addition, one may consider other observables such as kinematic distributions. Including these observables may improve the bounds on some of the WCs in the future and close flat directions in a global fit of decay data (see e.g. Ref.~\cite{Fleischer:2019wlx}).       

\section{\texorpdfstring{High-$p_T$}{High-pT} lepton production at the LHC}
\label{sec:LHC}

\subsection{Short-distance new physics in \texorpdfstring{high-$p_T$}{high-pT} tails}
\label{sec:tails}

The monolepton production in proton-proton collisions at high-energy, $\sqrt{s} \gg m_W$, is an excellent probe of new contact interactions between quarks and leptons.\footnote{There is a rich literature of NP exploration in neutral and charged Drell-Yan production, for an incomplete list see~\cite{Cirigliano:2012ab, deBlas:2013qqa, Gonzalez-Alonso:2016etj, Faroughy:2016osc, Greljo:2017vvb, Cirigliano:2018dyk, Greljo:2018tzh, Angelescu:2020uug, Farina:2016rws, Alioli:2017nzr, Raj:2016aky, Schmaltz:2018nls,Dawson:2018dxp,Brooijmans:2020yij}.} The final state in this process features missing energy plus a charged lepton of three possible flavors. In addition, there are five quark flavors accessible in the incoming protons whose composition is described by the corresponding parton distribution functions (PDF). Within the SMEFT, a total of 4 four-fermion operators contribute to this process at tree-level for each combination of quark and lepton flavors, see Eq.~\eqref{eq:SMEFTop4F}. Their contribution to the partonic cross section grows with energy as $\hat \sigma \propto s$, see Eq.~\eqref{eq:xsection}.  Other effects in the SMEFT include the chirality preserving (flipping) $W$-boson vertex corrections which scale as $\hat \sigma \propto s^{-1} (s^{0})$ and are negligible in the high-$p_T$ tails compared to the four-fermion interactions.\footnote{The modification of the $W$-boson propagator in the universal basis~\cite{Englert:2019zmt} through the $\hat W$ parameter is captured by the specific combination of the four-fermion contact interactions and vertex corrections in the Warsaw basis. For $\hat W$ searches in the high-$p_T$ lepton tails see Ref.~\cite{Farina:2016rws}.}

The numerical results derived in this work are based on the Monte Carlo simulations described in Section~\ref{sec:limits}. Here we present a (semi-)analytic understanding of the main physical effects. The tree-level unpolarized partonic differential cross section for $d^j(p_1)\,\bar u^i(p_2)\to e^\alpha(p_3)\bar \nu^\beta(p_4)$\,, induced by the SMEFT four-fermion operators in Eq.~\eqref{eq:SMEFTop4F}\,, expanded and matched to the notation of Eq.~\eqref{eq:LagCC}\,, is
\begin{align}
\frac{d\hat\sigma}{dt}&=\frac{G_F^2 |V_{ij}|^2}{6\pi s^2}\left[(s+t)^2\,\bigg|\delta^{\alpha \beta}\frac{m_W^2}{s}- \epsilon_{V_L}^{\alpha\beta i j} \bigg|^2+\frac{s^2}{4}\,\big(|\epsilon_{S_L}^{\alpha\beta i j}|^2+|\epsilon_{S_R}^{\alpha\beta i j}|^2\big)+4(s+2t)^2\,|\epsilon_{T}^{\alpha\beta i j}|^2\right.\nonumber\\
&\quad\left.-\,2s(s+2t)\,\mathrm{Re}\big(\epsilon_{S_L}^{*\,\alpha\beta i j}\,\epsilon_{T}^{\alpha\beta i j}\big)\right]\,,\label{eq:diffsec}
\end{align}
where $s \equiv (p_1 + p_2)^2$ and $t = (p_3 - p_1)^2$ are the corresponding Mandelstam variables. The interference with the SM is absent in the case of lepton flavor violation (LFV), i.e. $\alpha \neq \beta$. In the relativistic limit, chiral fermions act as independent particles with definite helicity. Therefore, the interference among operators is achieved only when the operators match the same flavor and chirality for all four fermions. Integrating over $t$, we find the partonic cross section
\begin{align}\label{eq:xsection}
\begin{aligned}
\hat\sigma (s) &=\frac{G_F^2 |V_{ij}|^2}{18\pi}\,s\left[\bigg|\delta^{\alpha \beta}\frac{m_W^2}{s}- \epsilon_{V_L}^{\alpha\beta i j} \bigg|^2+\frac{3}{4}\,\big(|\epsilon_{S_L}^{\alpha\beta i j}|^2+|\epsilon_{S_R}^{\alpha\beta i j}|^2 \big)+4\,|\epsilon_{T}^{\alpha\beta i j}|^2\right]\,,
\end{aligned}
\end{align}
as a function of the dilepton invariant mass $\sqrt{s}$. The interference with the SM is relevant for $|\epsilon_{V_L}| \sim m_W^2 / \textrm{TeV}^2$ or smaller. {This holds irrespective of the initial quark flavors in $d^j\bar u^i \to e^\alpha \bar\nu^\alpha$ ($i=1,2$ and $j=1,2,3$).} The results obtained in our numerical analysis (see Table~\ref{tab:WCsCChighE}) suggest that the quadratic term in $\epsilon_{V_L}$ dominates present limits. However, there is already a non-negligible correction from the interference term which will become prominent with more integrated luminosity. The lack of interference in the other cases tends to increase the cross section in the high-$p_T$ tails, and allows to extract bounds on several NP operators simultaneously.\footnote{The transverse mass distribution ($m_T \approx 2\, p_T^\ell$) also inherits negligible $\epsilon_{S_L} - \epsilon_T$ interference.} On the contrary, most of the bounds from $D_{(s)}$ mesons decays discussed in Section~\ref{sec:D} depend on interference terms among different WCs, and it becomes difficult to break flat directions without additional observables.

\begin{figure}[t]
\centering
\includegraphics[width=0.5\textwidth]{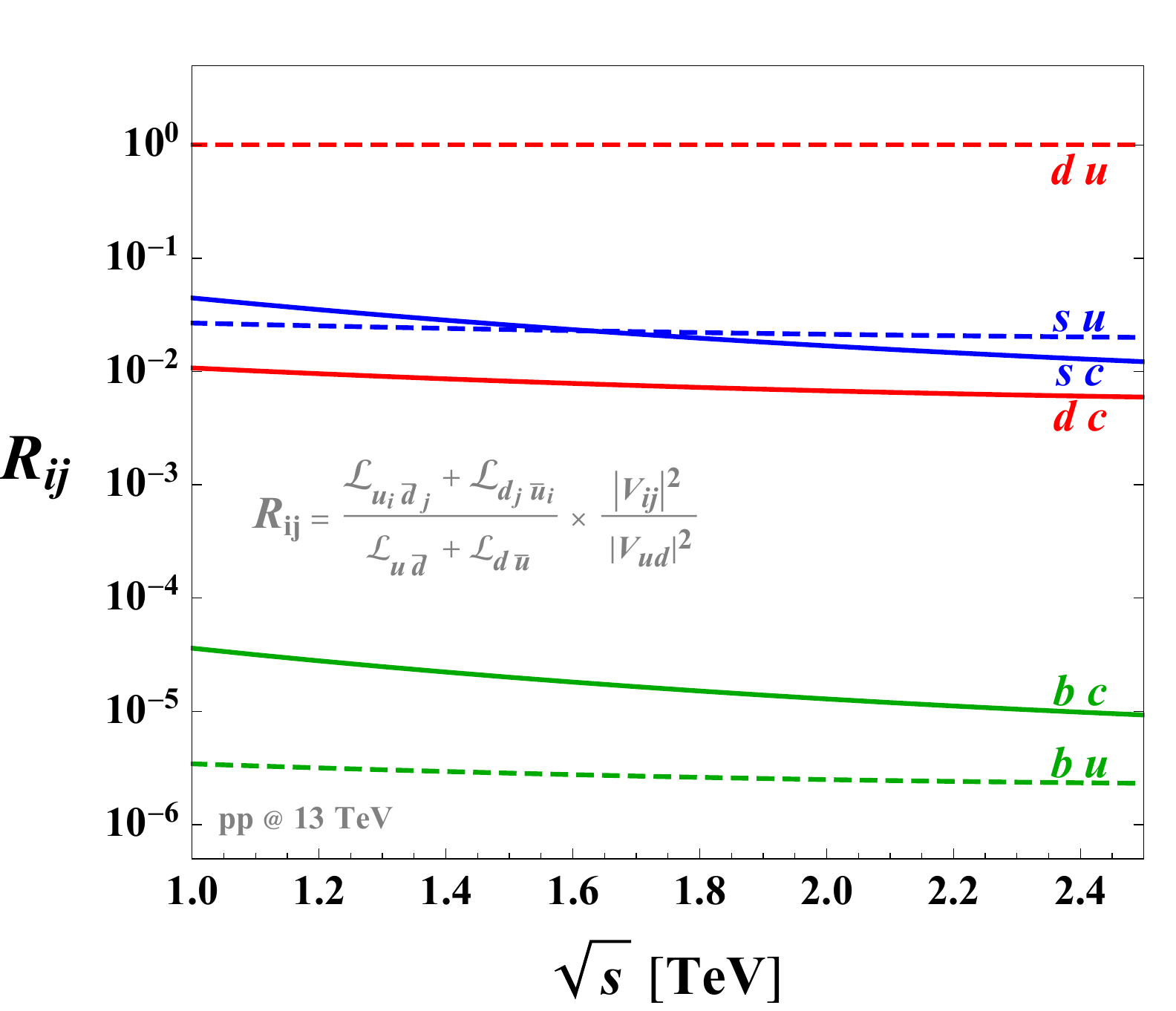}
\caption{{Suppression factors for the charged-current Drell-Yan cross section with different colliding quark flavors, $R_{ij}$,}
stemming from the PDF and the CKM matrix, see Eq.~\eqref{eq:pdfRatios}. 
\label{fig:plot-ratio}}
\end{figure}

While the energy growth of the amplitude enhances the signal, the PDF of the sea quarks reduce it. The parton luminosity for colliding flavors $i$ and $j$ is
\begin{equation}\label{eq:PDFlumi}
\mathcal{L}_{q_i \bar q_j}(\tau,\mu_F) = \int_{\tau}^1 \frac{d x}{x}~f_{q_i} (x,\mu_F) f_{\bar q_j} (\tau/x,\mu_F)\;,
\end{equation}
where $\tau = s / s_0$ and $\sqrt{s_0}$ is the collider energy (here set to 13~TeV).
The relative correction to the Drell-Yan cross section in the tails ($\sqrt{s} \gg m_W$) is
\begin{equation}
\frac{\Delta \sigma}{\sigma} \approx  R_{ij} \times \frac{d_X ~ \epsilon^2_X}{\left(m_W^2 / s \right)^2}\;, \label{eq:xtails}
\end{equation}
with $d_X = 1, \frac{3}{4}, 4$ for $X = V, S, T$ respectively, and
\begin{equation}\label{eq:pdfRatios}
R_{ij} \equiv  \frac{(\mathcal{L}_{u_i \bar d_j} + \mathcal{L}_{d_j \bar u_i}) \times |V_{ij}|^2}{(\mathcal{L}_{u \bar d} + \mathcal{L}_{d \bar u}) \times |V_{ud}|^2 } \;.
\end{equation}
We show in Figure~\ref{fig:plot-ratio} the ratios $R_{ij}$ for $d u$ (red dashed), $d c$ (red solid), $s u$ (blue dashed), $s c$ (blue solid), $b u$ (green dashed) and $b c$ (green solid) as a function of the dilepton invariant mass $\sqrt{s}$. Here we use the {\tt MMHT2014 NNLO}$_{188}$ PDF~\cite{Harland-Lang:2014zoa} with the factorization scale $\mu_F = \sqrt{s}$. The suppression from $R_{ij}$ is compensated by the energy enhancement $(\sqrt{s} / m_W)^4 \sim \mathcal{O}(10^5)$. Thus, a measurement of the cross section in the tails with $\mathcal{O}(10\%)$ precision would probe $c s$ and $c d$ at the level of $\epsilon_X \sim \mathcal{O}(10^{-2})$. The weak dependence on the energy across the most sensitive bins allows to rescale the limits for different flavor combinations provided the lepton cuts are sufficiently inclusive (see Section~\ref{sec:limits}).

The theoretical prediction for the signal rate is plagued by the uncertainties stemming from the missing high-order perturbative corrections, as well as the knowledge of the PDF of the colliding sea quarks. These have been studied in detail in~\cite{Greljo:2018tzh,Faroughy:2016osc}. More precisely, NLO QCD and PDF uncertainties are quantified in the supplemental material of Ref.~\cite{Greljo:2018tzh} for a $b c \to W'$ example (and in Ref.~\cite{Faroughy:2016osc} for $b b \to Z'$) as a function of the vector boson mass $m_{V'}$. These estimates are trivially applicable for the corresponding quark-lepton contact interactions when replacing $m_{V'}$ with the dilepton invariant mass $\sqrt{s}$. A relative uncertainty of $\sim 10\%$ is found on the differential cross section in the most sensitive bins. Another potential issue comes from the PDF extraction, as recent analyses also include Drell-Yan data, see e.g. Ref.~\cite{Ball:2017nwa}. While at the moment this data has a subleading impact on the PDF determination, it will become important at the HL-LHC~\cite{Khalek:2018mdn}. A proper approach would be to perform a combined SMEFT and PDF fit. First steps in this direction show discriminating power between EFT and PDF effects in the context of deep inelastic scattering~\cite{Carrazza:2019sec}.

\subsection{Recast of the existing experimental searches}\label{sec:limits}

We use the analyses reported by ATLAS and CMS collaborations with one lepton plus missing transverse momentum signature. For the $\tau+\nu$ channel, we recast the searches in Refs.~\cite{Aaboud:2018vgh,Sirunyan:2018lbg} using $36.1\,\mathrm{fb}^{-1}$ and $35.9\,\mathrm{fb}^{-1}$ of data, respectively. In the case of $\ell+\nu$ final state, we use the ATLAS $139\,\mathrm{fb}^{-1}$~\cite{Aad:2019wvl} and the CMS $35.9\,\mathrm{fb}^{-1}$~\cite{Sirunyan:2018mpc} analyses. The Monte Carlo (MC) simulation pipeline is as follows: we use {\tt FeynRules}~\cite{Alloul:2013bka} for the model generation, {\tt MadGraph5\_aMC@NLO}~\cite{Alwall:2011uj, Alwall:2014hca} for the partonic process simulation interfaced with {\tt Pythia 8}~\cite{Sjostrand:2014zea} to simulate the hadronic processes, and finally {\tt Delphes}~\cite{deFavereau:2013fsa} to get an estimate of the detector effects. We set a dynamical scale for renormalization and factorization scales, $\mu_{R/F}=m_T$. We use the ATLAS and CMS Delphes cards, respectively, when making the simulations for each experiment.  {\tt ROOT}~\cite{Brun:1997pa} is used to apply the selection criteria of each analysis to the corresponding Delphes output, and to obtain the expected yields for our signals in each bin of the reported transverse mass distributions. 

We validated our setup by producing MC samples for $W\to e^\alpha\nu + \mathrm{jets}$ in the SM, and comparing the yields with those reported by ATLAS and CMS. We reproduce their results within 10\% to 20\% accuracy. As we only use limited MC simulation capabilities, detector emulation via Delphes, and no experimental corrections from data, as done in the experimental analyses, we consider this level of agreement as an accurate reproduction of the experimental results from the phenomenological perspective. The same techniques have been used and reported in~\cite{Greljo:2018tzh}. Thus, the relative error on the limits derived here from the high-$p_T$ data is expected to be below $10\%$  ($\Delta \epsilon_X / \epsilon_X \approx 0.5 \, \Delta \sigma / \sigma$).

The limits on the WCs are obtained by comparing our simulated signal events for the transverse mass distributions to the background events in the corresponding collaboration analyses. For the statistical analysis, we use the modified frequentist CLs method~\cite{Read:2002hq}. We compute the CLs using the {\tt ROOT} package  {\tt Tlimit}~\cite{Junk:1999kv}, and exclude WC values with CLs < 0.05. In our statistical analysis, we include the SM background systematic and statistical errors (added in quadrature) provided by the collaborations for all bins. We ignore any possible correlation in the bin errors when combining the bins, since these are not provided. For the vector operator, both NP-squared and NP-SM interference contributions are computed. We do not include systematic errors for the signal simulation in our analysis, as they are expected to be subdominant compared to the overall signal normalization uncertainty stemming from the theoretical prediction of the cross section discussed in Section~\ref{sec:tails}.

\begin{table}[t]
\centering
\renewcommand*{\arraystretch}{1.1}
\begin{tabular}{c|c|cccccc}
\toprule
\multirow{2}{*}{$i$} & \multirow{2}{*}{$\alpha$} & \multirow{2}{*}{$\boldsymbol{\epsilon_{V_L}^{\alpha\alpha i}\times10^2}$} &
$\boldsymbol{|\epsilon_{V_L}^{\alpha\beta i}|\times10^2}$ &
\multicolumn{2}{c}{$\boldsymbol{|\epsilon_{S_{L,R}}^{\alpha\beta i}(\mu)|\times10^2}$} & \multicolumn{2}{c}{$\boldsymbol{|\epsilon_T^{\alpha\beta i}(\mu)|\times10^3}$}\\[2pt]
& & & $(\alpha\neq\beta)$ & $\mu=1$~TeV & $\mu=2$~GeV & $\mu=1$~TeV & $\mu=2$~GeV \\
\midrule
\multirow{3}{*}{$d$} & $e$ & $[-0.52,0.86]$ & $0.67\,(0.42)$ & $0.72\,(0.46)$ & $1.5\,(0.96)$ & $4.3\,(2.7)$ & $3.4\,(2.2)$ \\
& $\mu$ & $[-0.85,1.2]$ & $1.0\,(0.38)$ & $1.1\,(0.42)$ & $2.3\,(0.86)$ & $6.6\,(2.4)$ & $5.2\,(1.9)$\\
& $\tau$ & $[-1.4,1.8]$ & $1.6\,(0.68)$ & $1.5\,(0.55)$ &  $3.1\,(1.1)$ & $8.7\,(3.1)$ & $6.9\,(2.5)$\\
\midrule
\multirow{3}{*}{$s$ }& $e$ &  $[-0.28,0.59]$ & $0.42\,(0.26)$ & $0.43\,(0.28)$ & $0.91\,(0.57)$ & $2.8\,(1.5)$ & $2.2\,(1.2)$ \\
& $\mu$ &  $[-0.46,0.78]$ & $0.63\,(0.23)$ & $0.68\,(0.25)$ & $1.4\,(0.52)$ & $4.0\,(1.4)$& $3.1\,(1.1)$\\
& $\tau$ & $[-0.65,1.2]$ & $0.93\,(0.40)$ & $0.87\,(0.31)$ & $1.8\,(0.65)$ & $5.2\,(1.8)$ & $4.1\,(1.5)$\\
\bottomrule
\end{tabular}
  \caption{$95\%$~CL limits on the value of the WCs of the charged-current operators obtained from high-$p_T$ data ($\beta=e,\mu,\tau$). We also show in parenthesis the naive projections for the HL-LHC (3 ab$^{-1}$) on the expected limits, assuming that the error will be statistically dominated.}\label{tab:WCsCChighE}
\end{table}

Our results are reported in Table~\ref{tab:WCsCChighE} in terms of the WCs at two different scales $\mu = 1$~TeV and $\mu = 2$~GeV, respectively.\footnote{See Eq.~\eqref{eq:RGEops} for the RGE solutions. The difference between $S_L$ and $S_R$ is $\mathcal{O}(1\%)$ so we use a single column.}
The resulting limits qualitatively agree with the naive ratios in the absence of SM-NP interference,
\begin{align}
\label{eq:WCratios}
\epsilon_{V_L}^{\alpha\beta i} \; : \; \epsilon_{S_{L,R}}^{\alpha\beta i}  \; : \; \epsilon_{T}^{\alpha\beta i} \; \approx  \; 1:\frac{2}{\sqrt{3}}:\frac{1}{2}\,,
\end{align}
due to rather inclusive kinematics of the analysis. A dedicated future analysis should exploit the angular dependence in Eq.~\eqref{eq:diffsec} in order to differentiate among operators, and possibly further suppress the background. We also recommend separating future data by the lepton charge as a way to further enhance the signal over background discrimination. For instance, $ud$-induced monolepton production is asymmetric in lepton charge unlike $c s$.

 For the $\tau+\nu$ channel, the reported limits are well compatible with those obtained by naive rescaling via the $R_{ij}$ ratios in Eq.~\eqref{eq:pdfRatios} of the ones presented in 
 Ref.~\cite{Greljo:2018tzh} (neglecting the interference for $\epsilon_{V_L}$). In principle, this method can be used to estimate the limit on any $u^i \to d^j$ transition. Finally, the jackknife analysis performed in the supplemental material of~\cite{Greljo:2018tzh} suggests that the most sensitive bins in these types of searches to fall in the range between 1 and 1.5~TeV. This raises questions about the applicability of the high-$p_T$ bounds to the space of possible NP models modifying charged-current charm transitions, to which we turn next.

\subsection{Possible caveats within and beyond the EFT}\label{sec:caveats}

As shown in section~\ref{sec:limits}, most of the limits obtained from high-$p_T$ tails are stronger than their low-energy counterparts. However, one could argue that high-$p_T$ limits are not free of caveats, which would allow certain NP models to evade them while still yielding sizeable low-energy contributions. 

For concreteness, let us first remain within the realm of the SMEFT, where any new degree of freedom is well above the EW scale. The partonic cross-section for $\bar c d^i \to e^\alpha \bar \nu^\alpha$ scattering in the presence of dimension-six operators is given in Eq.~\eqref{eq:xsection}. As can be seen from this expression, the NP-squared piece receives an energy enhancement with respect to both the pure SM contribution and SM-NP interference. As a result, the limits shown in Table~\ref{tab:WCsCChighE} rely on dimension-six squared contributions. It could be argued that dimension-8 contributions that interfere with the SM are of the same order in the EFT, so their inclusion might significantly affect our results. To illustrate this point, let us work in a specific example involving both dimension-6 and dimension-8 operators,
\begin{align}\label{eq:toyEFT}
\mathcal{L}_{\rm EFT}\supset-\frac{4G_F}{\sqrt{2}}V_{ci}\left[\epsilon_{V_L}^{(6)}\,(\bar e_L^\alpha \gamma_\mu \nu_L^\alpha)(\bar c_L \gamma^\mu d_L^i) -\frac{1}{M_{\rm{NP}}^2}\epsilon_{V_L}^{(8)}\,(\bar e_L^\alpha \gamma_\mu \nu_L^\alpha)\partial^2(\bar c_L\gamma^\mu d_L^i)\right] + \rm{h.c.}\,,
\end{align}
with the normalization chosen such that $\epsilon^{(6,8)}_{V_L}$ are adimensional, and $M_{\rm{NP}}$ is the NP mass threshold.
The corresponding partonic cross section including both SM and the EFT contributions in Eq.~\eqref{eq:toyEFT} is given by
\begin{align}
\begin{aligned}
\hat\sigma (s) &=\frac{G_F^2 |V_{ci}|^2}{18\pi}\,s\, \bigg|\frac{m_W^2}{s}- \epsilon_{V_L}^{(6)} - \frac{s}{M_{\rm{NP}}^2}\,\epsilon_{V_L}^{(8)}  \bigg|^2 \\
&= \frac{G_F^2 |V_{ci}|^2}{18\pi}\,s\left[\frac{m_W^4}{s^2} - 2 \frac{m_W^2}{s} \mathrm{Re}(\epsilon_{V_L}^{(6)}) + |\epsilon_{V_L}^{(6)}|^2 - 2 \frac{m_W^2}{M_{\rm{NP}}^2}\,\mathrm{Re}(\epsilon_{V_L}^{(8)}) \right]  +\mathcal{O}\left(\frac{1}{M_{\rm{NP}}^{6}}\right) \,,
\end{aligned}
\end{align}
where, in the second line, we neglected the dimension-8 squared term. As already mentioned, the experimental limits in Table~\ref{tab:WCsCChighE} are dominated by the $|\epsilon_{V_L}^{(6)}|^2$ term, with a small correction from the term proportional to $\mathrm{Re}(\epsilon_{V_L}^{(6)})$. The term proportional to $\mathrm{Re}(\epsilon_{V_L}^{(8)})$ is even smaller than the dimension-6 interference if $|\mathrm{Re}(\epsilon_{V_L}^{(8)})| \leq |\mathrm{Re}(\epsilon_{V_L}^{(6)})|$, since $M_{\rm{NP}}^2 > s$ by construction. To give an example of explicit UV realization, a single tree-level $s$-channel resonance exchange predicts $\epsilon^{(6)}_{V_L} = \epsilon^{(8)}_{V_L}$. A significant cancellation between dimension-6 and 8 contributions would require a peculiar NP scenario.

Another possible way to evade our limits within the SMEFT regime would consist in including a semileptonic operator mediating $u\bar d\to \bar e^\alpha \nu^\alpha$ transitions which negatively interferes with the dominant SM background. One could then enforce a tuning between NP contributions to reduce the number of NP events in the tails. Even with this tuning, the different $\sqrt{s}$ dependence of each contribution would not allow for an exact cancellation between the two.

The EFT is no longer valid if a new mass threshold is at or below the typical energy of the process. Indeed, inverting the obtained limits on the WCs ($v/\sqrt{\epsilon_x} \approx$~few~TeV) and invoking perturbative unitarity suggests that the largest scales currently probed are at most $\mathcal{O}(10$~TeV) for strongly coupled theories. Any suppression in the matching, such as loop, weak coupling, or flavor spurion, brings the actual NP mass scale down. Clearly, the EFT approach has a significantly reduced scope in the high-$p_T$ lepton tails compared to charmed meson decays. Outside the EFT realm, one may wonder how well our limits approximate the correct values. Charged mediators responsible for generating charged currents at low energies, cannot be arbitrarily light since they would be directly produced at colliders by (at least) the EW pair production mechanism. Here, the signal yield is robustly determined in terms of the particle mass and known SM gauge couplings. A sizeable effect in low energy transitions also means sizeable decay branching ratio to usual final state with jets, leptons, etc, that has been searched for. Thus, charged mediators at or below the EW scale receive strong constraint from direct searches, yielding ${M_{\rm{NP}}} \gtrsim \mathcal{O}(100\,$GeV$)$.  

One could think of possible mediators that satisfy this bound, but still have a mass within the energy range invalidating the SMEFT, since the energy in the high-$p_T$ tails is around the TeV. At tree-level, there are a finite number of possible mediators, either colorless $s$-channel or colored $t$ ($u$)-channel resonances. In the case of $s$-channel mediators, the high-$p_T$ limits derived in the EFT are overly conservative, due to the resonance enhancement (see e.g. Figure~5 in \cite{Greljo:2017vvb}). On the other hand, for $t$ ($u$)-channel mediators, the EFT limits are typically (slightly) stronger than the real limits, but they serve as a good estimate (see e.g. Figure~3 in \cite{Greljo:2018tzh}). In addition, these latter mediators, known as leptoquarks, are copiously produced at the LHC by QCD (see e.g.~\cite{Dorsner:2018ynv}), and direct exclusion limits push their mass above the TeV. One could advocate for tuned scenarios where the high-$p_T$ contributions of a $t$-channel resonance is cancelled against a very wide $s$-channel resonance, while still yielding a sizable low-energy contribution (see example in Section 6.1 of~\cite{Buttazzo:2016kid}). As in the previous case, this requires a tuning of the NP contributions, and one can only achieve a partial cancellation. Finally, loop-induced contributions require the NP scale to be significantly lower (or the NP couplings to be strong) in order to generate the same effects at low energies. This translates into typically stronger high-$p_T$ limits than the ones considered here, either from non-resonant or from resonant production of the new mediators.

To conclude, the comparison of low-energy and high-$p_T$ within an EFT framework is useful even if the EFT validity is not guaranteed. If high-$p_T$ provides stronger limits relative to the ones derived from low-energies, this will also hold in a generic NP model barring tuned cancellations.

\section{Interplay between low and high energy}
\label{sec:interplay}

Once we have clarified possible caveats concerning high-$p_T$ limits on effective operators we are ready to compare low and high-energy results and discuss their complementarity. The comparison for scalar and tensor operators is quite direct because they receive contributions only from four-fermion operators in the SMEFT, cf. Eqs.~(\ref{eq:WCmatching}). 
Vector and axial operators, on the other hand, receive two types of SMEFT contributions from:
\textit{(i)} four-fermion operators, and \textit{(ii)} $W$  vertex corrections. As discussed in detail in Section~\ref{sec:LHC}, only \textit{(i)} experience the energy enhancement exploited by our analysis of the high-$p_T$ tails. {In the following, we discuss the interplay between low-energy and high $p_T$ bounds in four-fermion operators and then we obtain limits on $W$ vertex corrections.}

\subsection{Four-fermion interactions}

\begin{figure}[t]
\centering
\includegraphics[width=0.46\textwidth]{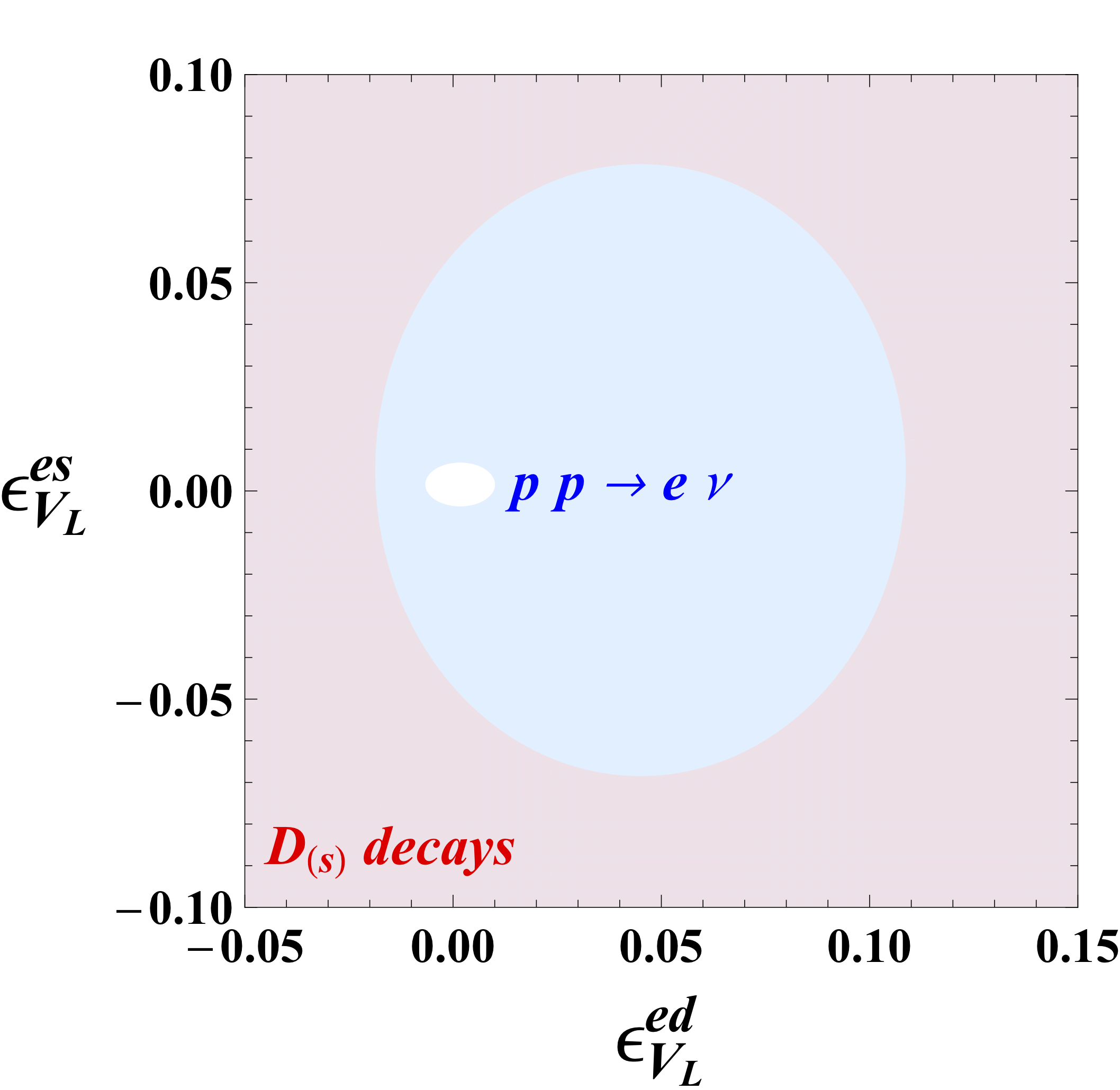} \;\; \includegraphics[width=0.45\textwidth]{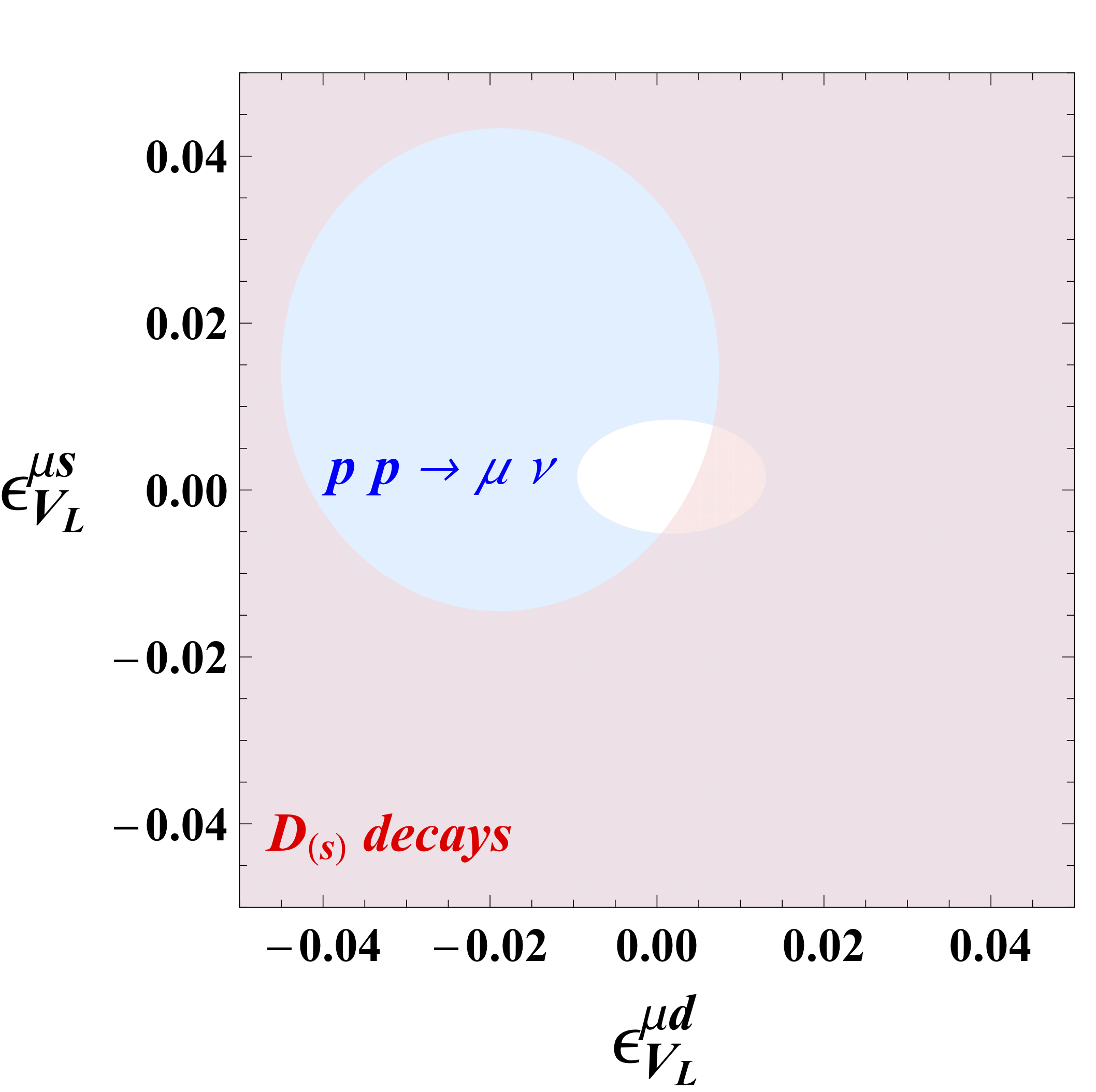}  \; \includegraphics[width=0.46\textwidth]{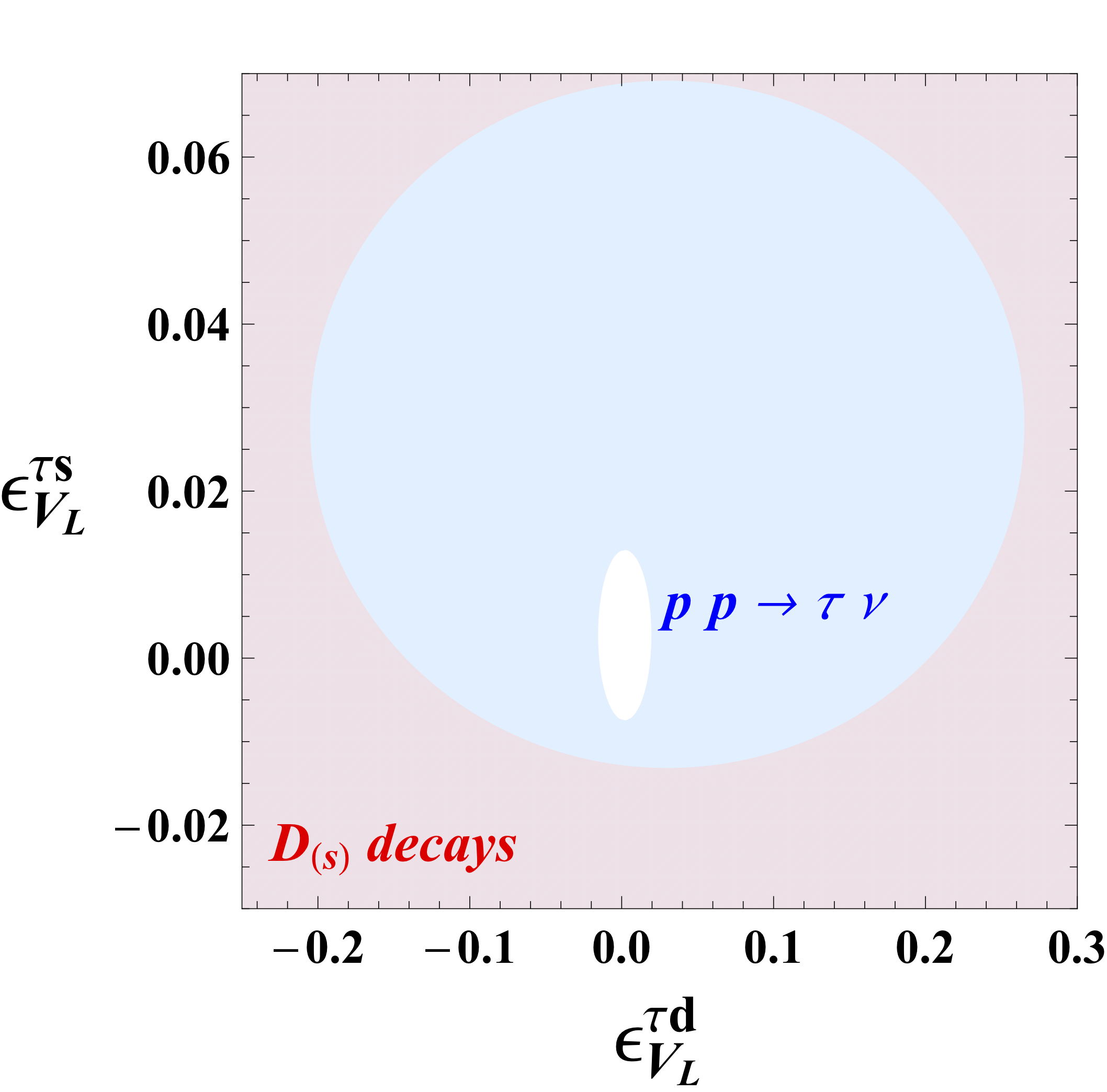}
\caption{Exclusion limits at 95\% CL on $c \to d (s) \bar e^\alpha \nu^\alpha$ transitions in $(\epsilon^{\alpha \alpha d}_{V_L}, \epsilon^{\alpha \alpha s}_{V_L})$ plane were $\alpha = e$ (top left), $\alpha =  \mu$ (top right), and $\alpha =  \tau$ (bottom). The region colored in pink is excluded by $D_{(s)}$ meson decays, while the region colored in blue is excluded by high-$p_T$ LHC.\label{fig:conclusions}}
\end{figure}

High-$p_T$ bounds on left-handed ($V-A$) four-fermion operators are almost an order of magnitude stronger than those derived from meson decays. In Figure~\ref{fig:conclusions}, we compare the regions excluded by charmed-meson decays (cf. Table~\ref{tab:WCsCClowE}) and high-$p_T$ monolepton tails (cf. Table~\ref{tab:WCsCChighE}) in the $(\epsilon^{\alpha \alpha d}_{V_L},~\epsilon^{\alpha \alpha s}_{V_L})$ plane, assuming NP only in the SMEFT operator $\mathcal{O}_{lq}^{(3)}$\,. The three plots are for each lepton flavor conserving combination $\alpha=\beta$, while for $\alpha \neq \beta$  the improvement with respect to charm decays is even more significant. These comparisons provide a striking illustration of the LHC potential to probe new flavor violating interactions at high-$p_T$.

\begin{figure}[t]
\centering
\begin{tabular}{ccc}
\includegraphics[width=0.3\textwidth]{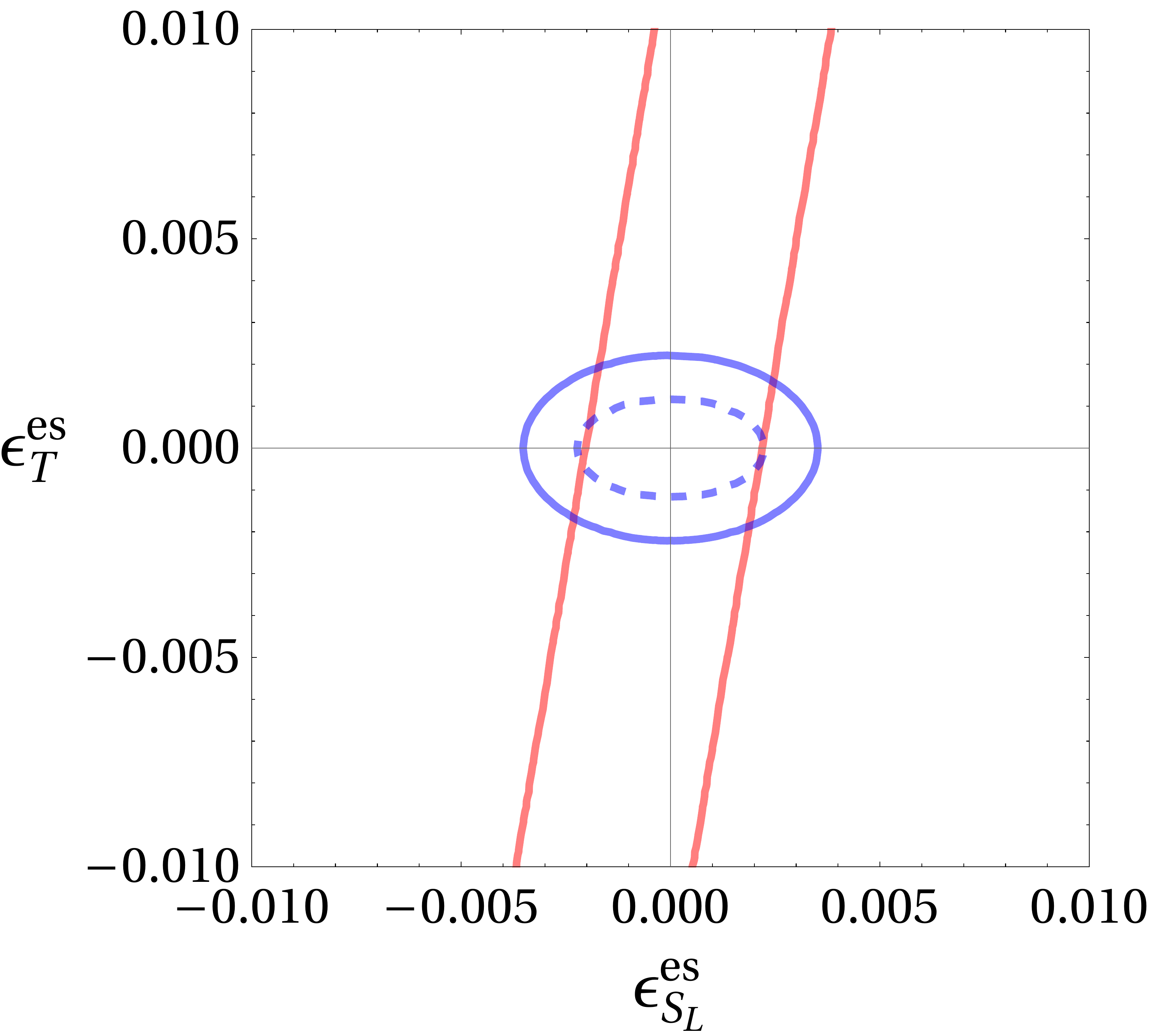} & \includegraphics[width=0.305\textwidth]{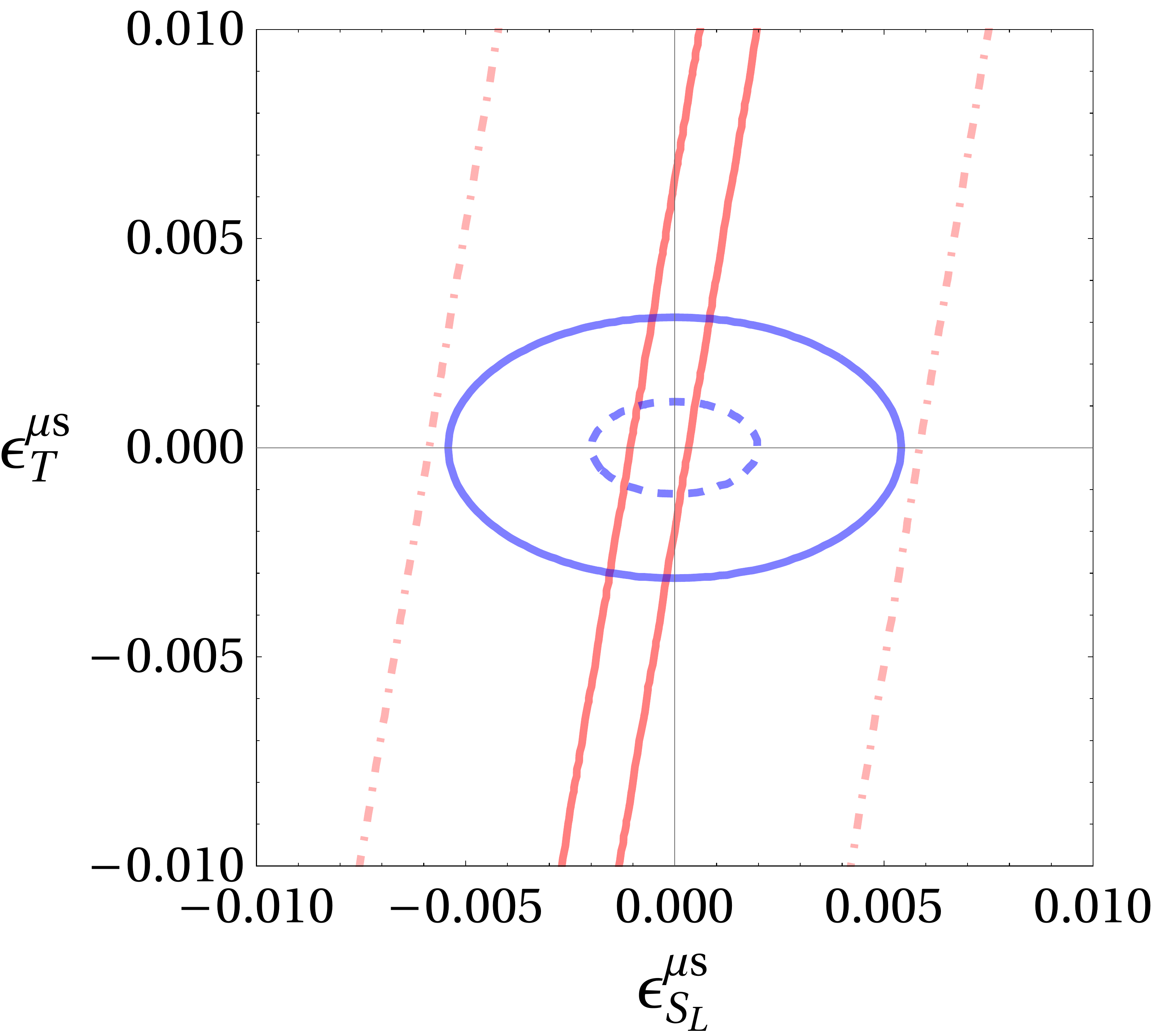} & \includegraphics[width=0.295\textwidth]{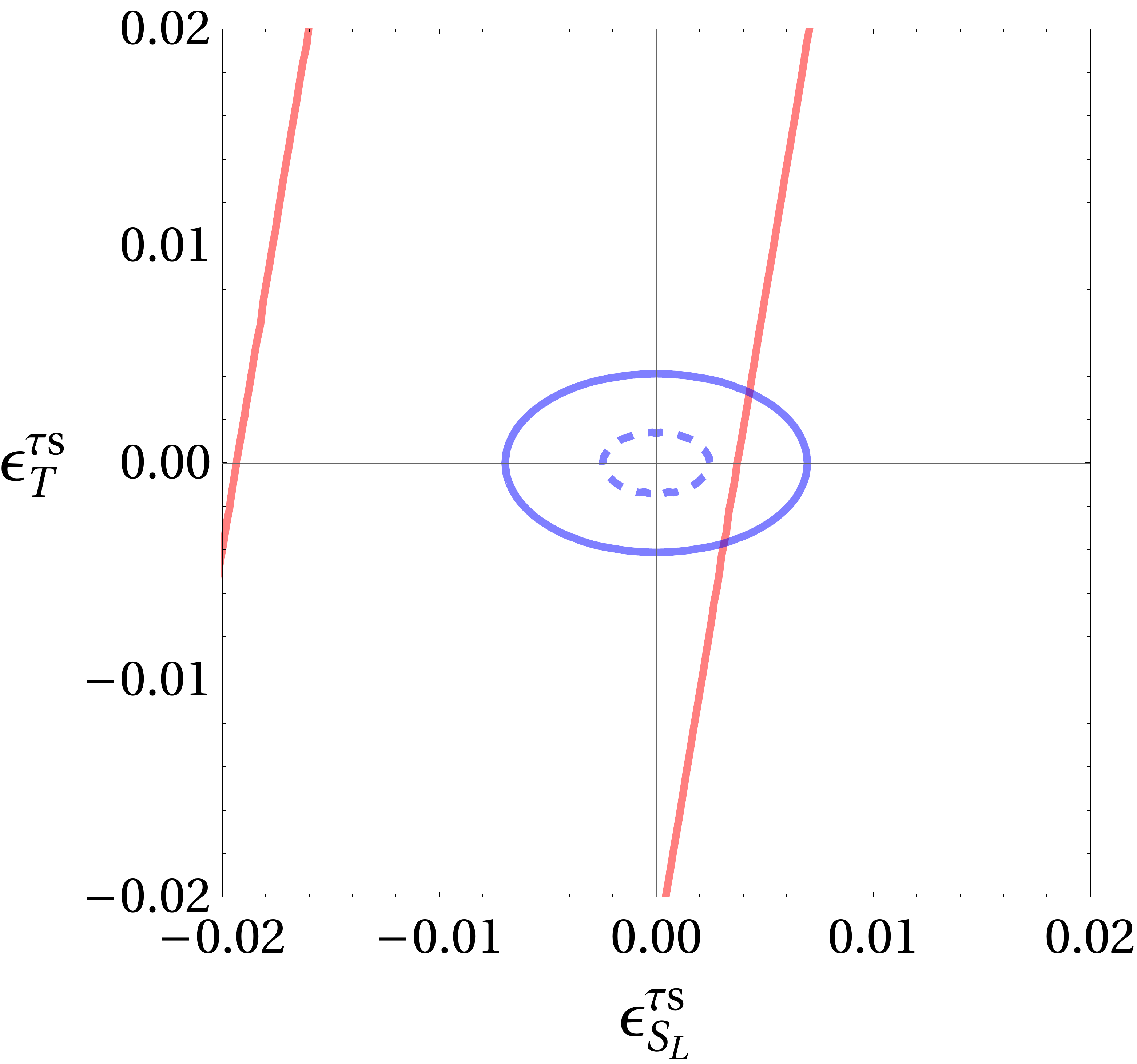} \\
\includegraphics[width=0.3\textwidth]{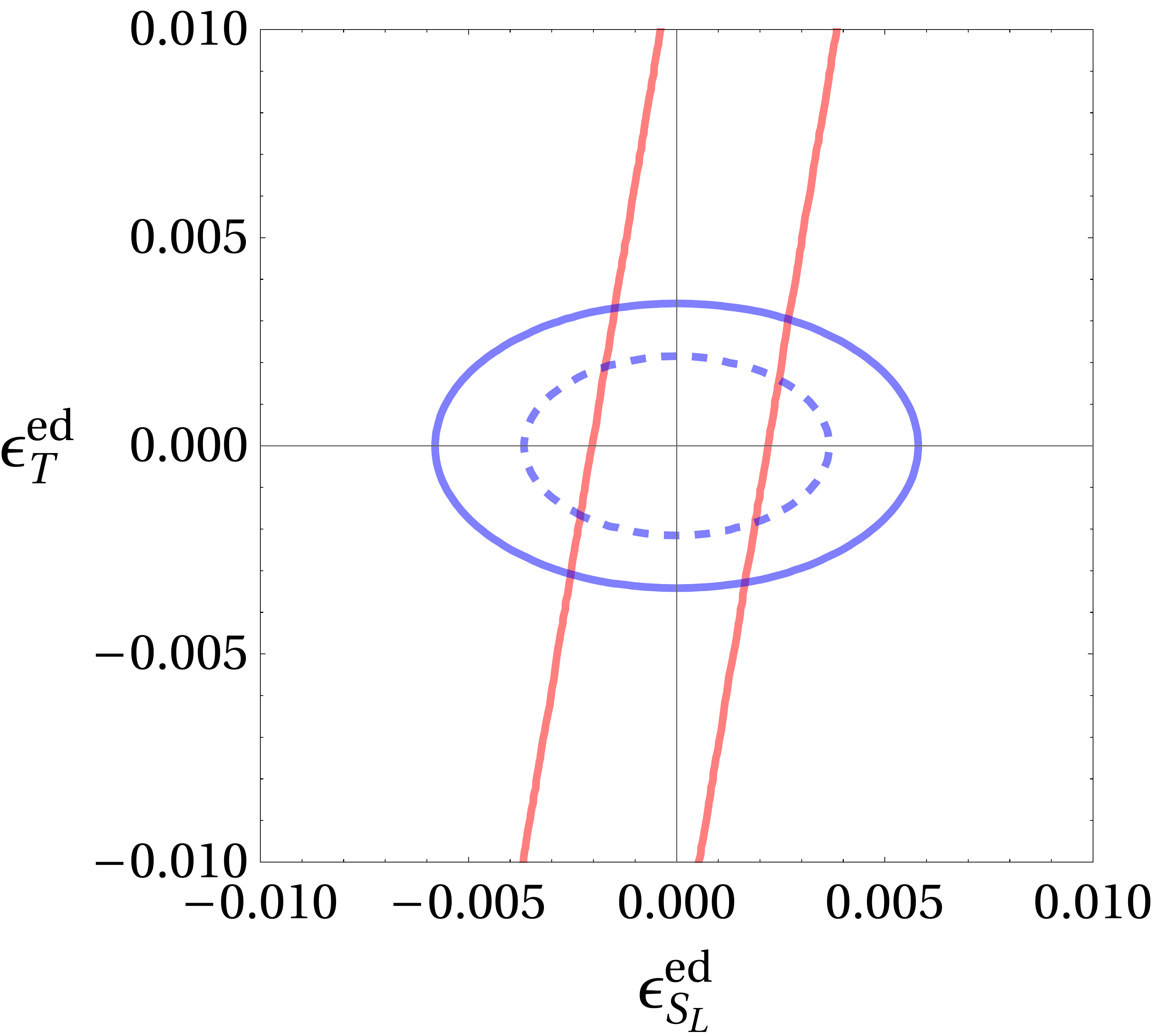} & \includegraphics[width=0.305\textwidth]{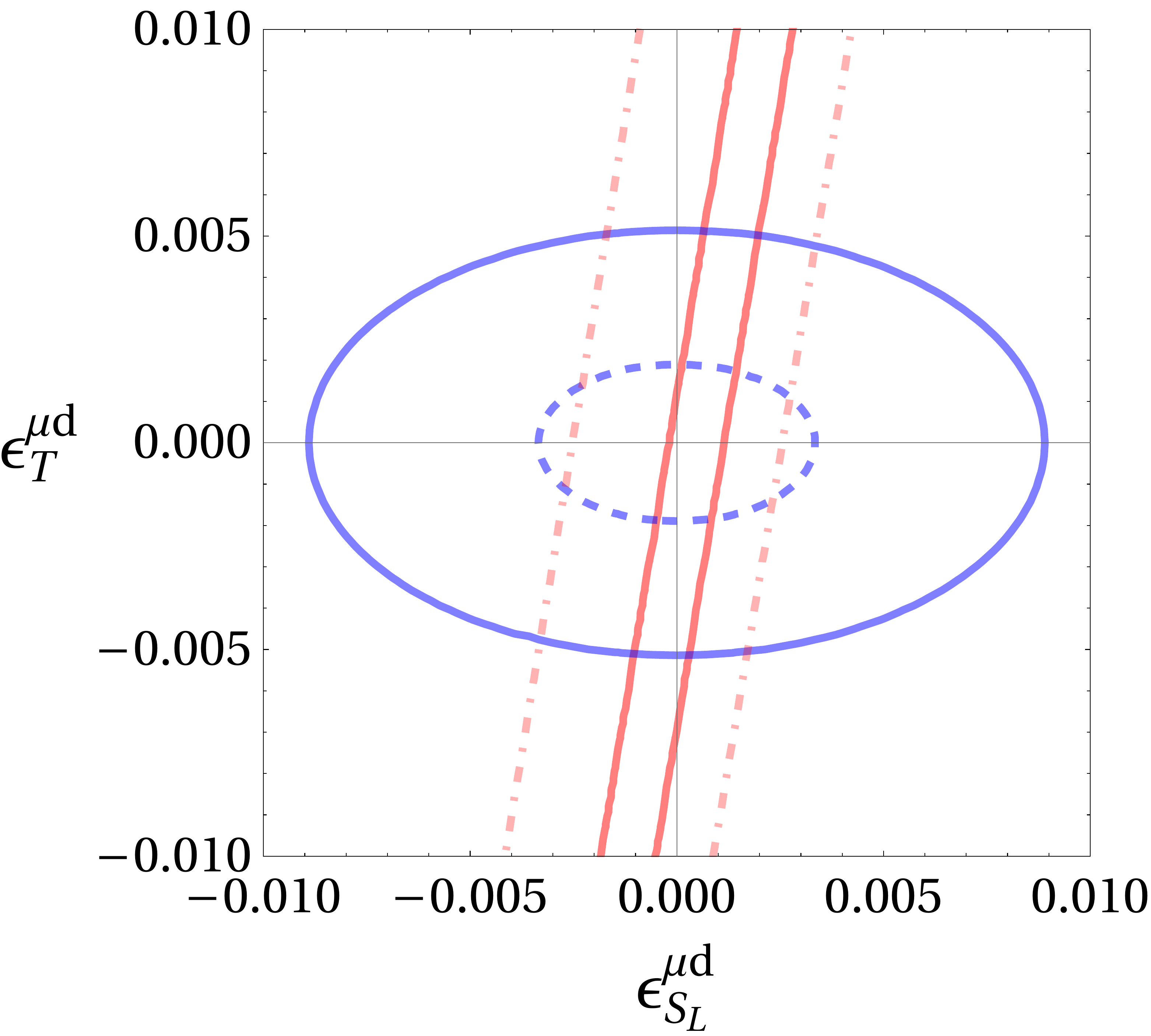} & \includegraphics[width=0.295\textwidth]{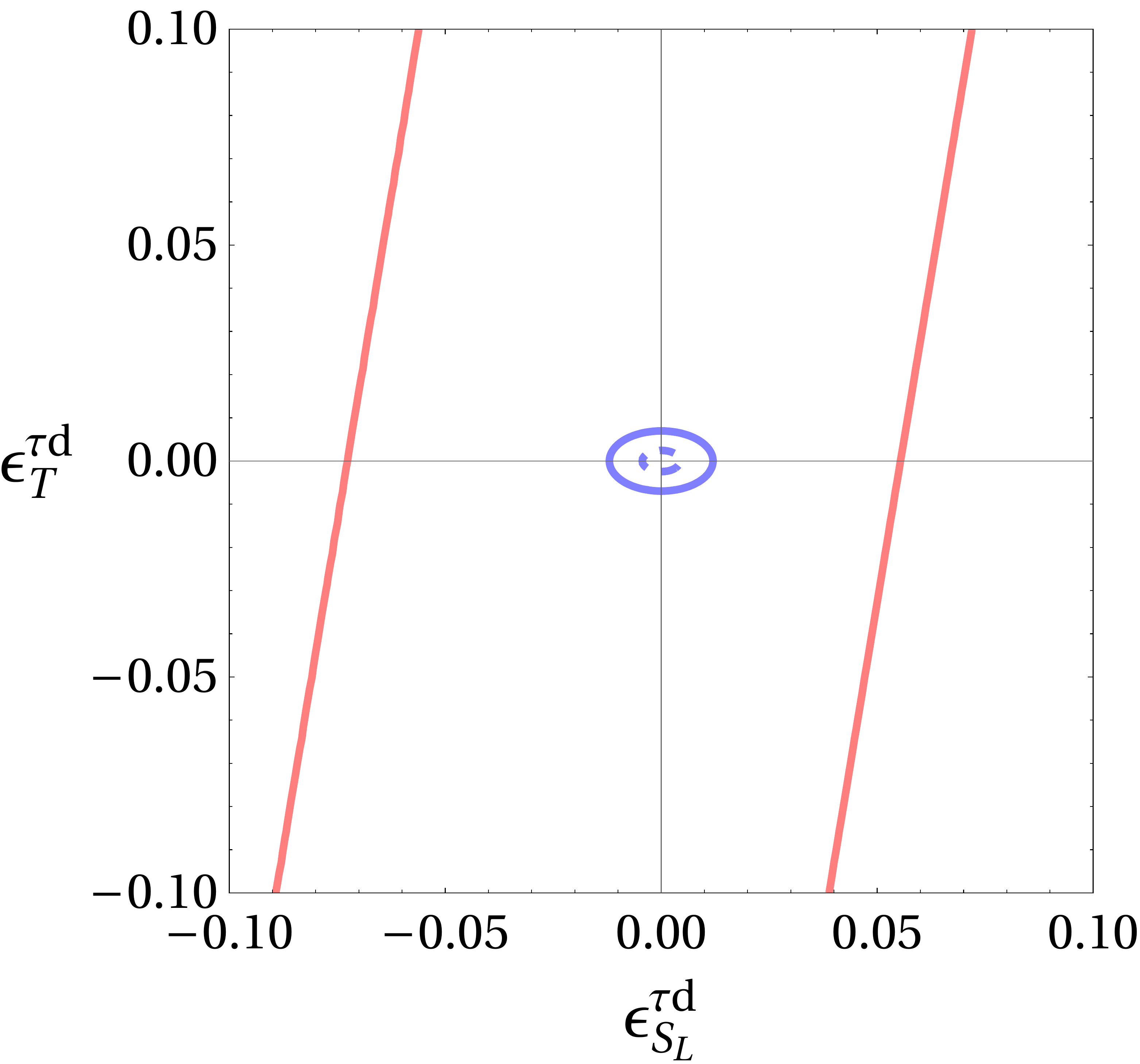}\\
\end{tabular}
\caption{95\% CL regions for {the combined fits of $\epsilon_{S_L}^{\alpha\beta i}$ and $\epsilon_{T}^{\alpha\beta i}$ to the charmed-meson decay data with $\beta=\alpha$ (red solid line) or $\beta\neq\alpha$ (light-red dash-dotted line) and to monolepton LHC data (blue solid line). Projections for the high-luminosity phase of the LHC (3 ab$^{-1}$), obtained by rescaling the expected limits with luminosity, are represented by dashed ellipses.}   \label{fig:interplay}}  
\end{figure}

The high-$p_T$ LHC bounds are also stronger than those from $D_{(s)}$-meson decays in all channels and WCs except for the pseudoscalar operators, constrained by the electronic and muonic $D_{(s)}$ decays. As discussed in Section~\ref{sec:D} and shown in Table~\ref{tab:WCsCC1TeV}, the latter strongly constrains any NP producing a single scalar or tensor operator at the high-energy scale. Even in this scenario, high-$p_T$ LHC limits are stronger for the tauonic operators and for the electronic tensor operators. 

In NP scenarios where various operators 
{with the same flavor entries} are produced at the matching scale, the complementarity between high-$p_T$ LHC and meson decays becomes more pronounced. As discussed above, the quadratic contributions of NP dominate the high-$p_T$ limits, allowing one to extract bounds on several operators simultaneously (see e.g.~Figure~\ref{fig:conclusions}). On the other hand, the $D_{(s)}$ branching fractions depend on interference terms between WCs, and some combinations remain unconstrained (tauonic operators) or poorly bounded by the low-energy data. 

To illustrate this, we compare in Figure~\ref{fig:interplay} the constraints on the $(\epsilon_{S_L}^{\alpha \beta i},~\epsilon_{T}^{\alpha\beta i})$ planes for $\mu = 1$\,TeV  obtained from fits to low-energy and high-$p_T$ data. We also show projections for the HL-LHC (3 ab$^{-1}$) derived by rescaling the sensitivity of the corresponding monolepton {expected limits} with luminosity. The bound stemming from the leptonic decays can be clearly appreciated in these figures, while the orthogonal directions are only constrained at low-energies by the semileptonic decays (electron and muon) or remain unconstrained (tau). Also, as shown in Figure~\ref{fig:interplay}, the low-energy bounds relax for LFV transitions as these do not interfere with the SM. For the same reason, imaginary WCs or operators beyond the SMEFT with light right-handed neutrinos accessible in charm decays are better constrained from high-$p_T$ tails.

\subsection{\texorpdfstring{$W$}{W} vertex corrections}
\label{sec:Wvertex}

\begin{table}[t]
\centering
\begin{tabular}{c|c|cc}
\toprule
$i$& $\alpha$ &$\boldsymbol{\delta g_L^{ci}\times10^2}$ &$\boldsymbol{\delta g_R^{ci}\times 10^2}$\\
\midrule
\multirow{4}{*}{$d$}&$e$&$[-2,~11]$&$[-2,~11]$\\
                    &$\mu$&$[-6,1.8]$&$[-1.2,~6]$\\
                    &$\tau$&$[-21,~27]$&$[-27,~21]$\\[2pt]
                    \cline{2-4}
                    &Av.&$[-4,~3]$&$[-0.4,~6]$\\
\midrule
\multirow{4}{*}{$s$}&$e$&$[-7,~8]$&$[-7,~8]$\\
                    &$\mu$&$[-3,~5]$&$[-6,~2]$\\
                    &$\tau$&$[-1.8,~7]$&$[-7,~1.4]$\\[2pt]
                    \cline{2-4}
                    &Av.&$[-1,~4]$&$[-5,~0.7]$\\
\bottomrule                 
\end{tabular}
  \caption{$95\%$~CL limits on $W c d_i$ vertex corrections assuming only one coupling active at a time. See Section~\ref{sec:Wvertex} for details.}\label{tab:vertexcorrs}
\end{table}

Right-handed contributions of the type $\epsilon_{V_R}^{\alpha i}$ can be generated at $\mathcal O(v^2/\Lambda^2)$ only as a vertex correction  to the quarks by the operator $\mathcal O_{\phi u d}$ (universal for all lepton flavors $\alpha$). Thus, $\epsilon_{V_R}^{\alpha i}$ does not receive a bound from our analysis of the LHC data. In case of the left-handed operator a combination of meson decay and high-$p_T$ LHC bounds constrains simultaneously vertex and four-fermion corrections. We define vertex corrections to $W$ couplings to quarks as
\begin{equation}\label{eq:modWcoups}
\mathcal L_W \supset \frac{g}{\sqrt{2}}V_{ci}\left(1+\delta g_L^{ci}\right)\bar c \gamma^\mu P_L d_i \,W^+_\mu+\frac{g}{\sqrt{2}}V_{ci}\,\delta g_R^{ci}\,\bar c \gamma^\mu P_R d_i\, W^+_\mu+{\rm h.c.} \,,
\end{equation}
where in terms of the conventions of Section~\ref{sec:eft} imply,
\begin{equation}\label{eq:modWcoupsSMEFT}
 \delta g_L^{ci}=\frac{V_{ji}}{V_{ci}}[\mathcal{C}_{\phi q}^{(3)}]_{2j},\qquad  \delta g_R^{ci}=\frac{1}{2V_{ci}}\,[\mathcal{C}_{\phi ud}]_{2i},
\end{equation}
with a sum over $j$ implicit in the first equation. These coupling modifications are typically constrained by LEP and LHC on-shell vector boson production~\cite{Efrati:2015eaa,Falkowski:2015jaa,Falkowski:2017pss}. Still, our analysis of charm transitions can play an important role to fully constrain some of these couplings or give a handle to disentangle the contributions of different operators. 

In Table~\ref{tab:vertexcorrs}, we show the limits on the vertex corrections obtained by combining the low and high energy bounds in Tables~\ref{tab:WCsCClowE} and~\ref{tab:WCsCChighE}, respectively. The results for $\delta g_L^{ci}$ are obtained subtracting and profiling over the maximal contribution of four-fermion operators allowed by the high-$p_T$ tails. We have assumed that the analogous vertex corrections in the couplings of the $W$ to the leptons are absent, so the bounds in the different channels can be combined ("Av." in the table). In addition, for each lepton channel we are assuming that only one of the two possible corrections (left-handed or right-handed) are active at a time. This is not needed for the muon channels where leptonic and semileptonic decays lead to comparable limits such that both couplings can be simultaneously constrained. However for the electron (tau) channel only the bound from the semileptonic (leptonic) decay is relevant and there are blind directions in the corresponding $(\delta g_L^{ci},~\delta g_R^{ci})$ planes. It is remarkable that the combination of charm decays and high-$p_T$ monolepton tails leads to a determination of $W$ vertex corrections competitive to LEP and LHC on-shell $W$ production~\cite{Efrati:2015eaa,Falkowski:2015jaa,Falkowski:2017pss}.

\section{Neutral currents}
\label{sec:NC}

\subsection{Theoretical framework: \texorpdfstring{$c \to u \, e^\alpha \bar e^\beta$}{c->ull}}

As a rule of thumb, flavor changing neutral currents (FCNC) probe scales far beyond the reach of current high energy colliders. However, FCNC in charmed meson decays seem to be an exception to a large extent. In this section, we perform a combined analysis of low- and high-$p_T$ data in the context of $c \to u \, e^\alpha \bar e^\beta$ transitions. The relevant dimension-six effective Lagrangian is
\begin{align}\label{eq:LagNC}
\mathcal{L}_{\rm NC}=\frac{4G_F}{\sqrt{2}}\frac{\alpha}{4\pi}\lambda_c\sum_{k,\alpha,\beta}\epsilon_k^{\alpha\beta}\mathcal{O}_k^{\alpha\beta}+{\rm h.c.} 
\end{align}
The most general set of four-fermion operators compatible with $SU(3)_c\times U(1)_{\rm em}$ is 
\begin{align}\label{eq:OpsNC}
\begin{aligned}
\mathcal{O}_{V_{LL}}^{\alpha\beta}&=(\bar e_L^\alpha\gamma_\mu e_L^\beta)(\bar u_L\gamma^\mu c_L)\,, \qquad\qquad&
 \mathcal{O}_{V_{RR}}^{\alpha\beta}&=(\bar e_R^\alpha\gamma_\mu e_R^\beta)(\bar u_R\gamma^\mu c_R)\,,\\
\mathcal{O}_{V_{LR}}^{\alpha\beta}&=(\bar e_L^\alpha\gamma_\mu e_L^\beta)(\bar u_R\gamma^\mu c_R)\,, &
 \mathcal{O}_{V_{RL}}^{\alpha\beta}&=(\bar e_R^\alpha\gamma_\mu e_R^\beta)(\bar u_L\gamma^\mu c_L)\,,\\
\mathcal{O}_{S_{LL}}^{\alpha\beta}&=(\bar e_R^\alpha e_L^\beta)(\bar u_R c_L)\,, & 
\mathcal{O}_{S_{RR}}^{\alpha\beta}&=(\bar e_L^\alpha e_R^\beta)(\bar u_L c_R)\,,\\
\mathcal{O}_{S_{LR}}^{\alpha\beta}&=(\bar e_R^\alpha e_L^\beta)(\bar u_L c_R)\,, &
 \mathcal{O}_{S_{RL}}^{\alpha\beta}&=(\bar e_L^\alpha e_R^\beta)(\bar u_R c_L)\,,\\
\mathcal{O}_{T_L}^{\alpha\beta}&=(\bar e_R^\alpha \sigma_{\mu\nu} e_L^\beta)(\bar u_R \sigma^{\mu\nu} c_L)\,, &
 \mathcal{O}_{T_R}^{\alpha\beta}&=(\bar e_L^\alpha \sigma_{\mu\nu} e_R^\beta)(\bar u_L \sigma^{\mu\nu} c_R)\,,
\end{aligned}
\end{align}
with $\alpha,\beta$ being lepton flavor indices. Note that mixed chirality tensor operators are zero by Lorentz invariance. The matching to the SMEFT in Eq.~\eqref{eq:SMEFTlag} yields the following relations,
\begin{align}
\begin{aligned}\label{eq:CN-match}
\epsilon_{V_{LL}}^{\alpha\beta}&=\frac{2\pi}{\alpha\lambda_c}\,\big([\mathcal{C}_{lq}^{(1)}]_{\alpha\beta12}-[\mathcal{C}_{lq}^{(3)}]_{\alpha\beta12}\big)\,,  \qquad& 
\epsilon_{V_{RR}}^{\alpha\beta}&=\frac{2\pi}{\alpha\lambda_c}\,[\mathcal{C}_{eu}]_{\alpha\beta12}\,, \\
\epsilon_{V_{LR}}^{\alpha\beta}&=\frac{2\pi}{\alpha\lambda_c}\,[\mathcal{C}_{lu}]_{\alpha\beta12}\,, & 
\epsilon_{V_{RL}}^{\alpha\beta}&=\frac{2\pi}{\alpha\lambda_c}\,[\mathcal{C}_{qe}]_{12\alpha\beta}\,, \\
\epsilon_{S_{LL}}^{\alpha\beta}&=-\frac{2\pi}{\alpha\lambda_c}\,[\mathcal{C}_{lequ}^{(1)}]^*_{\beta\alpha21}\,, &
\epsilon_{S_{RR}}^{\alpha\beta}&=-\frac{2\pi}{\alpha\lambda_c}\,[\mathcal{C}_{lequ}^{(1)}]_{\alpha\beta12}\,, \\
\epsilon_{S_{LR}}^{\alpha\beta}&=0\,, &
\epsilon_{S_{RL}}^{\alpha\beta}&=0\,, \\
\epsilon_{T_L}^{\alpha\beta}&=-\frac{2\pi}{\alpha\lambda_c}\,[\mathcal{C}_{lequ}^{(3)}]^*_{\beta\alpha21}\,, &
\epsilon_{T_R}^{\alpha\beta}&=-\frac{2\pi}{\alpha\lambda_c}\,[\mathcal{C}_{lequ}^{(3)}]_{\alpha\beta12}\,,
\end{aligned}
\end{align}
at the matching scale $\mu = m_W$ and in the up-quark mass basis. We only consider the four-fermion operators in the \textit{Warsaw basis} (see Table 3 of Ref.~\cite{Grzadkowski:2010es}) and neglect other effects such as the $Z$-boson vertex modification. The operators $\mathcal{O}_{S_{LR}}$ and $\mathcal{O}_{S_{RL}}$ are not generated in the SMEFT due to gauge invariance and are consistently neglected in our analysis. As discussed before, the RGE running from $\mu=1$~TeV down to $\mu=2$~GeV yields sizable effects in scalar and tensor operators, while the vector operators remain practically unchanged. In particular, using Refs.~\cite{Jenkins:2017dyc,Gonzalez-Alonso:2017iyc} we {obtain}
\begin{align}\label{eq:RGEopsNC}
\epsilon_{S_X}(2\,\mathrm{GeV})&\approx 2.1\,\epsilon_{S_X}(\mathrm{TeV})-0.5\,\epsilon_{T_X}(\mathrm{TeV})\,, &
\epsilon_{T_X}(2\,\mathrm{GeV})&\approx 0.8\,\epsilon_{T_X}(\mathrm{TeV})\,, 
\end{align}
where $X$ stands for the same chirality pairs, either $LL$ or $RR$.

\subsection{Rare charm decays}
\label{sec:NCconst}

Short-distance SM contributions to $c\to u\ell\ell^{(\prime)}$ transitions are strongly suppressed by the GIM mechanism. As a result, the main SM contributions to rare $D$-meson decay amplitudes are due to long-distance effects~\cite{Burdman:2001tf,deBoer:2016dcg,deBoer:2015boa,Fajfer:2015zea,Feldmann:2017izn,Bause:2019vpr}. While this will be a limiting factor once the experimental measurements become more precise, at present one can obtain bounds on short-distance NP entering in $D^0\to\ell\ell^{(\prime)}$ and $D_{(s)}\to P\ell\ell^{(\prime)}$ by assuming that the experimental limits are saturated by short-distance NP contributions~\cite{Bause:2019vpr,deBoer:2015boa,Fajfer:2015zea}. The short-distance contributions to the leptonic rare $D$ decay rate read
\begin{align}
\begin{aligned}
\mathcal{B}(D^0\to \ell_\alpha^- \ell_\beta^+) &= \frac{\tau_{D_0}}{256\pi^3}\frac{\alpha^2\, G_F^2\,f_D^2\, \lambda_c^2}{m_{D_0}^2}\, \lambda^{1/2}(m_{D_0}^2,m_{\ell_\alpha}^2,m_{\ell_\beta}^2)\\
&\times\Bigg{\lbrace} [m_{D_0}^2-(m_{\ell_\alpha}-m_{\ell_\beta})^2]\, \Bigg|(m_{\ell_\alpha}+m_{\ell_\beta})\,\epsilon_A^{\alpha\beta}-\frac{m_{D_0}^2}{m_c+m_u}\,\epsilon_P^{\alpha\beta} \Bigg{|}^2 \\
&\;\;\;+ [m_{D_0}^2-(m_{\ell_\alpha}+m_{\ell_\beta})^2]\,\Bigg|(m_{\ell_\alpha}-m_{\ell_\beta})\,\epsilon_{A^\prime}^{\alpha\beta}-\frac{m_{D_0}^2}{m_c+m_u}\,\epsilon_{P^\prime}^{\alpha\beta} \Bigg{|}^2\Bigg{\rbrace},
\end{aligned}
\end{align}
with $\lambda(a,b,c)=(a-b-c)^2+4bc$ and where we used the following WC redefinitions
\begin{align}
\epsilon_{A,A^\prime}^{\alpha\beta}&=(\epsilon_{V_{LR}}^{\alpha\beta}-\epsilon_{V_{LL}}^{\alpha\beta})\mp(\epsilon_{V_{RR}}^{\alpha\beta}-\epsilon_{V_{RL}}^{\alpha\beta})\,,&&&
\epsilon_{P,P^\prime}^{\alpha\beta}&=(\epsilon_{S_{LR}}^{\alpha\beta}-\epsilon_{S_{LL}}^{\alpha\beta})\mp(\epsilon_{S_{RR}}^{\alpha\beta}-\epsilon_{S_{RL}}^{\alpha\beta})\,.
\end{align}
As already discussed in Section~\ref{sec:D}, leptonic decays are unable to probe parity-even scalar, vector and tensor quark currents. Moreover, axial vector quark currents are chirally suppressed. This suppression is particularly strong for the dielectron channel, making current limits from leptonic decays not competitive. In these cases, better limits are found using semileptonic transitions. The differential branching ratios for $D\to\pi\ell\ell^{(\prime)}$ and $D_s\to K\ell\ell^{(\prime)}$ decays are studied in Refs.~\cite{Bause:2019vpr,deBoer:2015boa}. Currently, the best limits are obtained using $D^+\to\pi^+\ell\ell^{(\prime)}$ decays, for which an expression analogous to that in Eq.~\eqref{eq:3bodyparam} can be found in Ref.~\cite{Bause:2019vpr}.\footnote{Note that we use a different EFT basis compared to Refs.~\cite{Bause:2019vpr}. The relation between our WCs and those in this reference are
\begin{align}
\begin{aligned}
C_{9,10}&=\frac{\lambda_c}{2}\,(\epsilon_{V_{RL}}\pm\epsilon_{V_{LL}})\,,&
C_{9^\prime,10^\prime}&=\frac{\lambda_c}{2}\,(\epsilon_{V_{RR}}\pm\epsilon_{V_{LR}})\,,&
C_{T,T5}&=\lambda_c\,(\epsilon_{T_R}\pm\epsilon_{T_L})\,,\\
C_{S,P}&=\frac{\lambda_c}{2}\,(\epsilon_{S_{RR}}\pm\epsilon_{S_{LR}})\,,&
C_{S^\prime,P^\prime}&=\frac{\lambda_c}{2}\,(\epsilon_{S_{RL}}\pm\epsilon_{S_{LL}})\,.
\end{aligned}
\end{align}
The SMEFT matching in Eq.~\eqref{eq:CN-match} imply $C_S=C_P$ and $C_{S^\prime}=-C_{P^\prime}$. This is analogous to the relations for neutral currents in the down sector found in~\cite{Alonso:2014csa}. The main difference is that tensor operators are not generated in the down sector when matching to the SMEFT.} Barring cancellations among WCs, we derive the following $95\%$ CL limits at the charm-mass scale\footnote{For the $ee$ channel, we use the same hadronic coefficients as the ones provided in Ref.~\cite{Bause:2019vpr} for the LFV case, given that both experimental limits are obtained from the same BaBar analysis~\cite{Lees:2011hb} using the same kinematical regimes and that lepton mass effects are negligible.}
\begin{align}
\begin{aligned}
|\epsilon_{V_i}^{ee}|&\lesssim42 
\,,&
|\epsilon_{S_{LL,RR}}^{ee}|&\lesssim1.5\,,&
|\epsilon_{T_{L,R}}^{ee}|&\lesssim66 
\,,\\[5pt]
|\epsilon_{V_i}^{\mu\mu}|&\lesssim8\,,&
|\epsilon_{S_{LL,RR}}^{\mu\mu}|&\lesssim0.4\,,&
|\epsilon_{T_{L,R}}^{\mu\mu}|&\lesssim9\,,\\[5pt]
|\epsilon_{V_i}^{e\mu,\mu e}|&\lesssim16\,,&
|\epsilon_{S_{LL,RR}}^{e\mu,\mu e}|&\lesssim0.6\,,&
|\epsilon_{T_{L,R}}^{e\mu,\mu e}|&\lesssim110\,,
\end{aligned}
\end{align}
with $i=LL,RR,LR,RL$. These low-energy limits have flat directions in WC space, which could significantly weaken these bounds in given NP scenarios. Moreover, there are no limits on tau leptons, since $D$ decays involving taus are either kinematically forbidden or have not been searched for like in the $D^0\to e \tau$ case. In comparison with the light lepton case, the strong phase space suppression in $D^0\to e \tau$ is compensated by the lack of chiral suppression for the axial current. If an experimental limit on the $\mathrm{BR}(D^0\to e^\pm \tau^\mp)$ at the level of the one for the $\mathrm{BR}(D^0\to e^\pm\mu^\mp)$ existed, we would obtain a bound of $|\epsilon_{V_i,S_{LL,RR}}^{e\tau,\tau e}|\lesssim10$.

\subsection{High-\texorpdfstring{$p_T$}{pT} dilepton tails}

Following the footsteps of Section~\ref{sec:LHC} we perform the high-$p_T$ Drell-Yan analysis to extract the limits on the WCs. We focus on the lepton flavor conserving cases while the limits on LFV can be found in the recent Ref.~\cite{Angelescu:2020uug}. The partonic level cross section formula can be trivially obtained from Eq.~\eqref{eq:xsection}. The notable difference is that the SM contribution to $\bar u c \to e^\alpha \bar e^\alpha$ scattering is loop and GIM-suppressed. As a result, the interference of NP with the SM can be completely neglected. As discussed in Section~\ref{sec:tails}, the interference among different WCs is negligible for an inclusive angular analysis. These two statements have an important implication. Namely, the high-$p_T$ tails can set a bound on the sum of absolute values of WCs featuring different Lorentz structures. 

We set up a simulation pipeline and analysis procedure analogous to the one discussed in Section~\ref{sec:LHC} to recast the experimental searches. For $ee$ and $\mu\mu$ channels, we recast the analysis from the CMS collaboration in Ref.~\cite{CMS:2019tbu}, using $140~\mathrm{fb}^{-1}$ of $13~\mathrm{TeV}$ data. For the $\tau\tau$ channel, we use the search by ATLAS~\cite{Aaboud:2017sjh} with $36.1~\mathrm{fb}^{-1}$ of $13~\mathrm{TeV}$ data. In all cases, we validate the simulation procedure against the MC samples for the SM $Z\to e^\alpha\bar e^\alpha$ process provided by the experimental collaborations. In $ee$ and $\mu\mu$ channels we achieved a 10\% level of agreement with respect to the experimental results. For the $\tau\tau$ channel, we additionally validate our analysis against the CMS simulation of the sequential SM $Z^\prime$ signal. In this case, the level of agreement achieved is around 20\%. The $95\%$~CL limits on the neutral-current WCs are shown in Table~\ref{tab:WCsNChighE}. 
As in the charged-current case, these limits are provided both at the high-energy and at the low-energy scale, using the expressions in Eq.~\eqref{eq:RGEopsNC}.

\begin{table}[t]
\centering
\begin{tabular}{c|ccccc}
\toprule
\multirow{2}{*}{$\alpha$} & \multirow{2}{*}{$\boldsymbol{|\epsilon_{V_i}^{\alpha\alpha}|}$} & \multicolumn{2}{c}{$\boldsymbol{|\epsilon_{S_{LL,RR}}^{\alpha\alpha}(\mu)|}$} & \multicolumn{2}{c}{$\boldsymbol{|\epsilon_{T_{L,R}}^{\alpha\alpha}(\mu)|}$}\\
& & $\mu=1$~TeV & $\mu=2$~GeV & $\mu=1$~TeV & $\mu=2$~GeV \\
\midrule
$e$ & $13\,(3.9)$ & $15\,(4.5)$ & $32\,(9.5)$ & $6.5\,(2.0)$ & $5.2\,(1.6)$ \\
$\mu$ & $7.0\,(3.4)$ & $8.1\,(3.9)$ & $17\,(8.3)$ & $3.5\,(1.7)$ & $2.8\,(1.4)$\\
$\tau$ & $25\,(12)$ & $29\,(13)$ & $60\,(28)$ & $14\,(6.6)$ & $11\,(5.2)$\\
\bottomrule
\end{tabular}
  \caption{$95\%$~CL limits on the neutral-current WCs from $p p \to e^\alpha \bar e^\alpha$ at the LHC, with $i=LL,RR,LR,RL$. We also show in parenthesis the naive projections of the expected limits for the HL-LHC (3 ab$^{-1}$), assuming that the error will be statistically dominated.}\label{tab:WCsNChighE}
\end{table}

\begin{figure}[t]
\centering
\includegraphics[width=0.5\textwidth]{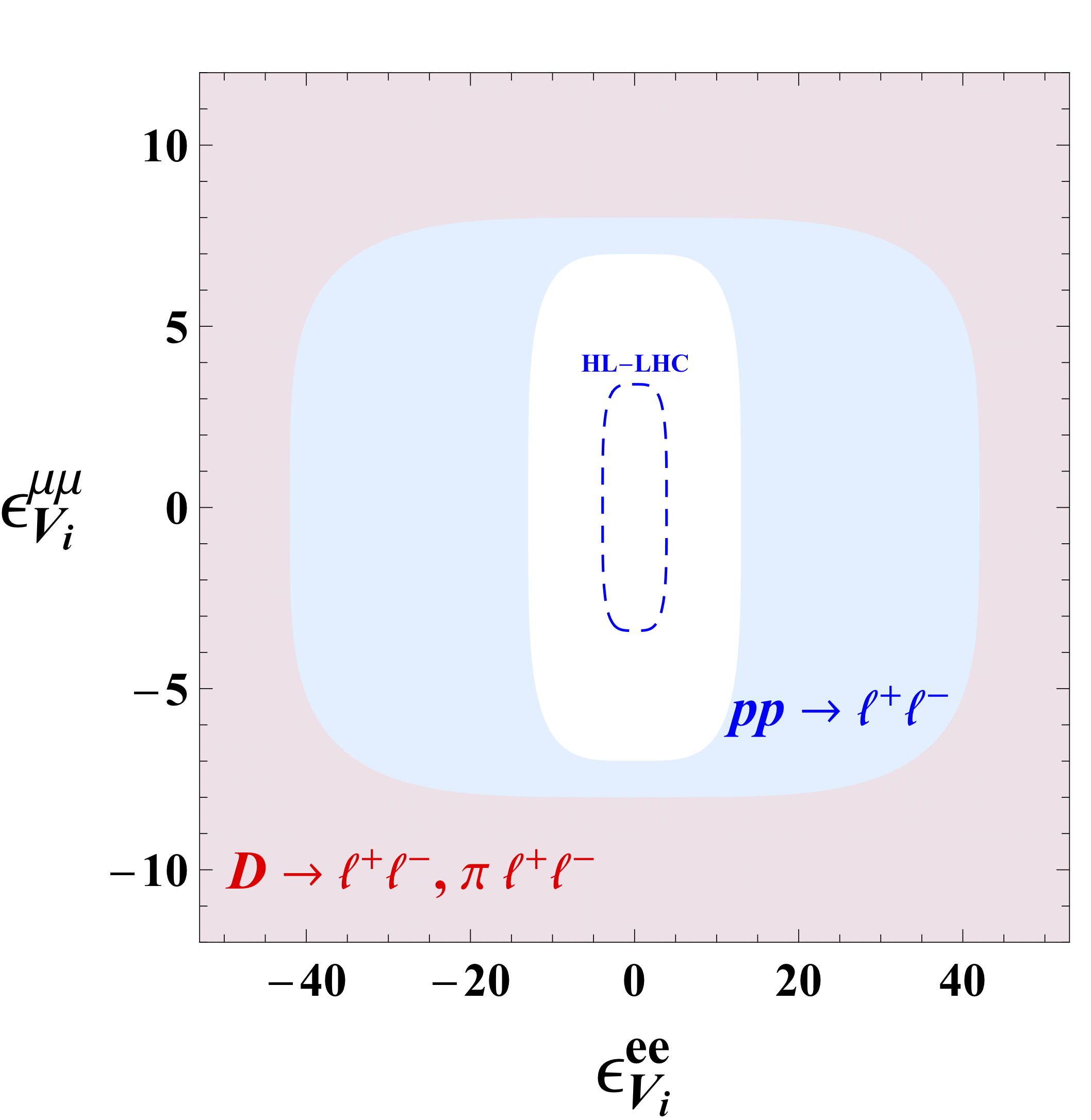}
\caption{Exclusion limits at 95\% CL on $c \to u \ell^+ \ell^-$ transitions in the $(\epsilon^{e e}_{V_i}, \epsilon^{\mu \mu}_{V_i})$ plane, where $i = LL, RR, LR, RL$. The region outside the red contour is excluded by $D$ meson decays, while the region outside the blue contour is excluded by high-$p_T$ LHC. \label{fig:NCinterplay}}
\end{figure}

As anticipated, the high-$p_T$ limits obtained for these transitions compete in most instances with those found at low-energies. This is particularly well illustrated in Figure~\ref{fig:NCinterplay} for the vector operators. In this case, our high-$p_T$ limits are stronger than (comparable to) those obtained from low-energy data for the electron (muon) channel. For the tensor operators, high-$p_T$ offers a better probe, while the scalar operators are better constrained by leptonic charm decays, since they receive a large chiral enhancement in $D \to \ell^+ \ell^-$ compared to the corresponding SM contribution. Furthermore, the $c\to u\tau^+\tau^-$ transition is only accessible at high-$p_T$, since the corresponding low-energy decays are  kinematically forbidden. Similar conclusions have been reached in the LFV channels~\cite{Angelescu:2020uug}. Namely, the high-$p_T$ bounds on the $\mu e$ channel are stronger than those from low-energy, with the exception of the scalar operators, while for $\tau e$ and $\tau\mu$ channels, high-$p_T$ tails offer the only available limits.

Concerning the possible caveats to the high-$p_T$ limits, there are two major differences with respect to the discussion for charge currents in Section~\ref{sec:caveats}. Firstly, the $c\to u\ell^+\ell^-$ SM amplitude is extremely suppressed, as mentioned before. Thus, the dimension-8 interference with the SM is negligible and unable to affect the leading dimension-6 squared contribution, even though the two are formally of the same order in the EFT expansion. Nonetheless, semileptonic operators with flavor-diagonal quark couplings which negatively interfere with the SM background can be used to tune a (partial) cancellation between NP contributions in the tails. Secondly, most UV completions of the relevant SMEFT operators feature mediators that are charged and (or) colored, such as leptoquarks or extra Higgses. The neutral components of these representations, which mediate $c\to u \ell^+ \ell^-$ transitions, cannot be  significantly lighter than other $SU(2)_L$ components due to electroweak precision tests. As an exception, and unlike the charged-current case, it is now possible to have an $s$-channel tree-level mediator which is a complete SM gauge singlet, a vector $Z^\prime$. Being a SM singlet, pair production limits are not robust and (a priori) the mediator could be very light. In this case, one would require a dedicated study for a low-mass dilepton resonance taking into account stringent limits from $D$ meson oscillation induced at tree-level.

\subsection{Comments on \texorpdfstring{$\Delta S = 1$}{DS=1} and  \texorpdfstring{$\Delta B = 1$}{DB=1} rare transitions}
\label{sec:Down}

In Section~\ref{sec:LHC}, we showed how to translate the high-$p_T$ monolepton bounds to other initial quark flavor combinations by simply rescaling with PDF, and validated this method against existing simulations of $b c \to \tau \nu$. The reasoning behind this procedure is that the signal acceptance is similar for other initial quark combinations, since the analyses are largely inclusive in angular cuts and the invariant mass distributions have a similar shape across the small range of most sensitive bins in the tails. In analogy, the results of the high-$p_T$ dilepton analysis reported in the context of $\Delta C = 1$ transitions in Table~\ref{tab:WCsNChighE} can be used to estimate the bounds on other flavor violating transitions.

\begin{figure}[t]
\centering
\includegraphics[width=0.5\textwidth]{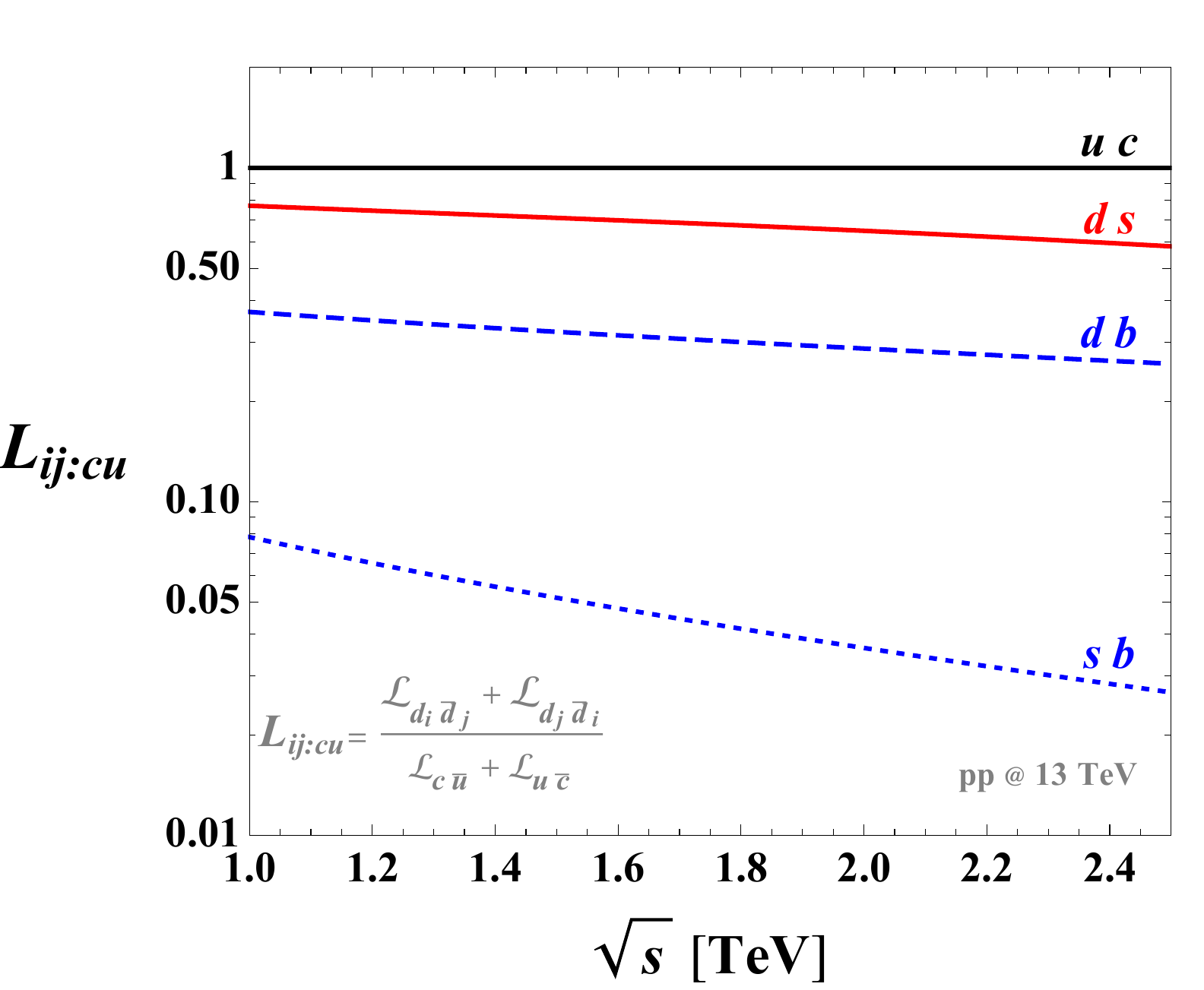}
\caption{Parton luminosity ratios defined in Eq.~\eqref{eq:ratiorescale} as a function of dilepton invariant mass $\sqrt{s}$.  \label{fig:plot-nc-ratio}}
\end{figure}

To illustrate this point, here we constrain $\Delta S = 1$ and $\Delta B = 1$ rare transitions from high-$p_T$ dilepton tails. More precisely, we derive limits on $s \to d$, $b \to d$, and $b \to s$ transitions starting from $c \to u$ limits. The low-energy Lagrangian for $d_i \to d_j$ with $i > j$ is given by Eqs.~\eqref{eq:LagNC} and \eqref{eq:OpsNC} after replacing $\lambda_c$ with $V_{t i} V^*_{t j}$\,, $\bar u$ with $\bar d_j$ and $c$ with $d_i$. Note that, tensor operators are absent for these transitions, see Ref.~\cite{Alonso:2014csa} for the SMEFT matching in the down sector. 
By equating the hadronic cross sections in the tails, we find
\begin{equation}\label{eq:rescale}
    |\epsilon_X^{\alpha \beta j i}| = |\epsilon_X^{\alpha \beta u c}| \, \frac{\lambda_c}{|V_{t i} V^*_{t j}| \sqrt{L_{ij:cu}}}\,, 
\end{equation}
where
\begin{equation}\label{eq:ratiorescale}
   L_{ij:cu} = \frac{\mathcal{L}_{d_i \bar d_j} + \mathcal{L}_{d_j \bar d_i}}{\mathcal{L}_{c \bar u} + \mathcal{L}_{u \bar c}}\,.
\end{equation}
The parton luminosity functions $\mathcal{L}_{q_i \bar q_j}$ are defined in Eq.~\eqref{eq:PDFlumi} and evaluated in the most sensitive bin $\sqrt{s} \sim [1-1.5]$\,TeV. The luminosity ratio $L_{ij:cu}$ as a function of the dilepton invariant mass is shown in Figure~\ref{fig:plot-nc-ratio} for all $i,j$ combinations. Some kinematic dependence is present when comparing valence and see quarks, which limits the accuracy of the method to $\mathcal{O}(10\%)$. Using Eq.~\eqref{eq:rescale}, we find approximate limits on the WCs of  $s \to d$, $b \to d$, and $b \to s$ to be $700$, $40$, and $20$ times the $c \to u$ limits in Table~\ref{tab:WCsNChighE}, respectively.

Rare $\Delta S = 1$ and $\Delta B = 1$ decays to light leptons clearly outperform the high-$p_T$ searches. On the other hand, tauonic modes are difficult at low energies and experimental limits on e.g. $b \to s \tau^+ \tau^-$  are far above the SM prediction, leaving plenty of room for NP. Assuming nonzero $\epsilon_{V_{LL}}^{\tau \tau sb} = \epsilon_{V_{RL}}^{\tau \tau sb}$, the BaBar search on $B \to K \tau^+\tau^-$~\cite{TheBaBar:2016xwe} imposes a limit $| \epsilon_{V_{LL}}^{\tau \tau sb} | < 990$ at 95\% CL. However, this transition is better probed in $p p \to \tau^+ \tau^-$ high invariant mass tail, $| \epsilon_{V_{LL}}^{\tau \tau sb} | < 420$. In addition, future prospects at the HL-LHC are more competitive than the prospects on $B \to K \tau^+\tau^-$ rescattering contribution to $B \to K \mu^+\mu^-$ derived in Ref.~\cite{Cornella:2020aoq}. 

Finally, a generic NP model correlates $\Delta F = 1$ operators to flavor non-universal flavor conserving $q^i \bar q^i \to e^{\alpha} \bar e^{\alpha}$ processes, which are the leading signatures if the flavor structure is MFV-like~\cite{Faroughy:2016osc, Greljo:2017vvb}. In fact, in many explicit models, $b \bar b \to \tau^+ \tau^-$ dominates over $V_{cb}$ suppressed $b \bar s \to \tau^+ \tau^-$, see Ref.~\cite{Faroughy:2016osc}.

\section{Constraints from \texorpdfstring{$SU(2)_L$}{SU(2)} gauge invariance}
\label{sec:SU2}

Imposing $SU(2)_L$ gauge invariance yields strong constraints on the WCs entering in charm decays by relating them to other transitions, such as $K$, $\pi$ or $\tau$ decays. We discuss the impact of these correlated constraints here. To keep the $SU(2)_L$ relations as generic as possible, in this section we use a different flavor basis in which the $SU(2)_L$ doublets are defined as
\begin{align}\label{eq:SU2Lorientation}
q_L^i&=\begin{pmatrix} V_u^{ij} u_L^j \\ V_d^{ij}\,d_L^j\end{pmatrix}\,,&
l_L^\alpha&=\begin{pmatrix} \nu_L^\alpha \\ e_L^\alpha\end{pmatrix}\,,
\end{align}
with the CKM matrix being $V=V_u^\dagger\, V_d$, while the right-handed fermions are already in their mass-eigenstate basis. Furthermore, whenever we do not impose down-quark alignment ($V_d\approx\mathbb{1}$) or up-quark alignment ($V_u\approx\mathbb{1}$) we assume that both $V_u$ and $V_d$ exhibit the same hierarchies as the CKM matrix.

\subsection{Charged currents}

We find the following complementary constraints:
\begin{itemize}
\item {\bf $\mathcal{O}_{V_L}^{\alpha\beta i}$:} We can decompose the SMEFT operators $\mathcal{O}_{lq}^{(1,3)}$ as
\begin{align}\label{eq:Olqdecomp}
{[\mathcal{O}_{lq}^{(3)}]}^{\alpha\beta ij}&=2\,({V_u^*}^{ik}\,V_d^{jl}\,[\mathcal{O}_{V_L}]^{\alpha\beta kl} + {V_d^*}^{ik}\,V_u^{jl}\,[\mathcal{O}_{V_L}^\dagger]^{\beta\alpha lk})\nonumber\\
&\quad+{V_u^*}^{ik}\,V_u^{jl}\left[(\bar\nu_L^\alpha\gamma^\mu\nu_L^\beta)(\bar u_L^k\gamma_\mu u_L^l)-(\bar e_L^\alpha\gamma^\mu e_L^\beta)(\bar u_L^k\gamma_\mu u_L^l)\right]\nonumber\\
&\quad-{V_d^*}^{ik}\,V_d^{jl}\left[(\bar\nu_L^\alpha\gamma^\mu\nu_L^\beta)(\bar d_L^k\gamma_\mu d_L^l)-(\bar e_L^\alpha\gamma^\mu e_L^\beta)(\bar d_L^k\gamma_\mu 
d_L^l)\right]\,,\\[5pt]
[\mathcal{O}_{lq}^{(1)}]^{\alpha\beta ij}&={V_u^*}^{ik}\,V_u^{jl}\left[(\bar\nu_L^\alpha\gamma^\mu\nu_L^\beta)(\bar u_L^k\gamma_\mu u_L^l)+(\bar e_L^\alpha\gamma^\mu e_L^\beta)(\bar u_L^k\gamma_\mu u_L^l)\right]\nonumber\\
&\quad+{V_d^*}^{ik}\,V_d^{jl}\left[(\bar\nu_L^\alpha\gamma^\mu\nu_L^\beta)(\bar d_L^k\gamma_\mu d_L^l)+(\bar e_L^\alpha\gamma^\mu e_L^\beta)(\bar d_L^k\gamma_\mu 
d_L^l)\right]\,,\nonumber
\end{align}
with $\mathcal{O}_{V_L}^{\alpha\beta i}=\mathcal{O}_{V_L}^{\alpha\beta 2i}$. Clearly, by imposing $SU(2)_L$ invariance, one obtains new operator structures that lead to additional observables. From Eq.~\eqref{eq:Olqdecomp}, we find correlated relations with the following observables:
\begin{itemize}
\item[$i)$] Charged-current $d_i\to u\ell\nu$ and $\tau\to d_iu\nu$ transitions (1st line)\,,
\item[$ii)$] Neutral-current $c\to u\ell\ell^{(\prime)}$, $\tau\to\ell uu$ {decays} and $\mu u \to e u$ {conversion}  (2nd line)\,,
\item[$iii)$] Neutral-current $s\to d\ell\ell^{(\prime)}$, $s\to d\nu\nu$, $\tau\to\ell d_i d_j$ {decays} and $\mu d_i\to e d_i$ {conversion} (3rd line)\,,
\end{itemize}
where $\ell =e, \mu$\,. {Adjusting the coefficients of singlet and triplet operators in Eqs.~\eqref{eq:Olqdecomp} and adopting up- or down-quark alignment, one can in principle avoid some of these correlations. However, one cannot always escape all of them simultaneously, as we will discuss in the following.}

Assuming the CKM-like structure for $V_d$, $K\to\pi\nu\nu$ decays impose $|\epsilon_{V_L}^{\alpha\beta i}|\lesssim10^{-4}$, independently of the quark and lepton flavors. These bounds are significantly stronger than both charm and high-$p_T$ limits (see Sections~\ref{sec:D} and~\ref{sec:limits}). However, they can be alleviated by enforcing the relation $\mathcal{C}_{lq}^{(3)}\approx \mathcal{C}_{lq}^{(1)}$, or by assuming down-alignment and a diagonal flavor structure (nonzero WCs only for $i=j$).
Irrespective of these assumptions, the combination of $K \to \pi \nu \nu$, $K_L \to e \mu$ and $\mu-e$ conversion in nuclei set the robust bound $|\epsilon_{V_L}^{e\mu i}|\lesssim 10^{-4}$. 

For the $\tau\ell$ channel, LFV tau decays always offer {bounds stronger than those from charm decays or high-$p_T$.} To alleviate these, together with those from $K\to\pi\nu\nu$, one needs to enforce $\mathcal{C}_{lq}^{(3)}\approx-\mathcal{C}_{lq}^{(1)}$ {to cancel the contribution to tau decays} plus the down-quark aligned flavor structure described above {to avoid the bound from kaon decays}. Even in that tuned scenario, {the contribution to} $\tau\to\ell\rho$ {remains unsuppressed, and the corresponding bounds} are better than those from charm decays but comparable to the high-$p_T$ limits. 

For the $\ell\ell$ channel, the $K\to\pi\nu\nu$ and $K\to\ell\ell$ decays give the constraints  $|\epsilon_{V_L}^{eei}|\lesssim10^{-3}$, $|\epsilon_{V_L}^{\mu\mu i}|\lesssim10^{-4}$, even if we allow for cancellations between the singlet and triplet operators. For the $c\to s$ case, it is possible to avoid these constraints by enforcing down alignment and a diagonal flavor structure with non-zero $i=j=2$ entry.
In this limit, the bounds from $K\to\ell\nu$ are stronger than charged-current charm decays, and comparable to those from high-$p_T$ monolepton tails. Likewise, for the $c\to d$ decays obtained by demanding down alignment and a flavor structure with a non-zero $i=j=1$ WC, one would enter in conflict with $\pi\to \ell \nu$ decays or high $p_T$.

Finally, the only relevant neutral-current constraint for the $\tau\tau$ channel is $K\to\pi\nu\nu$, which can be removed by $\mathcal{C}_{lq}^{(3)}\approx \mathcal{C}_{lq}^{(1)}$. Still, charged-current $\tau$ decays provide comparable limits to those from charm decays, and can be alleviated with a mild alignment to the up eigenbasis.

\item {\bf $\mathcal{O}_{S_R}^{\alpha\beta i}$:} The SMEFT operator $\mathcal{O}_{ledq}$ decomposes as
\begin{align}
[\mathcal{O}_{ledq}^{\,\dagger}]^{\beta \alpha ij}=V_u^{*\,jk}\,\mathcal{O}_{S_R}^{\alpha\beta ki}+V_d^{*\,jk}\,(\bar e_R^\alpha \,e_L^\beta)(\bar d_L^k\, d_R^i)\,,
\end{align} 
with  $\mathcal{O}_{S_R}^{\alpha\beta i} \equiv \mathcal{O}_{S_R}^{\alpha\beta 2i}$. Both LFV and lepton flavor conserving transitions involving first- and second-generation leptons are better probed in kaon decays than in charm decays. In general, the correlated neutral-current transitions $K_L \to \ell^+ \ell^-$ set constraints on the corresponding WCs that are orders of magnitude stronger. One can evade this bound by imposing a strong down-alignment. 
Even in this case, $d_i\to u\ell \nu$ transitions provide stronger bounds than those from charged-current charm decays. Moving to $\tau$, LFV combinations are better constrained by the correlated neutral-current $\tau \to \ell P$ ($P=K,\phi$) decays. On the other hand, for $\alpha=\beta=3$ no constraints from the neutral-current operators are obtained. However, for $i=2$ the bounds from the charged-current $\tau \to K \nu$ decays are stronger unless one imposes a mild alignment to the up basis.

\item {\bf $\mathcal{O}_{S_L}^{\alpha\beta i}$ and $\mathcal{O}_{T}^{\alpha\beta i}$:} We have the following decomposition for the SMEFT operators $\mathcal{O}_{lequ}^{(1,3)}$:
\begin{align}
\begin{aligned}
{[\mathcal{O}_{lequ}^{(1)\,\dagger}]}^{\beta\alpha i2}&=V_d^{ik}\,\mathcal{O}_{S_L}^{\alpha\beta\,k}+V_u^{ik}\,(\bar e_R^\alpha \, e_L^\beta)(\bar c_R\, u_L^k)\,,\\
[\mathcal{O}_{lequ}^{(3)\,\dagger}]^{\beta\alpha i2}&=V_d^{ik}\,\mathcal{O}_{T}^{\alpha\beta\,k}+V_u^{ik}\,(\bar e_R^\alpha\sigma^{\mu\nu} e_L^\beta)(\bar c_R\sigma_{\mu\nu} u_L^k)\,,
\end{aligned}
\end{align}
which yields additional operators that generically contribute to $c\to u\ell\ell^{(\prime)}$ transitions.  This is not relevant for transitions involving $\tau$ leptons, since the bounds are absent due to kinematics. More precisely, $c \to u \tau \tau$ and $c \to u \mu \tau$ are forbidden while $c \to u e \tau$ is suppressed. On the other hand, neutral-current charm decays provide stronger constraints than their charged-current counterpart for $\alpha,\beta = 1,2$, unless the contributions to these transitions are suppressed by enforcing an approximate up alignment. While the scalar operators are better constrained at low energies for $\alpha,\beta = 1,2$, the tensor operators receive more stringent bounds from high-$p_T$.
In this case, monolepton and dilepton bounds are comparable.

\end{itemize}

\begin{figure}[t]
\centering
\includegraphics[width=0.9\textwidth]{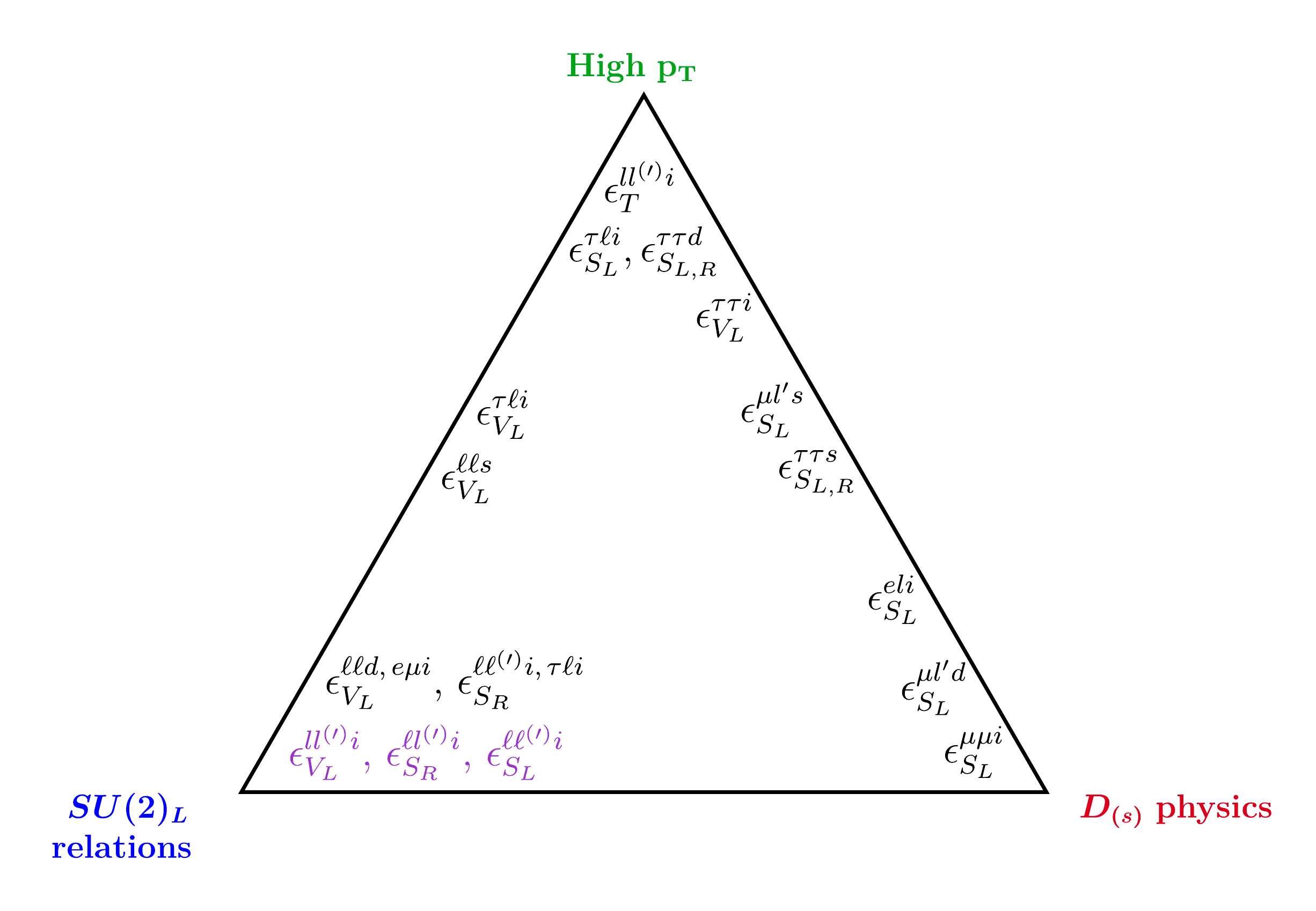}
\caption{Interplay between charm physics, high-$p_T$ LHC, and $SU(2)_L$ relations for the charged-current case ($\ell=e,\mu$ and $l=e,\mu,\tau$). The proximity of the WCs to a particular vertex of the triangle is determined, approximately, by the relative strength of the corresponding constraints. In purple, those constraints that can be avoided by a particular flavor structure and/or WC combination.}\label{fig:trianglePlotCC}
\end{figure}

The interplay between charged-current charm decays, high-$p_T$ lepton tails, and $SU(2)_L$ relations is shown in Figure~\ref{fig:trianglePlotCC} for $\mu = 2$\,GeV. While it is possible to evade some of the constraints obtained by $SU(2)_L$ gauge invariance, either by taking specific flavor structures and/or by having appropriate WC combinations, this typically requires tuning in most UV completions. Moreover, the required conditions are not radiatively stable, and one should in general consider loop-induced misalignments in a given NP scenario. Going beyond this analysis, explicit models typically generate $\Delta F = 2$ transitions which are severely constrained by neutral meson oscillations. This can be particularly problematic in the context of the up alignment, and it represents a challenge for model building. Another possible avenue beyond the SMEFT framework is to introduce a new light right-handed neutrino accessible in charm decays, yielding a new class of operators of the form $\mathcal{O}_{V_R}^{\nu_R} = (\bar e_R \gamma^\mu \nu_R) (\bar c_R \gamma_\mu d_R^i)$. However, explicit UV completions of this operator are not completely free from $SU(2)_L$ relations (see Refs.~\cite{Greljo:2018ogz,Asadi:2018wea}).

\subsection{Neutral currents}

Imposing $SU(2)_L$ gauge invariance in the neutral-current case also yields strong correlated constraints for the operators with left-handed fermions. Focusing on  $\mathcal{O}_{V_{LL}}^{\alpha\beta}$, it suffices to consider first only the contribution of the isosinglet SMEFT operator, $\mathcal O^{(1)}_{lq}$, avoiding correlations with charged-current decays. Assuming the CKM-like structure for $V_d$ one obtains the limit $\epsilon_{V_{LL}}^{\alpha\beta}<0.2$ for any lepton flavor from $K_L\to\pi\nu\bar\nu$ decays. This bound, which is considerably stronger than those from neutral-current charm, can however be alleviated by enforcing down-quark alignment. Even in this case, the LFV combinations receive better constraints than those from neutral charm decays (or high-$p_T$ dilepton production) by using the correlated bounds from $\mu-e$ conversion in nuclei and LFV tau decays. On the other hand, charm decays and high-$p_T$ dilepton tails give stronger constraints for the lepton-flavor conserving operators with $\alpha=\beta=1,~2$ if down alignment is enforced. However, in models producing also the isotriplet SMEFT operator, $\mathcal O^{(3)}_{lq}$, kaon semileptonic decays can provide similar (muon) or better (electron) bounds compared to charm rare decays.

Similarly, for the $\mathcal{O}_{V_{RL}}^{\alpha\beta}$ operator, the correlated bounds from $\mu-e$ conversion and LFV tau decays offer the best limits for the LFV channels, independently of the quark flavor assumptions. For $\alpha=\beta=1,2$, the related limits from $K_L\to\ell^+\ell^-$ yield bounds that are several orders of magnitude stronger than those from neutral charm decays, unless one imposes down alignment. On the other hand, the $\tau\tau$ channel remains unconstrained at low energies, even when considering the $SU(2)_L$ relations. Finally, no $SU(2)_L$ constraint can be derived for $\mathcal{O}_{V_{LR}}^{\alpha\beta}$ since $D \to P \nu \bar \nu$ have not been searched for.

\begin{figure}[t]
\centering
\includegraphics[width=0.9\textwidth]{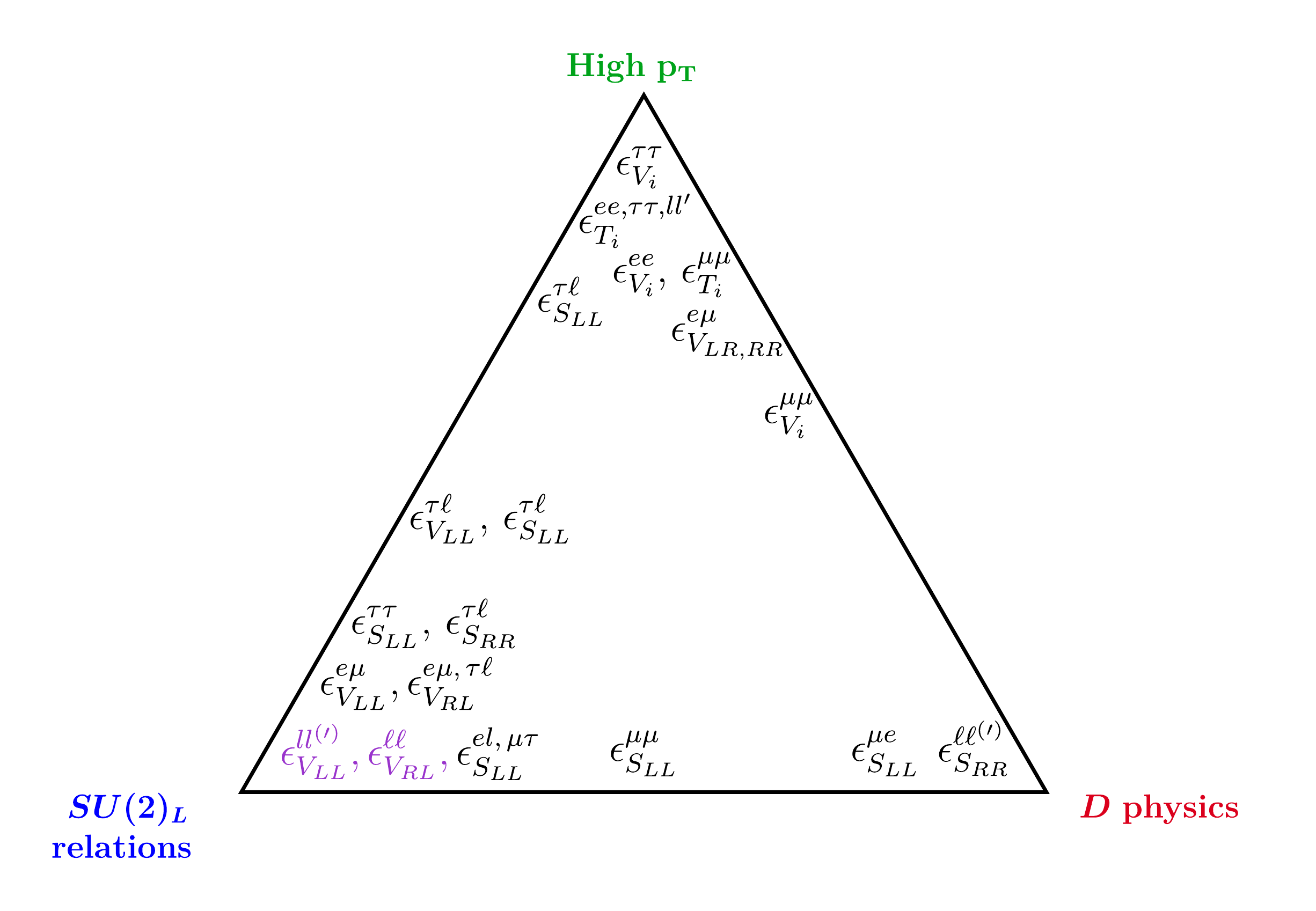}
\caption{Interplay between charm physics, high-$p_T$, and $SU(2)_L$ relations for the neutral-current case ($\ell=e,\mu$ and $l=e,\mu,\tau$). The proximity of the WCs to a particular vertex of the triangle is determined, approximately, by the relative strength of the corresponding constraints. In purple, those constraints that can be avoided by a particular flavor structure and/or WC combination.}\label{fig:trianglePlotNC}
\end{figure}

The $\mathcal{O}_{S_{LL}}^{\alpha\beta}$ and $\mathcal{O}_{T_L}^{\alpha\beta}$ operators are related by $SU(2)_L$ invariance to $d_i\to u\ell\nu$ and $\tau\to ud_i\nu$ transitions. The ordering of indices in the chirality-flipping operator is relevant, since the second index refers to the left-handed lepton, and thus it is the one connected to the neutrino flavor.
The related constraints are several orders of magnitude stronger than neutral charm for $\mathcal{O}_{S_{LL}}^{e\alpha}$, stronger than  high-$p_T$ dilepton tails for $\mathcal{O}_{S_{LL}}^{\mu\tau,\,\tau\tau}$, and comparable to those from neutral charm for $\mathcal{O}_{S_{LL}}^{\mu\mu}$. For $\mathcal{O}_{T_L}^{\alpha\beta}$, the $SU(2)_L$-correlated low-energy bounds are not competitive with the ones from high-$p_T$ dilepton tails. However, the analysis of the high-$p_T$ monolepton tails produced by $\bar u d_i\to e^\alpha \bar\nu^\beta$ give a marginal improvement compared to those. 
The $\mathcal{O}_{S_{RR}}^{\alpha\beta}$ and $\mathcal{O}_{T_R}^{\alpha\beta}$ operators receive correlated bounds from charged-current charm decays. These are only relevant for the lepton channels involving the tau flavor, since they are not constrained by the corresponding neutral currents. In this case, however, high-$p_T$ dilepton production offers the best bounds, with the exception of $\mathcal{O}_{S_{RR}}^{\tau\ell}$ that is better constrained by $D\to\ell\nu_\tau$. The interplay between charm decays, high-$p_T$ dilepton tails, and $SU(2)_L$ related constraints for the neutral-current case is summarized in Figure~\ref{fig:trianglePlotNC} for $\mu = 2$\,GeV.

\section{Conclusions}
\label{sec:Conc}

Charm is a cornerstone of the SM; a unique arena for QCD and flavor, with a bright experimental future ahead.
But how unique is the charm sector as a probe of new physics within the zoo of flavor and collider phenomenology? In other words, what is the role of charm in a broader quest for a microscopic theory beyond the SM?

In this work, we performed a detailed phenomenological analysis of new physics affecting charm $\Delta C = 1$ leptonic and semileptonic flavor transitions. We used effective field theory methods to establish a model-independent interplay between low- and high-energy experimental data, under the assumption of short-distance new-physics above the electroweak scale. The classic flavor-physics program consists in measuring and predicting the $D_{(s)}$ meson decays with high precision. In the context of charged currents, we have focused on the pure leptonic decays $D_{(s)}\to\ell\nu$ and the semileptonic decays $D\to P\ell\nu$ ($P=\pi,K$), for which accurate and robust predictions from lattice QCD exist and the most precise measurements have been reported. The main results are summarized in Table~\ref{tab:WCsCClowE}, while the analogous limits on neutral currents are reported in Section~\ref{sec:NCconst}. 

On the other hand, the analysis of high-$p_T$ lepton tails in $pp$ collisions at the LHC provides complementary constraints. Heavy flavors are virtually present in the 
proton and contribute to the Drell-Yan production with an amplitude which is connected by crossing symmetry to the one entering charmed meson decays. In fact, the energy-growing behavior of the EFT scattering amplitudes with respect to the SM, compensates for the lower partonic luminosities and lower statistics, eventually leading to strong constraints in the high-$p_T$ tails. The main results of our recast of recent ATLAS and CMS searches are reported in Tables~\ref{tab:WCsCChighE} and \ref{tab:WCsNChighE} for charged and neutral currents, respectively. A primary concern of the analysis is the EFT validity, discussed at length in Section~\ref{sec:caveats}.

We find a striking complementarity between charm decays and high-$p_T$ lepton tails. The reason behind this is that QCD selects the parity basis of fermionic currents at low energy, while at high-$p_T$, chiral fermions act as independent asymptotic states. This is best illustrated for scalar and tensor operators, where the combination of the two datasets is crucial to set optimal constraints, see Figure~\ref{fig:interplay}. For some scenarios, high-$p_T$ lepton tails offer the most competitive probe. As highlighted in Figure~\ref{fig:conclusions}, NP in four-fermion vector operators is by an order of magnitude better constrained in high-$p_T$ monolepton tails than in charm decays for all $c \to d_i \bar e^\alpha \nu^\beta$ transitions. Somewhat surprisingly, even for rare FCNC transitions $c \to u \ell^+ \ell^-$, we find $p p \to \ell^+ \ell^-$ high invariant mass tails to compete well with $D \to \ell^+ \ell^-$, see Figure~\ref{fig:NCinterplay}. The results presented here are applicable even beyond charm physics. In particular, in Section~\ref{sec:Down} we reinterpret the high-$p_T$ analysis in terms of limits on $b \to s \tau^+ \tau^-$ transitions to show that these are more stringent that the ones from $B \to K \tau^+ \tau^-$ searches.

Embedding the low energy effective theory in the $SU(2)_L\times U(1)_Y$ gauge invariant SMEFT, implies powerful model-independent correlations among observables in different sectors. Specific connections usually require to select a specific set of operators and its flavor structure as a remnant of a particular class of dynamics and symmetries in the UV. Nonetheless, one can assess the level of tuning required to avoid certain constraints or even find that avoiding all of them is not possible. An exhaustive map of $SU(2)_L$ correlations of charm decays with $K$, $\tau$ and $\pi$ decays is presented in Section~\ref{sec:SU2}, see also Figures~\ref{fig:trianglePlotCC} and \ref{fig:trianglePlotNC}. In conclusion, and to answer the question posed in the first paragraph of this Section, the bounds from $D_{(s)}$ decays, high-$p_T$ lepton tails and $SU(2)$ relations chart the space of all SMEFT operators affecting semi(leptonic) charm flavor transitions.

\acknowledgments

We thank Nud\v zeim Selimovi\' c for  carefully reading  the manuscript. The work of JFM has received funding from the Swiss National Science Foundation (SNF) under contract 200021-159720, and from the Generalitat Valenciana under contract SEJI/2018/033. The work of JFM and AG is partially supported by the European Research Council (ERC) under the European Union’s Horizon 2020 research and innovation programme, grant agreement 833280 (FLAY).
JMC acknowledges support from the Spanish MINECO through the ``Ram\'on y Cajal'' program RYC-2016-20672 and the grant  PGC2018-102016-A-I00. J. D. R.-\'{A}. gratefully acknowledges the support of the Colombian Science Ministry and Sostenibilidad-UdeA.

\bibliographystyle{JHEP}
\bibliography{references}

\providecommand{\href}[2]{#2}\begingroup\raggedright\begin{thebibliography}{100}

\bibitem{Aaij:2019kcg}
{\bf LHCb} Collaboration, R.~Aaij et~al., {\it {Observation of CP Violation in
  Charm Decays}},  {\em Phys. Rev. Lett.} {\bf 122} (2019), no.~21 211803,
  [\href{http://arxiv.org/abs/1903.08726}{{\tt arXiv:1903.08726}}].

\bibitem{Ablikim:2019hff}
{\bf BESIII} Collaboration, M.~Ablikim et~al., {\it {White Paper on the Future
  Physics Programme of BESIII}},  \href{http://arxiv.org/abs/1912.05983}{{\tt
  arXiv:1912.05983}}.

\bibitem{Cerri:2018ypt}
A.~Cerri et~al., {\it {Report from Working Group 4}},  {\em CERN Yellow Rep.
  Monogr.} {\bf 7} (2019) 867--1158,
  [\href{http://arxiv.org/abs/1812.07638}{{\tt arXiv:1812.07638}}].

\bibitem{Kou:2018nap}
{\bf Belle-II} Collaboration, W.~Altmannshofer et~al., {\it {The Belle II
  Physics Book}},  \href{http://arxiv.org/abs/1808.10567}{{\tt
  arXiv:1808.10567}}.

\bibitem{Amhis:2019ckw}
{\bf HFLAV} Collaboration, Y.~S. Amhis et~al., {\it {Averages of $b$-hadron,
  $c$-hadron, and $\tau$-lepton properties as of 2018}},
  \href{http://arxiv.org/abs/1909.12524}{{\tt arXiv:1909.12524}}.

\bibitem{Aoki:2019cca}
{\bf Flavour Lattice Averaging Group} Collaboration, S.~Aoki et~al., {\it {FLAG
  Review 2019}},  \href{http://arxiv.org/abs/1902.08191}{{\tt
  arXiv:1902.08191}}.

\bibitem{Barranco:2013tba}
J.~Barranco, D.~Delepine, V.~Gonzalez~Macias, and L.~Lopez-Lozano, {\it
  {Constraining New Physics with D meson decays}},  {\em Phys. Lett.} {\bf
  B731} (2014) 36--42, [\href{http://arxiv.org/abs/1303.3896}{{\tt
  arXiv:1303.3896}}].

\bibitem{Fajfer:2015ixa}
S.~Fajfer, I.~Nisandzic, and U.~Rojec, {\it {Discerning new physics in charm
  meson leptonic and semileptonic decays}},  {\em Phys. Rev.} {\bf D91} (2015),
  no.~9 094009, [\href{http://arxiv.org/abs/1502.07488}{{\tt
  arXiv:1502.07488}}].

\bibitem{Fleischer:2019wlx}
R.~Fleischer, R.~Jaarsma, and G.~Koole, {\it {Testing Lepton Flavour
  Universality with (Semi)-Leptonic $D_{(s)}$ Decays}},
  \href{http://arxiv.org/abs/1912.08641}{{\tt arXiv:1912.08641}}.

\bibitem{Carrasco:2014poa}
N.~Carrasco et~al., {\it {Leptonic decay constants $f_{K},f_{D},$ and
  $f_{{D}_{s}}$ with $N_{f} = 2+1+1$ twisted-mass lattice QCD}},  {\em Phys.
  Rev.} {\bf D91} (2015), no.~5 054507,
  [\href{http://arxiv.org/abs/1411.7908}{{\tt arXiv:1411.7908}}].

\bibitem{Bazavov:2017lyh}
A.~Bazavov et~al., {\it {$B$- and $D$-meson leptonic decay constants from
  four-flavor lattice QCD}},  {\em Phys. Rev.} {\bf D98} (2018), no.~7 074512,
  [\href{http://arxiv.org/abs/1712.09262}{{\tt arXiv:1712.09262}}].

\bibitem{Aubin:2004ej}
{\bf Fermilab Lattice, MILC, HPQCD} Collaboration, C.~Aubin et~al., {\it
  {Semileptonic decays of D mesons in three-flavor lattice QCD}},  {\em Phys.
  Rev. Lett.} {\bf 94} (2005) 011601,
  [\href{http://arxiv.org/abs/hep-ph/0408306}{{\tt hep-ph/0408306}}].

\bibitem{Na:2010uf}
H.~Na, C.~T.~H. Davies, E.~Follana, G.~P. Lepage, and J.~Shigemitsu, {\it {The
  $D \rightarrow K, l \nu$ Semileptonic Decay Scalar Form Factor and $|V_{cs}|$
  from Lattice QCD}},  {\em Phys. Rev.} {\bf D82} (2010) 114506,
  [\href{http://arxiv.org/abs/1008.4562}{{\tt arXiv:1008.4562}}].

\bibitem{Na:2011mc}
H.~Na, C.~T.~H. Davies, E.~Follana, J.~Koponen, G.~P. Lepage, and
  J.~Shigemitsu, {\it {$D \rightarrow \pi, l \nu$ Semileptonic Decays,
  $|V_{cd}|$ and 2$^{nd}$ Row Unitarity from Lattice QCD}},  {\em Phys. Rev.}
  {\bf D84} (2011) 114505, [\href{http://arxiv.org/abs/1109.1501}{{\tt
  arXiv:1109.1501}}].

\bibitem{Lubicz:2017syv}
{\bf ETM} Collaboration, V.~Lubicz, L.~Riggio, G.~Salerno, S.~Simula, and
  C.~Tarantino, {\it {Scalar and vector form factors of $D \to \pi(K) \ell \nu$
  decays with $N_f=2+1+1$ twisted fermions}},  {\em Phys. Rev.} {\bf D96}
  (2017), no.~5 054514, [\href{http://arxiv.org/abs/1706.03017}{{\tt
  arXiv:1706.03017}}]. [erratum: Phys. Rev.D99,no.9,099902(2019)].

\bibitem{Lubicz:2018rfs}
{\bf ETM} Collaboration, V.~Lubicz, L.~Riggio, G.~Salerno, S.~Simula, and
  C.~Tarantino, {\it {Tensor form factor of $D \to \pi(K) \ell \nu$ and $D \to
  \pi(K) \ell \ell$ decays with $N_f=2+1+1$ twisted-mass fermions}},  {\em
  Phys. Rev.} {\bf D98} (2018), no.~1 014516,
  [\href{http://arxiv.org/abs/1803.04807}{{\tt arXiv:1803.04807}}].

\bibitem{Burdman:2001tf}
G.~Burdman, E.~Golowich, J.~L. Hewett, and S.~Pakvasa, {\it {Rare charm decays
  in the standard model and beyond}},  {\em Phys. Rev.} {\bf D66} (2002)
  014009, [\href{http://arxiv.org/abs/hep-ph/0112235}{{\tt hep-ph/0112235}}].

\bibitem{Paul:2011ar}
A.~Paul, I.~I. Bigi, and S.~Recksiegel, {\it {On $D\to X_u l^+ l^-$ within the
  Standard Model and Frameworks like the Littlest Higgs Model with T Parity}},
  {\em Phys. Rev.} {\bf D83} (2011) 114006,
  [\href{http://arxiv.org/abs/1101.6053}{{\tt arXiv:1101.6053}}].

\bibitem{Cappiello:2012vg}
L.~Cappiello, O.~Cata, and G.~D'Ambrosio, {\it {Standard Model prediction and
  new physics tests for $D^0 \to h^+ h^- \ell^+ \ell^- (h=\pi,K:
  \ell=e,\mu)$}},  {\em JHEP} {\bf 04} (2013) 135,
  [\href{http://arxiv.org/abs/1209.4235}{{\tt arXiv:1209.4235}}].

\bibitem{deBoer:2015boa}
S.~de~Boer and G.~Hiller, {\it {Flavor and new physics opportunities with rare
  charm decays into leptons}},  {\em Phys. Rev.} {\bf D93} (2016), no.~7
  074001, [\href{http://arxiv.org/abs/1510.00311}{{\tt arXiv:1510.00311}}].

\bibitem{Fajfer:2015mia}
S.~Fajfer and N.~Košnik, {\it {Prospects of discovering new physics in rare
  charm decays}},  {\em Eur. Phys. J.} {\bf C75} (2015), no.~12 567,
  [\href{http://arxiv.org/abs/1510.00965}{{\tt arXiv:1510.00965}}].

\bibitem{Bause:2019vpr}
R.~Bause, M.~Golz, G.~Hiller, and A.~Tayduganov, {\it {The New Physics Reach of
  Null Tests with $D \to \pi \ell \ell$ and $D_s \to K \ell \ell $ Decays}},
  {\em Eur. Phys. J.} {\bf C80} (2020), no.~1 65,
  [\href{http://arxiv.org/abs/1909.11108}{{\tt arXiv:1909.11108}}].

\bibitem{Fajfer:2015zea}
S.~Fajfer, {\it {Theoretical perspective on rare and radiative charm decays}},
  in {\em {Proceedings, 7th International Workshop on Charm Physics, CHARM
  2015: Detroit, USA, May 18-22, 2015}}, 2015.
\newblock \href{http://arxiv.org/abs/1509.01997}{{\tt arXiv:1509.01997}}.

\bibitem{Silvestrini:2015kqa}
L.~Silvestrini, {\it {CHARM-2015 Theory Summary}},  in {\em {Proceedings, 7th
  International Workshop on Charm Physics, CHARM 2015: Detroit, USA, May 18-22,
  2015}}, 2015.
\newblock \href{http://arxiv.org/abs/1510.05797}{{\tt arXiv:1510.05797}}.

\bibitem{Fajfer:1997bh}
S.~Fajfer and P.~Singer, {\it {Long distance $c\to u\gamma$ effects in weak
  radiative decays of $D$ mesons}},  {\em Phys. Rev.} {\bf D56} (1997)
  4302--4310, [\href{http://arxiv.org/abs/hep-ph/9705327}{{\tt
  hep-ph/9705327}}].

\bibitem{Fajfer:1998dv}
S.~Fajfer, S.~Prelovsek, and P.~Singer, {\it {Long distance contributions in
  $D\to V\gamma$ decays}},  {\em Eur. Phys. J.} {\bf C6} (1999) 471--476,
  [\href{http://arxiv.org/abs/hep-ph/9801279}{{\tt hep-ph/9801279}}].

\bibitem{deBoer:2017que}
S.~de~Boer and G.~Hiller, {\it {Rare radiative charm decays within the standard
  model and beyond}},  {\em JHEP} {\bf 08} (2017) 091,
  [\href{http://arxiv.org/abs/1701.06392}{{\tt arXiv:1701.06392}}].

\bibitem{Aaij:2013qta}
{\bf LHCb} Collaboration, R.~Aaij et~al., {\it {Measurement of
  Form-Factor-Independent Observables in the Decay $B^{0} \to K^{*0} \mu^+
  \mu^-$}},  {\em Phys. Rev. Lett.} {\bf 111} (2013) 191801,
  [\href{http://arxiv.org/abs/1308.1707}{{\tt arXiv:1308.1707}}].

\bibitem{Aaij:2014ora}
{\bf LHCb} Collaboration, R.~Aaij et~al., {\it {Test of lepton universality
  using $B^{+}\rightarrow K^{+}\ell^{+}\ell^{-}$ decays}},  {\em Phys. Rev.
  Lett.} {\bf 113} (2014) 151601, [\href{http://arxiv.org/abs/1406.6482}{{\tt
  arXiv:1406.6482}}].

\bibitem{Aaij:2015oid}
{\bf LHCb} Collaboration, R.~Aaij et~al., {\it {Angular analysis of the $B^{0}
  \to K^{*0} \mu^{+} \mu^{-}$ decay using 3 fb$^{-1}$ of integrated
  luminosity}},  {\em JHEP} {\bf 02} (2016) 104,
  [\href{http://arxiv.org/abs/1512.04442}{{\tt arXiv:1512.04442}}].

\bibitem{Aaij:2017vbb}
{\bf LHCb} Collaboration, R.~Aaij et~al., {\it {Test of lepton universality
  with $B^{0} \rightarrow K^{*0}\ell^{+}\ell^{-}$ decays}},  {\em JHEP} {\bf
  08} (2017) 055, [\href{http://arxiv.org/abs/1705.05802}{{\tt
  arXiv:1705.05802}}].

\bibitem{Aaij:2019wad}
{\bf LHCb} Collaboration, R.~Aaij et~al., {\it {Search for lepton-universality
  violation in $B^+\to K^+\ell^+\ell^-$ decays}},  {\em Phys. Rev. Lett.} {\bf
  122} (2019), no.~19 191801, [\href{http://arxiv.org/abs/1903.09252}{{\tt
  arXiv:1903.09252}}].

\bibitem{Lees:2012xj}
{\bf BaBar} Collaboration, J.~P. Lees et~al., {\it {Evidence for an excess of
  $\bar{B} \to D^{(*)} \tau^-\bar{\nu}_\tau$ decays}},  {\em Phys. Rev. Lett.}
  {\bf 109} (2012) 101802, [\href{http://arxiv.org/abs/1205.5442}{{\tt
  arXiv:1205.5442}}].

\bibitem{Huschle:2015rga}
{\bf Belle} Collaboration, M.~Huschle et~al., {\it {Measurement of the
  branching ratio of $\bar{B} \to D^{(\ast)} \tau^- \bar{\nu}_\tau$ relative to
  $\bar{B} \to D^{(\ast)} \ell^- \bar{\nu}_\ell$ decays with hadronic tagging
  at Belle}},  {\em Phys. Rev.} {\bf D92} (2015), no.~7 072014,
  [\href{http://arxiv.org/abs/1507.03233}{{\tt arXiv:1507.03233}}].

\bibitem{Aaij:2015yra}
{\bf LHCb} Collaboration, R.~Aaij et~al., {\it {Measurement of the ratio of
  branching fractions $\mathcal{B}(\bar{B}^0 \to
  D^{*+}\tau^{-}\bar{\nu}_{\tau})/\mathcal{B}(\bar{B}^0 \to
  D^{*+}\mu^{-}\bar{\nu}_{\mu})$}},  {\em Phys. Rev. Lett.} {\bf 115} (2015),
  no.~11 111803, [\href{http://arxiv.org/abs/1506.08614}{{\tt
  arXiv:1506.08614}}]. [Erratum: Phys. Rev. Lett.115,no.15,159901(2015)].

\bibitem{Aaij:2017tyk}
{\bf LHCb} Collaboration, R.~Aaij et~al., {\it {Measurement of the ratio of
  branching fractions
  $\mathcal{B}(B_c^+\,\to\,J/\psi\tau^+\nu_\tau)$/$\mathcal{B}(B_c^+\,\to\,J/\psi\mu^+\nu_\mu)$}},
  {\em Phys. Rev. Lett.} {\bf 120} (2018), no.~12 121801,
  [\href{http://arxiv.org/abs/1711.05623}{{\tt arXiv:1711.05623}}].

\bibitem{Aaij:2017uff}
{\bf LHCb} Collaboration, R.~Aaij et~al., {\it {Measurement of the ratio of the
  $B^0 \to D^{*-} \tau^+ \nu_{\tau}$ and $B^0 \to D^{*-} \mu^+ \nu_{\mu}$
  branching fractions using three-prong $\tau$-lepton decays}},  {\em Phys.
  Rev. Lett.} {\bf 120} (2018), no.~17 171802,
  [\href{http://arxiv.org/abs/1708.08856}{{\tt arXiv:1708.08856}}].

\bibitem{Abdesselam:2019dgh}
{\bf Belle} Collaboration, A.~Abdesselam et~al., {\it {Measurement of
  $\mathcal{R}(D)$ and $\mathcal{R}(D^{\ast})$ with a semileptonic tagging
  method}},  \href{http://arxiv.org/abs/1904.08794}{{\tt arXiv:1904.08794}}.

\bibitem{Cirigliano:2012ab}
V.~Cirigliano, M.~Gonzalez-Alonso, and M.~L. Graesser, {\it {Non-standard
  Charged Current Interactions: beta decays versus the LHC}},  {\em JHEP} {\bf
  02} (2013) 046, [\href{http://arxiv.org/abs/1210.4553}{{\tt
  arXiv:1210.4553}}].

\bibitem{Gonzalez-Alonso:2016etj}
M.~González-Alonso and J.~Martin~Camalich, {\it {Global Effective-Field-Theory
  analysis of New-Physics effects in (semi)leptonic kaon decays}},  {\em JHEP}
  {\bf 12} (2016) 052, [\href{http://arxiv.org/abs/1605.07114}{{\tt
  arXiv:1605.07114}}].

\bibitem{Cirigliano:2018dyk}
V.~Cirigliano, A.~Falkowski, M.~González-Alonso, and A.~Rodríguez-Sánchez,
  {\it {Hadronic $\tau$ Decays as New Physics Probes in the LHC Era}},  {\em
  Phys. Rev. Lett.} {\bf 122} (2019), no.~22 221801,
  [\href{http://arxiv.org/abs/1809.01161}{{\tt arXiv:1809.01161}}].

\bibitem{Faroughy:2016osc}
D.~A. Faroughy, A.~Greljo, and J.~F. Kamenik, {\it {Confronting lepton flavor
  universality violation in B decays with high-$p_T$ tau lepton searches at
  LHC}},  {\em Phys. Lett.} {\bf B764} (2017) 126--134,
  [\href{http://arxiv.org/abs/1609.07138}{{\tt arXiv:1609.07138}}].

\bibitem{Greljo:2017vvb}
A.~Greljo and D.~Marzocca, {\it {High-$p_T$ dilepton tails and flavor
  physics}},  {\em Eur. Phys. J.} {\bf C77} (2017), no.~8 548,
  [\href{http://arxiv.org/abs/1704.09015}{{\tt arXiv:1704.09015}}].

\bibitem{Altmannshofer:2017poe}
W.~Altmannshofer, P.~Bhupal~Dev, and A.~Soni, {\it {$R_{D^{(*)}}$ anomaly: A
  possible hint for natural supersymmetry with $R$-parity violation}},  {\em
  Phys.Rev.D} {\bf 96} (2017), no.~9 095010,
  [\href{http://arxiv.org/abs/1704.06659}{{\tt arXiv:1704.06659}}].

\bibitem{Greljo:2018tzh}
A.~Greljo, J.~Martin~Camalich, and J.~D. Ruiz-Álvarez, {\it {Mono-$\tau$
  Signatures at the LHC Constrain Explanations of $B$-decay Anomalies}},  {\em
  Phys. Rev. Lett.} {\bf 122} (2019), no.~13 131803,
  [\href{http://arxiv.org/abs/1811.07920}{{\tt arXiv:1811.07920}}].

\bibitem{Baker:2019sli}
M.~J. Baker, J.~Fuentes-Martín, G.~Isidori, and M.~König, {\it {High- $p_T$
  signatures in vector–leptoquark models}},  {\em Eur. Phys. J.} {\bf C79}
  (2019), no.~4 334, [\href{http://arxiv.org/abs/1901.10480}{{\tt
  arXiv:1901.10480}}].

\bibitem{Bhattacharya:2018ryy}
B.~Bhattacharya, R.~Morgan, J.~Osborne, and A.~A. Petrov, {\it {Studies of
  Lepton Flavor Violation at the LHC}},  {\em Phys.Lett.B} {\bf 785} (2018)
  165--170, [\href{http://arxiv.org/abs/1802.06082}{{\tt arXiv:1802.06082}}].

\bibitem{Angelescu:2020uug}
A.~Angelescu, D.~A. Faroughy, and O.~Sumensari, {\it {Lepton Flavor Violation
  and Dilepton Tails at the LHC}},  \href{http://arxiv.org/abs/2002.05684}{{\tt
  arXiv:2002.05684}}.

\bibitem{Buchmuller:1985jz}
W.~Buchmuller and D.~Wyler, {\it {Effective Lagrangian Analysis of New
  Interactions and Flavor Conservation}},  {\em Nucl. Phys.} {\bf B268} (1986)
  621--653.

\bibitem{Grzadkowski:2010es}
B.~Grzadkowski, M.~Iskrzynski, M.~Misiak, and J.~Rosiek, {\it {Dimension-Six
  Terms in the Standard Model Lagrangian}},  {\em JHEP} {\bf 10} (2010) 085,
  [\href{http://arxiv.org/abs/1008.4884}{{\tt arXiv:1008.4884}}].

\bibitem{Greljo:2018ogz}
A.~Greljo, D.~J. Robinson, B.~Shakya, and J.~Zupan, {\it {$R(D^{(*)})$ from
  $W^\prime$ and right-handed neutrinos}},  {\em JHEP} {\bf 09} (2018) 169,
  [\href{http://arxiv.org/abs/1804.04642}{{\tt arXiv:1804.04642}}].

\bibitem{Descotes-Genon:2018foz}
S.~Descotes-Genon, A.~Falkowski, M.~Fedele, M.~González-Alonso, and J.~Virto,
  {\it {The CKM parameters in the SMEFT}},  {\em JHEP} {\bf 05} (2019) 172,
  [\href{http://arxiv.org/abs/1812.08163}{{\tt arXiv:1812.08163}}].

\bibitem{Jenkins:2017dyc}
E.~E. Jenkins, A.~V. Manohar, and P.~Stoffer, {\it {Low-Energy Effective Field
  Theory below the Electroweak Scale: Anomalous Dimensions}},  {\em JHEP} {\bf
  01} (2018) 084, [\href{http://arxiv.org/abs/1711.05270}{{\tt
  arXiv:1711.05270}}].

\bibitem{Alonso:2013hga}
R.~Alonso, E.~E. Jenkins, A.~V. Manohar, and M.~Trott, {\it {Renormalization
  Group Evolution of the Standard Model Dimension Six Operators III: Gauge
  Coupling Dependence and Phenomenology}},  {\em JHEP} {\bf 04} (2014) 159,
  [\href{http://arxiv.org/abs/1312.2014}{{\tt arXiv:1312.2014}}].

\bibitem{Gonzalez-Alonso:2017iyc}
M.~Gonz{\'a}lez-Alonso, J.~Martin~Camalich, and K.~Mimouni, {\it
  {Renormalization-group evolution of new physics contributions to
  (semi)leptonic meson decays}},  {\em Phys. Lett.} {\bf B772} (2017) 777--785,
  [\href{http://arxiv.org/abs/1706.00410}{{\tt arXiv:1706.00410}}].

\bibitem{Grinstein:2015aua}
B.~Grinstein and J.~Martin~Camalich, {\it {Weak Decays of Excited B Mesons}},
  {\em Phys.Rev.Lett.} {\bf 116} (2016), no.~14 141801,
  [\href{http://arxiv.org/abs/1509.05049}{{\tt arXiv:1509.05049}}].

\bibitem{Khodjamirian:2015dda}
A.~Khodjamirian, T.~Mannel, and A.~A. Petrov, {\it {Direct probes of
  flavor-changing neutral currents in e$^+$ e$^-$-collisions}},  {\em JHEP}
  {\bf 11} (2015) 142, [\href{http://arxiv.org/abs/1509.07123}{{\tt
  arXiv:1509.07123}}].

\bibitem{Tanabashi:2018oca}
{\bf Particle Data Group} Collaboration, M.~Tanabashi et~al., {\it {Review of
  Particle Physics}},  {\em Phys. Rev.} {\bf D98} (2018), no.~3 030001.

\bibitem{Eisenstein:2008aa}
{\bf CLEO} Collaboration, B.~I. Eisenstein et~al., {\it {Precision Measurement
  of $B(D^+\to\mu^+\nu)$ and the Pseudoscalar Decay Constant f(D+)}},  {\em
  Phys. Rev.} {\bf D78} (2008) 052003,
  [\href{http://arxiv.org/abs/0806.2112}{{\tt arXiv:0806.2112}}].

\bibitem{Ablikim:2013uvu}
{\bf BESIII} Collaboration, M.~Ablikim et~al., {\it {Precision measurements of
  $B(D^+ \rightarrow \mu^+ \nu_{\mu})$, the pseudoscalar decay constant
  $f_{D^+}$, and the quark mixing matrix element $|V_{\rm cd}|$}},  {\em Phys.
  Rev.} {\bf D89} (2014), no.~5 051104,
  [\href{http://arxiv.org/abs/1312.0374}{{\tt arXiv:1312.0374}}].

\bibitem{Zupanc:2013byn}
{\bf Belle} Collaboration, A.~Zupanc et~al., {\it {Measurements of branching
  fractions of leptonic and hadronic $D_{s}^{+}$ meson decays and extraction of
  the $D_{s}^{+}$ meson decay constant}},  {\em JHEP} {\bf 09} (2013) 139,
  [\href{http://arxiv.org/abs/1307.6240}{{\tt arXiv:1307.6240}}].

\bibitem{Ablikim:2016duz}
{\bf BESIII} Collaboration, M.~Ablikim et~al., {\it {Measurement of the $D_s^+
  \to \ell^+\nu_\ell$ branching fractions and the decay constant $f_{D_s^+}$}},
   {\em Phys. Rev.} {\bf D94} (2016), no.~7 072004,
  [\href{http://arxiv.org/abs/1608.06732}{{\tt arXiv:1608.06732}}].

\bibitem{delAmoSanchez:2010jg}
{\bf BaBar} Collaboration, P.~del Amo~Sanchez et~al., {\it {Measurement of the
  Absolute Branching Fractions for $D^-_s\!\rightarrow\!\ell^-\bar{\nu}_{\ell}$
  and Extraction of the Decay Constant $f_{D_s}$}},  {\em Phys. Rev.} {\bf D82}
  (2010) 091103, [\href{http://arxiv.org/abs/1008.4080}{{\tt
  arXiv:1008.4080}}]. [Erratum: Phys. Rev.D91,no.1,019901(2015)].

\bibitem{Alexander:2009ux}
{\bf CLEO} Collaboration, J.~P. Alexander et~al., {\it {Measurement of $B(D_s^+
  \to \ell^+ \nu)$ and the decay constant $f_{D_s^+}$ from 600 $pb^{-1}$ of
  $e^\pm$ annihilation data near 4170 MeV}},  {\em Phys. Rev.} {\bf D79} (2009)
  052001, [\href{http://arxiv.org/abs/0901.1216}{{\tt arXiv:0901.1216}}].

\bibitem{Naik:2009tk}
{\bf CLEO} Collaboration, P.~Naik et~al., {\it {Measurement of the pseudoscalar
  decay constant $f_{D(s)}$ using $D_{(s)}^+\to\tau^+\nu$,
  $\tau^+\to\rho^+\bar\nu$ decays}},  {\em Phys. Rev.} {\bf D80} (2009) 112004,
  [\href{http://arxiv.org/abs/0910.3602}{{\tt arXiv:0910.3602}}].

\bibitem{Onyisi:2009th}
{\bf CLEO} Collaboration, P.~U.~E. Onyisi et~al., {\it {Improved measurement of
  absolute branching fraction of $D_{(s)}^+\to\tau^+\nu_{(\tau)}$}},  {\em
  Phys. Rev.} {\bf D79} (2009) 052002,
  [\href{http://arxiv.org/abs/0901.1147}{{\tt arXiv:0901.1147}}].

\bibitem{Abe:2005nq}
{\bf Belle} Collaboration, U.~Bitenc et~al., {\it {Search for $D_0 - \bar D_0$
  mixing using semileptonic decays at Belle}},  {\em Phys. Rev.} {\bf D72}
  (2005) 071101, [\href{http://arxiv.org/abs/hep-ex/0507020}{{\tt
  hep-ex/0507020}}].

\bibitem{Ablikim:2018evp}
{\bf BESIII} Collaboration, M.~Ablikim et~al., {\it {Study of the $D^0\to
  K^-\mu^+\nu_\mu$ dynamics and test of lepton flavor universality with $D^0\to
  K^-\ell^+\nu_\ell$ decays}},  {\em Phys. Rev. Lett.} {\bf 122} (2019), no.~1
  011804, [\href{http://arxiv.org/abs/1810.03127}{{\tt arXiv:1810.03127}}].

\bibitem{Widhalm:2006wz}
{\bf Belle} Collaboration, L.~Widhalm et~al., {\it {Measurement of
  $D_0\to\pi\ell\nu\,(K\ell\nu)$ Form Factors and Absolute Branching
  Fractions}},  {\em Phys. Rev. Lett.} {\bf 97} (2006) 061804,
  [\href{http://arxiv.org/abs/hep-ex/0604049}{{\tt hep-ex/0604049}}].

\bibitem{Ablikim:2015ixa}
{\bf BESIII} Collaboration, M.~Ablikim et~al., {\it {Study of Dynamics of $D^0
  \to K^- e^+ \nu_{e}$ and $D^0\to\pi^- e^+ \nu_{e}$ Decays}},  {\em Phys.
  Rev.} {\bf D92} (2015), no.~7 072012,
  [\href{http://arxiv.org/abs/1508.07560}{{\tt arXiv:1508.07560}}].

\bibitem{Besson:2009uv}
{\bf CLEO} Collaboration, D.~Besson et~al., {\it {Improved measurements of D
  meson semileptonic decays to pi and K mesons}},  {\em Phys. Rev.} {\bf D80}
  (2009) 032005, [\href{http://arxiv.org/abs/0906.2983}{{\tt
  arXiv:0906.2983}}].

\bibitem{Ablikim:2018frk}
{\bf BESIII} Collaboration, M.~Ablikim et~al., {\it {Measurement of the
  branching fraction for the semi-leptonic decay $D^{0(+)}\to
  \pi^{-(0)}\mu^+\nu_\mu$ and test of lepton universality}},  {\em Phys. Rev.
  Lett.} {\bf 121} (2018), no.~17 171803,
  [\href{http://arxiv.org/abs/1802.05492}{{\tt arXiv:1802.05492}}].

\bibitem{deBlas:2013qqa}
J.~de~Blas, M.~Chala, and J.~Santiago, {\it {Global Constraints on Lepton-Quark
  Contact Interactions}},  {\em Phys. Rev.} {\bf D88} (2013) 095011,
  [\href{http://arxiv.org/abs/1307.5068}{{\tt arXiv:1307.5068}}].

\bibitem{Farina:2016rws}
M.~Farina, G.~Panico, D.~Pappadopulo, J.~T. Ruderman, R.~Torre, and A.~Wulzer,
  {\it {Energy helps accuracy: electroweak precision tests at hadron
  colliders}},  {\em Phys. Lett.} {\bf B772} (2017) 210--215,
  [\href{http://arxiv.org/abs/1609.08157}{{\tt arXiv:1609.08157}}].

\bibitem{Alioli:2017nzr}
S.~Alioli, M.~Farina, D.~Pappadopulo, and J.~T. Ruderman, {\it {Catching a New
  Force by the Tail}},  {\em Phys. Rev. Lett.} {\bf 120} (2018), no.~10 101801,
  [\href{http://arxiv.org/abs/1712.02347}{{\tt arXiv:1712.02347}}].

\bibitem{Raj:2016aky}
N.~Raj, {\it {Anticipating nonresonant new physics in dilepton angular spectra
  at the LHC}},  {\em Phys. Rev.} {\bf D95} (2017), no.~1 015011,
  [\href{http://arxiv.org/abs/1610.03795}{{\tt arXiv:1610.03795}}].

\bibitem{Schmaltz:2018nls}
M.~Schmaltz and Y.-M. Zhong, {\it {The leptoquark Hunter’s guide: large
  coupling}},  {\em JHEP} {\bf 01} (2019) 132,
  [\href{http://arxiv.org/abs/1810.10017}{{\tt arXiv:1810.10017}}].

\bibitem{Dawson:2018dxp}
S.~Dawson, P.~Giardino, and A.~Ismail, {\it {Standard model EFT and the
  Drell-Yan process at high energy}},  {\em Phys.\ Rev.\ D} {\bf 99} (2019),
  no.~3 035044, [\href{http://arxiv.org/abs/1811.12260}{{\tt
  arXiv:1811.12260}}].

\bibitem{Brooijmans:2020yij}
G.~Brooijmans et~al., {\it {Les Houches 2019 Physics at TeV Colliders: New
  Physics Working Group Report}},  in {\em {11th Les Houches Workshop on
  Physics at TeV Colliders: PhysTeV Les Houches (PhysTeV 2019) Les Houches,
  France, June 10-28, 2019}}, 2020.
\newblock \href{http://arxiv.org/abs/2002.12220}{{\tt arXiv:2002.12220}}.

\bibitem{Englert:2019zmt}
C.~Englert, G.~F. Giudice, A.~Greljo, and M.~Mccullough, {\it {The
  $\hat{H}$-Parameter: An Oblique Higgs View}},  {\em JHEP} {\bf 09} (2019)
  041, [\href{http://arxiv.org/abs/1903.07725}{{\tt arXiv:1903.07725}}].

\bibitem{Harland-Lang:2014zoa}
L.~A. Harland-Lang, A.~D. Martin, P.~Motylinski, and R.~S. Thorne, {\it {Parton
  distributions in the LHC era: MMHT 2014 PDFs}},  {\em Eur. Phys. J.} {\bf
  C75} (2015), no.~5 204, [\href{http://arxiv.org/abs/1412.3989}{{\tt
  arXiv:1412.3989}}].

\bibitem{Ball:2017nwa}
{\bf NNPDF} Collaboration, R.~D. Ball et~al., {\it {Parton distributions from
  high-precision collider data}},  {\em Eur. Phys. J.} {\bf C77} (2017), no.~10
  663, [\href{http://arxiv.org/abs/1706.00428}{{\tt arXiv:1706.00428}}].

\bibitem{Khalek:2018mdn}
R.~Abdul~Khalek, S.~Bailey, J.~Gao, L.~Harland-Lang, and J.~Rojo, {\it {Towards
  Ultimate Parton Distributions at the High-Luminosity LHC}},  {\em Eur. Phys.
  J.} {\bf C78} (2018), no.~11 962,
  [\href{http://arxiv.org/abs/1810.03639}{{\tt arXiv:1810.03639}}].

\bibitem{Carrazza:2019sec}
S.~Carrazza, C.~Degrande, S.~Iranipour, J.~Rojo, and M.~Ubiali, {\it {Can New
  Physics hide inside the proton?}},  {\em Phys. Rev. Lett.} {\bf 123} (2019),
  no.~13 132001, [\href{http://arxiv.org/abs/1905.05215}{{\tt
  arXiv:1905.05215}}].

\bibitem{Aaboud:2018vgh}
{\bf ATLAS} Collaboration, M.~Aaboud et~al., {\it {Search for High-Mass
  Resonances Decaying to $\tau\nu$ in pp Collisions at $\sqrt{s}$=13 TeV with
  the ATLAS Detector}},  {\em Phys. Rev. Lett.} {\bf 120} (2018), no.~16
  161802, [\href{http://arxiv.org/abs/1801.06992}{{\tt arXiv:1801.06992}}].

\bibitem{Sirunyan:2018lbg}
{\bf CMS} Collaboration, A.~M. Sirunyan et~al., {\it {Search for a $W'$ boson
  decaying to a $\tau$ lepton and a neutrino in proton-proton collisions at
  $\sqrt{s} =$ 13 TeV}},  {\em Phys. Lett.} {\bf B792} (2019) 107--131,
  [\href{http://arxiv.org/abs/1807.11421}{{\tt arXiv:1807.11421}}].

\bibitem{Aad:2019wvl}
{\bf ATLAS} Collaboration, G.~Aad et~al., {\it {Search for a heavy charged
  boson in events with a charged lepton and missing transverse momentum from
  $pp$ collisions at $\sqrt{s} = 13$ TeV with the ATLAS detector}},  {\em Phys.
  Rev.} {\bf D100} (2019), no.~5 052013,
  [\href{http://arxiv.org/abs/1906.05609}{{\tt arXiv:1906.05609}}].

\bibitem{Sirunyan:2018mpc}
{\bf CMS} Collaboration, A.~M. Sirunyan et~al., {\it {Search for high-mass
  resonances in final states with a lepton and missing transverse momentum at $
  \sqrt{s}=13 $ TeV}},  {\em JHEP} {\bf 06} (2018) 128,
  [\href{http://arxiv.org/abs/1803.11133}{{\tt arXiv:1803.11133}}].

\bibitem{Alloul:2013bka}
A.~Alloul, N.~D. Christensen, C.~Degrande, C.~Duhr, and B.~Fuks, {\it
  {FeynRules 2.0 - A complete toolbox for tree-level phenomenology}},  {\em
  Comput. Phys. Commun.} {\bf 185} (2014) 2250--2300,
  [\href{http://arxiv.org/abs/1310.1921}{{\tt arXiv:1310.1921}}].

\bibitem{Alwall:2011uj}
J.~Alwall, M.~Herquet, F.~Maltoni, O.~Mattelaer, and T.~Stelzer, {\it {MadGraph
  5 : Going Beyond}},  {\em JHEP} {\bf 06} (2011) 128,
  [\href{http://arxiv.org/abs/1106.0522}{{\tt arXiv:1106.0522}}].

\bibitem{Alwall:2014hca}
J.~Alwall, R.~Frederix, S.~Frixione, V.~Hirschi, F.~Maltoni, O.~Mattelaer,
  H.~S. Shao, T.~Stelzer, P.~Torrielli, and M.~Zaro, {\it {The automated
  computation of tree-level and next-to-leading order differential cross
  sections, and their matching to parton shower simulations}},  {\em JHEP} {\bf
  07} (2014) 079, [\href{http://arxiv.org/abs/1405.0301}{{\tt
  arXiv:1405.0301}}].

\bibitem{Sjostrand:2014zea}
T.~Sj{\"o}strand, S.~Ask, J.~R. Christiansen, R.~Corke, N.~Desai, P.~Ilten,
  S.~Mrenna, S.~Prestel, C.~O. Rasmussen, and P.~Z. Skands, {\it {An
  Introduction to PYTHIA 8.2}},  {\em Comput. Phys. Commun.} {\bf 191} (2015)
  159--177, [\href{http://arxiv.org/abs/1410.3012}{{\tt arXiv:1410.3012}}].

\bibitem{deFavereau:2013fsa}
{\bf DELPHES 3} Collaboration, J.~de~Favereau, C.~Delaere, P.~Demin,
  A.~Giammanco, V.~Lemaître, A.~Mertens, and M.~Selvaggi, {\it {DELPHES 3, A
  modular framework for fast simulation of a generic collider experiment}},
  {\em JHEP} {\bf 02} (2014) 057, [\href{http://arxiv.org/abs/1307.6346}{{\tt
  arXiv:1307.6346}}].

\bibitem{Brun:1997pa}
R.~Brun and F.~Rademakers, {\it {ROOT: An object oriented data analysis
  framework}},  {\em Nucl. Instrum. Meth.} {\bf A389} (1997) 81--86.

\bibitem{Read:2002hq}
A.~L. Read, {\it {Presentation of search results: The CL(s) technique}},  {\em
  J. Phys.} {\bf G28} (2002) 2693--2704. [,11(2002)].

\bibitem{Junk:1999kv}
T.~Junk, {\it {Confidence level computation for combining searches with small
  statistics}},  {\em Nucl. Instrum. Meth.} {\bf A434} (1999) 435--443,
  [\href{http://arxiv.org/abs/hep-ex/9902006}{{\tt hep-ex/9902006}}].

\bibitem{Dorsner:2018ynv}
I.~Doršner and A.~Greljo, {\it {Leptoquark toolbox for precision collider
  studies}},  {\em JHEP} {\bf 05} (2018) 126,
  [\href{http://arxiv.org/abs/1801.07641}{{\tt arXiv:1801.07641}}].

\bibitem{Buttazzo:2016kid}
D.~Buttazzo, A.~Greljo, G.~Isidori, and D.~Marzocca, {\it {Toward a coherent
  solution of diphoton and flavor anomalies}},  {\em JHEP} {\bf 08} (2016) 035,
  [\href{http://arxiv.org/abs/1604.03940}{{\tt arXiv:1604.03940}}].

\bibitem{Efrati:2015eaa}
A.~Efrati, A.~Falkowski, and Y.~Soreq, {\it {Electroweak constraints on
  flavorful effective theories}},  {\em JHEP} {\bf 07} (2015) 018,
  [\href{http://arxiv.org/abs/1503.07872}{{\tt arXiv:1503.07872}}].

\bibitem{Falkowski:2015jaa}
A.~Falkowski, M.~Gonzalez-Alonso, A.~Greljo, and D.~Marzocca, {\it {Global
  constraints on anomalous triple gauge couplings in effective field theory
  approach}},  {\em Phys. Rev. Lett.} {\bf 116} (2016), no.~1 011801,
  [\href{http://arxiv.org/abs/1508.00581}{{\tt arXiv:1508.00581}}].

\bibitem{Falkowski:2017pss}
A.~Falkowski, M.~Gonz{\'a}lez-Alonso, and K.~Mimouni, {\it {Compilation of
  low-energy constraints on 4-fermion operators in the SMEFT}},  {\em JHEP}
  {\bf 08} (2017) 123, [\href{http://arxiv.org/abs/1706.03783}{{\tt
  arXiv:1706.03783}}].

\bibitem{deBoer:2016dcg}
S.~de~Boer, B.~M{\"u}ller, and D.~Seidel, {\it {Higher-order Wilson
  coefficients for $c \to u$ transitions in the standard model}},  {\em JHEP}
  {\bf 08} (2016) 091, [\href{http://arxiv.org/abs/1606.05521}{{\tt
  arXiv:1606.05521}}].

\bibitem{Feldmann:2017izn}
T.~Feldmann, B.~Müller, and D.~Seidel, {\it {$D \to \rho \,\ell^+\ell^-$
  decays in the QCD factorization approach}},  {\em JHEP} {\bf 08} (2017) 105,
  [\href{http://arxiv.org/abs/1705.05891}{{\tt arXiv:1705.05891}}].

\bibitem{Alonso:2014csa}
R.~Alonso, B.~Grinstein, and J.~Martin~Camalich, {\it {$SU(2)\times U(1)$ gauge
  invariance and the shape of new physics in rare $B$ decays}},  {\em Phys.
  Rev. Lett.} {\bf 113} (2014) 241802,
  [\href{http://arxiv.org/abs/1407.7044}{{\tt arXiv:1407.7044}}].

\bibitem{Lees:2011hb}
{\bf BaBar} Collaboration, J.~Lees et~al., {\it {Searches for Rare or Forbidden
  Semileptonic Charm Decays}},  {\em Phys.\ Rev.\ D} {\bf 84} (2011) 072006,
  [\href{http://arxiv.org/abs/1107.4465}{{\tt arXiv:1107.4465}}].

\bibitem{CMS:2019tbu}
{\bf CMS} Collaboration, C.~Collaboration, {\it {Search for a narrow resonance
  in high-mass dilepton final states in proton-proton collisions using
  140$~\mathrm{fb}^{-1}$ of data at $\sqrt{s}=13~\mathrm{TeV}$}}, .

\bibitem{Aaboud:2017sjh}
{\bf ATLAS} Collaboration, M.~Aaboud et~al., {\it {Search for additional heavy
  neutral Higgs and gauge bosons in the ditau final state produced in 36
  fb$^{-1}$ of pp collisions at $ \sqrt{s}=13 $ TeV with the ATLAS detector}},
  {\em JHEP} {\bf 01} (2018) 055, [\href{http://arxiv.org/abs/1709.07242}{{\tt
  arXiv:1709.07242}}].

\bibitem{TheBaBar:2016xwe}
{\bf BaBar} Collaboration, J.~P. Lees et~al., {\it {Search for
  $B^{+}\rightarrow K^{+} \tau^{+}\tau^{-}$ at the BaBar experiment}},  {\em
  Phys. Rev. Lett.} {\bf 118} (2017), no.~3 031802,
  [\href{http://arxiv.org/abs/1605.09637}{{\tt arXiv:1605.09637}}].

\bibitem{Cornella:2020aoq}
C.~Cornella, G.~Isidori, M.~König, S.~Liechti, P.~Owen, and N.~Serra, {\it
  {Hunting for $B^+\to K^+ \tau^+\tau^-$ imprints on the $B^+ \to K^+
  \mu^+\mu^-$ dimuon spectrum}},  \href{http://arxiv.org/abs/2001.04470}{{\tt
  arXiv:2001.04470}}.

\bibitem{Asadi:2018wea}
P.~Asadi, M.~R. Buckley, and D.~Shih, {\it {It’s all right(-handed
  neutrinos): a new $W^\prime$ model for the $R_{D^{(*)}}$ anomaly}},  {\em
  JHEP} {\bf 09} (2018) 010, [\href{http://arxiv.org/abs/1804.04135}{{\tt
  arXiv:1804.04135}}].

\end{thebibliography}\endgroup

\end{document}